\pdfoutput=1
\documentclass[12pt]{article}
\usepackage[
top = 2.5cm, 
bottom = 2.5cm,  
left = 2.55cm,  
right = 2.55cm]{geometry}

\usepackage{subfigure}
\usepackage{caption}
\usepackage{latexsym} 
\usepackage{verbatim} 
\usepackage{tikz}
\usetikzlibrary{matrix}
\usepackage{color}
\usepackage{graphicx,amssymb,amsfonts,amsmath,amssymb,amscd,amstext, mathrsfs}
\usepackage{graphicx} 
\usepackage{dsfont}
\usepackage{xcolor}
\usepackage{amsmath}
\usepackage{empheq}
\usepackage{cite}

\numberwithin{equation}{section}

\newlength\dlf

\def\cO{\mathcal{O}}
\def\CT{\mathcal{T}}

\newcommand{\bw}{\begin{widetext}}
\newcommand{\ew}{\end{widetext}}
\newcommand{\bea}{\begin{eqnarray}}
\newcommand{\eea}{\end{eqnarray}}
\newcommand{\be}{\begin{equation}}
\newcommand{\ee}{\end{equation}}
\newcommand{\nn}{\nonumber}

\renewcommand{\bar}[1]{\overline{#1}}
\renewcommand{\tilde}[1]{\widetilde{#1}}
\renewcommand{\hat}[1]{\widehat{#1}}

\newcommand{\<}{\langle}
\renewcommand{\>}{\rangle}

\renewcommand{\cal}{\mathcal}
\newcommand{\CO}{\mathcal{O}}
\newcommand{\co}{\mathcal{O}}

\newcommand{\cn}{\mathcal{N}}

\newcommand{\CF}{\mathcal{F}}

\DeclareMathOperator{\csch}{csch}

\DeclareFontShape{OT1}{cmr}{mx}{n}{<->cmr10}{}
\newcommand{\titlefont}{\fontseries{mx}\selectfont}

\def\frac#1#2{{#1\over #2}}

\newcommand{\myRho}{{\mathfrak{r}}}

\usepackage{jheppub}

\begin{document}

\begin{titlepage}

\begin{flushright} 
\end{flushright}

\begin{center} 

\vspace{0.35cm}

{\fontsize{21.5pt}{0pt}{\titlefont 
Bootstrapping 2d $\phi^4$ Theory \\ with Hamiltonian Truncation Data 
}}

\vspace{1.6cm}  

{{Hongbin Chen$^1$, A. Liam Fitzpatrick$^1$,  Denis Karateev$^{2}$}}

\vspace{1cm} 

{{\it
$^1$Department of Physics, Boston University, 
Boston, MA  02215, USA
\\
\vspace{0.1cm}
$^2$Philippe Meyer Institute, Physics Department \'Ecole Normale Sup\'erieure (ENS), Universit\'e PSL24 rue Lhomond, F-75231 Paris, France
}}\\
\end{center}
\vspace{1.5cm}

{\noindent 
We combine the methods of Hamiltonian Truncation and the recently proposed generalisation of the S-matrix bootstrap that includes local operators to determine the two-particle scattering amplitude and the two-particle form factor of the stress tensor  at  $s>0$ in the 2d $\phi^4$ theory.  We use the form factor of the stress tensor at $s\le 0$ and its spectral density computed using  Lightcone Conformal Truncation (LCT), and inject them into the generalized S-matrix bootstrap set-up.  The obtained results for the scattering amplitude and the form factor are fully reliable only in the elastic regime. We independently construct the ``pure'' S-matrix bootstrap bounds (bootstrap without including matrix elements of local operators), and find that the sinh-Gordon model and its analytic continuation the ``staircase model'' saturate these bounds.  Surprisingly, the $\phi^4$ two-particle scattering amplitude also very nearly saturates these bounds, and moreover is extremely close to that of the sinh-Gordon/staircase model. 
}

\end{titlepage}

\tableofcontents

\section{Introduction} 
There is a set of powerful non-perturbative techniques to study quantum field theories (QFTs) commonly referred to as ``bootstrap'' methods. Such methods attempt to bound the space of QFTs using only basic principles such as symmetries, unitarity, crossing, etc. The most famous bootstrap technique is the numerical conformal bootstrap pioneered in \cite{Rattazzi:2008pe}. It allows one to derive precise bounds on the space of conformal field theories (CFTs), see \cite{Poland:2018epd} for a review. Another bootstrap technique which allows one to study QFTs with a mass gap was pioneered in \cite{Paulos:2016but,Paulos:2017fhb}. In this paper we will refer to it as the numerical $S$-matrix bootstrap. The $S$-matrix bootstrap gained further attention in recent years, see \cite{Paulos:2016fap,Doroud:2018szp,He:2018uxa,Cordova:2018uop,Guerrieri:2018uew,Homrich:2019cbt,EliasMiro:2019kyf,Cordova:2019lot,Bercini:2019vme,Correia:2020xtr,Bose:2020shm,Guerrieri:2020bto,Hebbar:2020ukp,He:2021eqn,Guerrieri:2021ivu,Miro:2021rof,Guerrieri:2020kcs,Guerrieri:2021tak}. The recent work \cite{Karateev:2019ymz,Karateev:2020axc} made a concrete proposal for how to extended the $S$-matrix bootstrap to accommodate form factors and spectral densities. We will refer to this approach as the numerical $S$-matrix/form factor bootstrap. 

A simultaneous advantage and disadvantage of  bootstrap methods is their model-independent nature. If one wants to study some particular model one generically has to inject additional model specific information. The amount of this additional information highly depends on the situation. For example one can solve numerically the 3d Ising model using the conformal bootstrap method by simply specifying that there are only two relevant operators in the spectrum, one is $Z_2$ even and one is $Z_2$ odd, see \cite{ElShowk:2012ht}.
Another example is the work \cite{Guerrieri:2018uew}, where the authors attempted to study the 4d QCD by using the $S$-matrix bootstrap injecting some known information from chiral perturbation theory. Another notable example in this spirit is the study of the 2d Ising Field Theory in \cite{Gabai:2019ryw}, where the authors injected the S-matrix of the theory in one kinematic regime to learn about its behavior more generally.

A great class of tools for obtaining non-perturbative results in a particular model are  Hamiltonian Truncation methods, which involve numerically diagonalizing the Hamiltonian in a finite dimensional subspace of the full Hilbert space. 
 This approach is a special case of more general variational methods,  so all else being equal the larger the truncation subspace, the more accurate the approximation to the eigenstates of the Hamiltonian.  There are various different ways one can try to implement Hamiltonian truncation for continuum QFT, the most well-known probably being the Truncated Conformal Space Approximation (TCSA) of Zamolodchikov and Yurov \cite{yurov1990truncated,yurov1991correlation}, see \cite{james2018non} for a recent review and guide to the literature. 
 One immediate output of such methods is the mass spectrum, which is just the set of eigenvalues of the Hamiltonian. Because one also obtains the eigenvectors of the Hamiltonian, one can compute spectral densities of local operators quite straightforwardly. By contrast, constructing multi-particle asymptotic states with Hamiltonian methods is much more subtle, since these are not just eigenstates of the Hamiltonian. Thus the computation of observables like the scattering amplitudes requires a more involved approach.

The main goal of this paper is to study non-perturbatively the 2d $\phi^4$ model (in the unbroken phase) and to compute as many observables as we can. In a companion paper \cite{truncffsd}, we used the Lightcone Conformal Truncation (LCT) method to compute the two-particle form factor of the stress tensor in the unphysical regime ($s\le 0$)  and its spectral density. In this paper, we will  inject this data into the S-matrix/form factor bootstrap program, and obtain the form factor at $s>0$ and also the elastic 2-to-2 scattering amplitude in the $\phi^4$ model.

The rest of the introduction is organized as follows. In section \ref{sec:models}, we discuss a wide class of scalar field theories in 2d, where we precisely define the $\phi^4$ model and discuss its relation with other models. In section \ref{sec:summary}, we provide an extended summary of our main results.

\subsection{Models in 2d}
\label{sec:models}
Let us consider the class of quantum field theories in 2d which consists of a single real scalar field $\phi(x)$ and is defined as the deformation of the free scalar field theory in the UV by a potential $V(\phi)$. The corresponding action reads
\begin{equation}
\label{eq:UV_theory}
S_{UV}=\int d^2x\left(-\frac{1}{2} (\partial\phi)^2-V(\phi)
\right).
\end{equation}
Notice that the field $\phi(x)$ has the mass dimension zero, $[\phi]=0$. This situation is special to 2d and allows for complicated potentials $V(\phi)$ not present in higher dimensions. In this paper we will further restrict our attention to potentials which are invariant under the following $Z_2$ transformation $\phi(x)\rightarrow -\phi(x)$.
The most generic potential then has the following form
\begin{equation}
\label{eq:potential_V}
V(\phi) = \frac{1}{2} m_0^2\phi^2+\sum_{n=2}^{\infty} g_{2n} \phi^{2n} +\text{counterterms}, 
\end{equation}
where $m_0$ is the mass-like parameter and $g_{2n}$ is an infinite set of coupling constants. We focus on the case when $m_0^2>0$ (unbroken phase).
Below we will define and discuss several potentials $V(\phi)$.  We will take the operators to be normal-ordered in order to remove divergences in the theory; this choice is equivalent to a hard cutoff with a particular choice for the counterterms above.

We start with two integrable models called the sine-Gordon and the sinh-Gordon models. They are given by the following potentials respectively
\begin{align}
\label{eq:sine-Gordon}
V_\text{sine-Gordon}(\phi) &\equiv -m_0^2\beta^{-2}\left(\cos(\beta \phi)-1\right)+\text{counterterms},\\
\label{eq:sinh-Gordon}
V_\text{sinh-Gordon}(\phi) &\equiv +m_0^2\beta^{-2}\left(\cosh(\beta \phi)-1\right)+\text{counterterms}.
\end{align}
Here $\beta$ is the single dimensionless parameter which specifies the models. Expanding these potentials around $\phi=0$ one can bring them to the form given by \eqref{eq:potential_V}, and thus express all the $g_{2n}$ coefficients in terms of $\beta$.  The two models are formally related by the replacement $\beta \leftrightarrow i \beta$.
These two models have been extensively studied in the literature. For a summary of the sine-Gordon results see, for example, section 4.1 in \cite{Karateev:2019ymz} and references therein. We will summarize the results for the sinh-Gordon model in section \ref{sec:sinh_gordon}.

Another interesting model is the $\phi^4$ model. It will play the central role in this paper. It is defined by the potential \eqref{eq:potential_V} with $g_4=\lambda/4!$ and $g_{2n}=0$ for all $n\geq 3$. Here  $\lambda\geq 0$ is the quartic coupling constant. This is possibly the simplest quantum field theory model one can think of. Let us write out its potential explicitly, it reads
\begin{equation}
\label{eq:phi4_definition}
V_{\phi^4} \equiv \frac{1}{2}m_0^2\,\phi^2+\frac{\lambda}{4!}\phi^4
+\frac{1}{2}\delta_m\phi^2.
\end{equation}
No counterterms are required for the coupling constant $\lambda$ in $d=2$.  Normal-ordering the interaction and setting $\delta_m=0$ is equivalent to choosing a hard cutoff $\Lambda_{\rm cutoff}$ and setting $\delta_m = - \frac{\lambda}{8 \pi} \log \Lambda_{\rm cutoff}^2/m_0^2$. 
The quartic coupling $\lambda$ has mass dimension $[\lambda]=2$; we define the dimensionless quartic coupling $\bar \lambda$  as\footnote{We caution the reader that this convention for $\bar{\lambda}$ differs from that in \cite{Anand:2020gnn};  $\bar \lambda_\text{here} = 4\pi \bar \lambda_\text{there}$.}
\begin{equation}
\label{eq:lambda_bar}
\bar \lambda \equiv m_0^{-2}\lambda.
\end{equation}
The $\phi^4$ model is non-integrable, and one needs numerical non-perturbative techniques in order to compute observables in this theory.\footnote{See e.g. \cite{Chabysheva:2015ynr,Burkardt:2016ffk,Schaich:2009jk,Milsted:2013rxa,Bosetti:2015lsa,Rychkov:2014eea,Rychkov:2015vap,Bajnok:2015bgw,Elliott:2014fsa,Chabysheva:2016ehd,Serone:2018gjo,Tilloy:2021hhb} for  various recent nonperturbative works on this model.} It was shown in \cite{Anand:2017yij} that the $\phi^4$ model in lightcone quantization\footnote{It is important to note that, due to the contribution from zero modes, the critical value of the coupling differs in equal-time and lightcone quantization.  See \cite{Burkardt,Burkardt2,Fitzpatrick:2018xlz} for details.} in the unbroken phase is in the following range
\begin{equation} 
\bar \lambda \in [0,\,23.1]. 
\end{equation}
The critical value $\bar \lambda \approx 23.1$ leads to the conformal IR fixed point given by the free massless Majorana fermion (which is the 2d Ising model).

Finally, we consider the 2d $O(N)$ model, which is the case when instead of a single field $\phi$, we have $N$ fields $\phi_1$, $\phi_2$, $\ldots$, $\phi_N$ with the same mass. Requiring the $O(N)$ symmetry we can write the following analogue of the pure $\phi^4$ theory
\begin{equation}
V_{\phi^4}^{O(N)}(\phi) \equiv \frac{1}{2}m_0^2 (\phi_i\phi_i )+ \frac{\lambda}{8 N} (\phi_i\phi_i) (\phi_j\phi_j)+\frac{1}{2}\delta_m(\phi_i\phi_i),
\end{equation}
where there is an implicit summation over the repeated indices. In the large $N$ limit when $N\rightarrow \infty$ the model becomes integrable. We will mostly use it in this paper to check our numerical procedures.

To conclude our brief discussion of 2d models let us clarify an important point. One can consider an infinite class of potentials \eqref{eq:potential_V} with $g_4=\lambda/4!$ and
\begin{equation}
m_0^{-2}g_{2n} \ll 1,\qquad n\geq 3.
\end{equation}
All such models will lead to observables very similar to the pure $\phi^4$ model \eqref{eq:phi4_definition}. Using non-perturbative techniques we can compute in practice observables only at finite precision and thus we will never be able to distinguish the pure $\phi^4$ model from this infinite class of models. In order to be pedantic we say that we compute observables for $\phi^4$-like theories. In this sense, the sinh-Gordon model belongs to the class of $\phi^4$-like models if $\beta^2=\bar\lambda$ at very small values of the coupling, $\bar\lambda\ll4\pi$.

\subsection{Summary of Main Results}
\label{sec:summary}

Given a model there are various observables one would like to compute. In this paper we will focus on three different observables: the two-particle form factor of the trace of the stress tensor  $\mathcal{F}_{2,0}^\Theta(s)$, the spectral density of the trace of the stress tensor  $\rho_\Theta(s)$ and the two-to-two scattering amplitude $\mathcal{S}(s)$. For their precise definitions see section \ref{sec:definitions}. Given the definition of the $\phi^4$ model in \eqref{eq:phi4_definition}, one would like to obtain the above observables in terms of the bare coupling $\bar\lambda$ and the mass-like parameter $m_0$ which simply sets the scale.\footnote{In practice we provide all our final expressions in terms of the physical mass $m$ which can also be computed in terms of $m_0$ for a given value of  $\bar\lambda$.} In the $\bar\lambda\ll4\pi$ regime,  one can use perturbation theory to do that.  For $\bar\lambda\gtrsim 4\pi$, one can use Hamiltonian Truncation methods instead. In the companion paper \cite{truncffsd}, we have computed the spectral density of the trace of the stress tensor  $\rho_\Theta(s)$ and  two-particle form factor $\mathcal{F}_{2,0}^\Theta(s)$ at $s\le 0$ for various values of $\bar\lambda$, see figure 4 and 12 therein.
The main goal of this paper is to compute $\mathcal{F}_{2,0}^\Theta(s)$ for $s>0$ and the scattering amplitude $\mathcal{S}(s)$ given the input of \cite{truncffsd}. Below we outline the main results.

We start by employing the pure $S$-matrix bootstrap to study the space of scattering amplitudes of a single $Z_2$ odd particle with the physical mass $m$. One can characterize such amplitudes for example by their value (and the value of their derivatives) at the crossing symmetric point $s=2m^2$. See \eqref{eq:Lambda_definition} and \eqref{eq:Lambda_derivative_definition} for details. Using crossing, analyticity and unitarity we construct a non-perturbative bound on a two-dimensional subspace of these parameters. The bound is presented in figure \ref{fig:bound}. We discover that the left tip describes the scattering of free bosons and the right tip describes the scattering of free Majorana fermions (i.e. the 2d Ising model). The two tips are connected by the lower and upper edges. The lower edge is saturated by the sinh-Gordon model and its analytic continuation the ``staircase model''.

We then propose a strategy which allows one to inject the Hamiltonian Truncation data of \cite{truncffsd} into the $S$-matrix/form factor bootstrap. This allows one to isolate a specific theory, instead of constructing generic bounds on the space of allowed theories. Using this strategy we numerically obtain  the form factor of the trace of the stress tensor  $\mathcal{F}_{2,0}^\Theta(s)$ with $s> 0$ and the scattering amplitude $\mathcal{S}(s)$. The results are presented in figures \ref{fig:T_re} - \ref{fig:FF_im} for various values of $\bar\lambda$. In the regime $\bar\lambda\ll4\pi$ they agree with perturbative expressions. Due to the limitations of the $S$-matrix/form factor bootstrap restricted to two-particle scattering states, we expect that our results are fully accurate only in the ``elastic'' regime for $s\leq 16m^2$. Given the numerical scattering amplitudes in the $\phi^4$ model, we can determine the position of the $\phi^4$ model with respect to the generic bound given in figure \ref{fig:bound}. It turns out that the $\phi^4$ model lies very close to the lower edge of this plot, see figure \ref{fig:boundWithData}. This means that the $\phi^4$ model is very similar to the sinh-Gordon/staircase model (which is exactly on the lower edge) if one only looks at the two-dimension subspace shown in this plot.  This fact calls for further investigation which we summarize in the next paragraph. Leaving this issue aside for a moment we observe that the $\phi^4$ model starts close to the free boson theory for small values of $\bar\lambda$ and monotonically moves towards the free fermion theory (2d Ising) along the lower edge when we increase $\bar\lambda$. This behaviour is in agreement with the fact that there is a critical value of $\bar\lambda$ when the $\phi^4$ theory flows to the 2d Ising fixed point. 

 The $\phi^4$ and the sinh-Gordon models are inherently  different. The former has particle production and the latter does not. In practice this difference will become evident for example if we look at the scattering amplitude $\mathcal{S}(s)$ in the ``non-elastic'' regime $s\geq 16m^2$. For small values of $\bar\lambda\ll4\pi$, the $\phi^4$ model is expected to be very similar to the sinh-Gordon model with $\beta^2=\bar\lambda$. To our surprise we have discovered that at strong coupling, the $\phi^4$ model still gives very similar observables in the ``elastic'' regime as the sinh-Gordon model with some value $\beta_*^2$. Notice however that $\beta_*^2\neq \bar\lambda$. For large values of $\bar\lambda$, the value of $\beta_*^2$ is allowed to become complex (describing the ``staircase model''). The comparison of the observables in the $\phi^4$ model and in the sinh-Gordon model is given in figure \ref{fig:SGandPhi4Comparison_1}, \ref{fig:SGandPhi4comparison_2} and \ref{fig:phi4VsSGTPlots}. There, one sees a striking similarity of the two models in the ``elastic'' regime and their small deviation in the non-elastic regime.

\paragraph{Outline of the paper}
We summarize basic definitions and set up the notation in section \ref{sec:definitions}. We summarize various analytic results in section \ref{sec:analytics}. For instance we discuss the sinh-Gordon model and its analytic continuation (the ``staircase model''),  the $\phi^4$ model in perturbation theory, and the 2d $O(N)$ in the large $N$ limit. In section \ref{sec:s-matrix_pure}, we construct a generic bound on the space of scattering amplitudes of $Z_2$ odd particles. In section \ref{sec:s-matrix}, we show how one can inject the LCT data into the $S$-matrix/form factor bootstrap and apply this strategy to the 2d $\phi^4$ model.
We discuss open questions and further directions in section \ref{sec:discussion}.

Some supplementary material is provided in appendices. We review the details of the 2d kinematics in appendix \ref{app:kinematics_2d}. We discuss the 2d $O(N)$ models and their large $N$ limit in appendix \ref{app:ON}. We provide details of perturbative and large $N$ computations in the $\phi^4$ model and the $O(N)$ model in appendix \ref{app:analytic}. Finally, in appendix \ref{app:sinh-gordonFF}, we discuss two- and four-particle form factors in the sinh-Gordon model.

\section{Basic Definitions and Notation}
\label{sec:definitions}

Let us start by carefully defining the most important objects for our work. We will first work with scalars and general number of dimensions $d$, and focus on $d=2$ in the second half of this section. We also define the scattering amplitudes for 2d Majorana fermions (which is used in later sections) at the end of this section. We will use the ``mostly plus'' Lorentzian metric
\begin{equation}
\eta^{\mu\nu} = \{-1,+1,+1,\ldots\}.
\end{equation}

We require that our quantum field theory contains the local stress tensor $T^{\mu\nu}(x)$ which obeys the following conditions
\begin{equation}
\label{eq:stress-tensor}
T^{\mu\nu}(x) = T^{\nu\mu}(x),\qquad
\partial_\mu T^{\mu\nu}(x) = 0. 
\end{equation}
We denote its trace by 
\begin{equation}
\label{eq:trace}
\Theta(x) \equiv \eta_{\mu\nu} T^{\mu\nu}(x).
\end{equation}
One of the simplest observables of any theory is  two-point function of the trace of the stress tensor. One can distinguish the Wightman two-point function
\begin{equation}
\< {\rm vac} |\Theta(x_1)\Theta(x_2)|{\rm vac}\>_W \equiv \lim_{\epsilon\rightarrow 0^+} \< {\rm vac} |\Theta(x_1^0-i\epsilon, \vec x_1)\Theta(x_2)|{\rm vac}\>
\end{equation}
and the time-ordered two-point function
\begin{equation}
\< {\rm vac} |\Theta(x_1)\Theta(x_2)|{\rm vac}\>_T \equiv
\theta(x_1^0-x_2^0 ) \< {\rm vac} |\Theta(x_1)\Theta(x_2)|{\rm vac}\>_W +
\theta(x_2^0-x_1^0 ) \< {\rm vac} |\Theta(x_2)\Theta(x_1)|{\rm vac}\>_W.
\end{equation}
Let us introduce the spectral density of the trace of the stress tensor $\rho_\Theta$ as the Fourier transformed Wightman two-point function. In the notation of \cite{Weinberg:1995mt} we have
\begin{equation}
2\pi\theta(p^0)\rho_\Theta(-p^2) \equiv \int d^dx e^{-i p\cdot x}
\< {\rm vac} |\Theta(x)\Theta(0)|{\rm vac}\>_W.
\end{equation}
One can express $\rho_\Theta$ also in terms of the time-ordered two-point function as follows\footnote{For a simple derivation see for example the beginning of section 3.2 in \cite{Karateev:2020axc}.}
\begin{equation}
\label{eq:SD_TimeOrderedCorrelator}
2\pi\theta(p^0)\rho_\Theta(-p^2) = 2\,\text{Re}
\int d^dx e^{-i p\cdot x}
\< {\rm vac} |\Theta(x)\Theta(0)|{\rm vac}\>_T.
\end{equation}

In quantum field theories with a ``mass gap'' there exist one-particle asymptotic in and out states denoted by
\begin{equation}
\label{eq:1PS}
|m,\vec p\,\>_\text{in}\qquad|m,\vec p\,\>_\text{out},
\end{equation}
where $m$ and $\vec p$ stand for the physical mass and the $(d-1)$-momentum of the asymptotic particles.
The standard normalization choice for them reads as
\begin{equation}
\label{eq:normalization_1PS}
{}_\text{in}\<m,\vec p_1\,|m,\vec p_2\,\>_\text{in} = 
{}_\text{out}\<m,\vec p_1\,|m,\vec p_2\,\>_\text{out} = 
2\sqrt{m^2+\vec p_1^{\,2}}\times (2\pi)^{d-1}\delta^{d-1}(\vec p_1-\vec p_2).
\end{equation}
Using the one-particle asymptotic states one can construct general $n$-particle asymptotic states with $n\geq 2$, for details see for example section 2.1.2 in \cite{Karateev:2019ymz}. Let us denote the two-particle in and out asymptotic states by 
\begin{equation}
\label{eq:2PS}
|m,\vec p_1;m,\vec p_2\,\>_\text{in}
\quad\text{and}\quad
|m,\vec p_1;m,\vec p_2\,\>_\text{out}.
\end{equation}
They are constructed in such a way that they obey the following normalization
\begin{multline}
\label{eq:normalization_2PS}
{}_\text{in}\<m,\vec k_1;m,\vec k_2|m,\vec p_1;m,\vec p_2\,\>_\text{in} =
{}_\text{out}\<m,\vec k_1;m,\vec k_2|m,\vec p_1;m,\vec p_2\,\>_\text{out} =\\
4\sqrt{m^2+\vec p_1^{\,2}}\sqrt{m^2+\vec p_2^{\,2}}\, (2\pi)^{2(d-1)}\delta^{(d-1)}(\vec p_1-\vec k_1)\delta^{(d-1)}(\vec p_2-\vec k_2)+
(\vec p_1 \leftrightarrow \vec p_2).
\end{multline}
Using the asymptotic states one can define another set of observables called the form factors. In this work we will use the following form factors of the trace of the stress tensor
\begin{equation}
\label{eq:form_factor}
\begin{aligned}
\mathcal{F}^\Theta_{1,1}(t)&\equiv
{}_\text{out}\<m,\vec p_1\,|\Theta(0)|m,\vec p_2\>_\text{in},\\
\mathcal{F}^\Theta_{2,0}(s)&\equiv
{}_\text{out}\<m,\vec p_1\,;m,\vec p_2\,|\Theta(0)|{\rm vac}\>.
\end{aligned}
\end{equation}
Here we have introduced the analogues of the Mandelstam variables for the form factors which are defined as
\begin{equation}
\label{eq:mandelstam_analogue}
s \equiv - (p_1+p_2)^2,\qquad
t \equiv - (p_1-p_2)^2,\qquad s+t=4m^2.
\end{equation}
The latter relation simply follows from the definitions of $s$ and $t$.
The two form factors in \eqref{eq:form_factor} are related by the crossing symmetry as follows\footnote{The crossing symmetry is the condition that the matrix elements in \eqref{eq:form_factor} remain invariant under the following change of the $d$-momenta $p^\mu_{2,\text{out}} \rightarrow -p^\mu_{2,\text{in}}$.}
\begin{equation}
\label{eq:relation_FF}
\mathcal{F}^\Theta_{2,0}(s) = \mathcal{F}^\Theta_{1,1}(s).
\end{equation}
The Ward identity imposes the following normalization condition
\begin{equation}
\label{eq:mass_definition}
\lim_{s\rightarrow 0} \mathcal{F}^\Theta_{2,0}(s) = -2m^2.
\end{equation}
See for example appendix G in \cite{Karateev:2020axc} for its derivation. The condition \eqref{eq:mass_definition} can be seen as the definition of the physical mass.

The last observable in which we are interested is the scattering amplitude $\mathcal{S}(s)$. We define it via the following matrix element
\begin{equation}
\label{eq:amplitude_S}
\mathcal{S}(s,t) \times 
(2\pi)^{(d-1)}\delta^{(d)}(p_1+p_2-k_1-k_2)\equiv
{}_\text{out}\<m,\vec k_1;m,\vec k_2|m,\vec p_1;m,\vec p_2\,\>_\text{in}.
\end{equation}
In this case,  the usual Mandelstam variables are defined as
\begin{equation}
\label{eq:mandelstam}
s\equiv -(p_1+p_2)^2,\quad
t\equiv -(p_1-k_1)^2,\quad
u\equiv -(p_1-k_2)^2,\quad
s+t+u=4m^2.
\end{equation}
The difference between \eqref{eq:mandelstam_analogue} and \eqref{eq:mandelstam} should be understood from the context. Instead of using the full scattering amplitude $\mathcal{S}(s)$, it is often very convenient to define the interacting part of the scattering amplitude $\mathcal{T}(s,t)$ as follows
\begin{multline}
\label{eq:amplitude_T} 
i\mathcal{T}(s,t) \times 
(2\pi)^{(d-1)}\delta^{(d)}(p_1+p_2-k_1-k_2)\equiv\\
{}_\text{out}\<m,\vec k_1;m,\vec k_2|m,\vec p_1;m,\vec p_2\,\>_\text{in}-{}_\text{out}\<m,\vec k_1;m,\vec k_2|m,\vec p_1;m,\vec p_2\,\>_\text{out}.
\end{multline}

We have defined the observables in general dimensions up to now. In the rest of this section, we focus on $d=2$. In the special case of $d=2$, the scattering amplitude takes a particularly simple form since it depends only on the single Mandelstam variable $s$. In our convention, $u=0$. This restriction is imposed via the Heaviside step function $\theta$ (not to confuse with the rapidity $\theta$ that is used in later sections) in the equations below. We provide details of 2d kinematics in appendix \ref{app:kinematics_2d}.\footnote{See also the end of section 2 in \cite{Zamolodchikov:1978xm} (in particular equations (2.9) - (2.11) and the surrounding discussion).}
In $d=2$ we define the scattering amplitude of identical particles as\footnote{Note that although we use the vector notation $\vec{p}$ for the momentum, it is really just a single number, since are are in 2d, and the step functions make sense.}${}^,$\footnote{In the left-hand of this equation and all similar equations below it is understood that the scattering amplitude implicitly contains the appropriate step functions. This is is because all our amplitudes are required to have $u=0$.}
\begin{multline}
\label{eq:amplitude_S_2d}
\mathcal{S}(s) \times 
(2\pi)\delta^{(2)}(p_1+p_2-k_1-k_2)\equiv\\
{}_\text{out}\<m,\vec k_1;m,\vec k_2|m,\vec p_1;m,\vec p_2\,\>_\text{in}\times \theta(\vec p_1-\vec p_2) \theta(\vec k_2-\vec k_1).
\end{multline}
The interacting part of the scattering amplitude in $d=2$ is then defined as
\begin{multline}
\label{eq:amplitude_S_2d}
\mathcal{T}(s) \times 
(2\pi)\delta^{(2)}(p_1+p_2-k_1-k_2)\equiv
\Big(
{}_\text{out}\<m,\vec k_1;m,\vec k_2|m,\vec p_1;m,\vec p_2\,\>_\text{in}\\
-{}_\text{out}\<m,\vec k_1;m,\vec k_2|m,\vec p_1;m,\vec p_2\,\>_\text{out}
\Big)\times \theta(\vec p_1-\vec p_2) \theta(\vec k_2-\vec k_1).
\end{multline}
In $d=2$, it is straightforward to rewrite \eqref{eq:normalization_2PS} in the following way
\begin{equation}
\label{eq:normalization_2PS_2d}
{}_\text{out}\langle m,\vec k_1;m,\vec k_2|m,\vec p_1;m,\vec p_2\>_\text{out}
=\mathcal{N}_2\times(2\pi)^{2}\delta^{(2)}(p_1+p_2-k_1-k_2),
\end{equation}
where we have defined
\begin{equation}
\mathcal{N}_2\equiv 2\sqrt{s}\sqrt{s-4m^2}.
\end{equation}
Combining \eqref{eq:amplitude_S}, \eqref{eq:amplitude_T} and \eqref{eq:normalization_2PS_2d}, we obtain the following simple relation between the full amplitude and its interacting part
\begin{equation}
\mathcal{S}(s)=\mathcal{N}_2+
i\mathcal{T}(s).
\end{equation}
It is also convenient to introduce the following amplitude (which can be seen as the analogue of the partial amplitudes in higher dimensions)
\begin{equation}
\label{eq:Shat}
\widehat{\mathcal{S}}(s)\equiv\mathcal{N}_2^{-1}\mathcal{S}(s) 
=1+i\,\mathcal{N}_2^{-1}\mathcal{T}(s).
\end{equation}
Unitarity imposes the following constraint:
\begin{equation}
\label{eq:unitarity_Shat}
\left| \widehat{\mathcal{S}}(s) \right|^2 \leq 1,\quad \text{for } s\geq 4m^2.
\end{equation}

One can define the non-perturbative quartic coupling $\Lambda$ via the interacting part of the physical amplitude as
 \begin{equation}
\label{eq:Lambda_definition}
\Lambda \equiv -\lim_{s\rightarrow 2m^2}\mathcal{T}(s).
\end{equation}
We can also define the following set of non-perturbative parameters
\begin{equation}
\label{eq:Lambda_derivative_definition}
\Lambda^{(n)}\equiv \lim_{s\rightarrow 2m^2}\partial_{s}^{n}\mathcal{T}(s).
\end{equation}
Note that there is no minus sign in the definition of $\Lambda^{(n)}$, which we find to be convenient. 
Due to the crossing symmetry $s \leftrightarrow 4m^2-s$, all the odd derivatives in $s$ at the crossing symmetric point vanish.
The infinite set of physical non-perturbative parameters $\Lambda$,  $\Lambda^{(2)}$, $\Lambda^{(4)}$, $\ldots$ can be chosen to fully describe any scattering process.

According to  \cite{Zamolodchikov:1986gt,Cardy:1988tj}, in $d=2$, one can define the $C$-function as
\begin{equation}
\label{eq:C-function}
C(s_0) \equiv 12\pi\int_{0}^{s_0} ds\, \frac{\rho_\Theta(s)}{s^2}.
\end{equation}
The UV central charge $c_{UV}$ is related to the $C$-function in the following simple way\footnote{Here we use the standard conventions for the central charge $c_{UV}$ in which the theory of a free scalar boson has $c_{UV}=1$. For a summary of the standard conventions see for example the end of appendix A in \cite{Karateev:2020axc}.}$^,$\footnote{For the derivation and further discussion see also \cite{Cappelli:1990yc} and section 5 of \cite{Karateev:2020axc}.}
\begin{equation}
\label{eq:central_charge}
c_{UV} = C(\infty).
\end{equation}
The full spectral density can be written as a sum as follows
\begin{equation}
\label{eq:SD_FF}
\rho_\Theta(s) = 
\rho^{(2)}_\Theta(s) \theta(s-4m^2)+
\rho^{(4)}_\Theta(s) \theta(s-16m^2)+
\rho^{(6)}_\Theta(s) \theta(s-36m^2)+\ldots
\end{equation}
Here the superscript $(2)$, $(4)$, $(6)$ stand for 2-, 4- and 6-particle part of the spectral density and $\ldots$ represent higher-particle contributions. Note that $\forall n$, $\rho^n_\Theta(s)\geq0$. In writing \eqref{eq:SD_FF} we assumed the absence of odd-number particle states due to the $Z_2$ symmetry. The two-particle part of the spectral density is related to the two-particle form factor as
\begin{equation}
\label{eq:SD_FF_rel}
\rho^{(2)}_\Theta(s) =
(2\pi \mathcal{N}_2)^{-1}|\mathcal{F}^\Theta_{2,0}(s)|^2.
\end{equation}
In the ``elastic'' regime $s\in[4m^2,16m^2]$ we also have  Watson's equation which reads
\begin{equation}
\label{eq:Watson}
\widehat{\mathcal{S}}(s) =\frac{\mathcal{F}^\Theta_{2,0}(s)}{\mathcal{F}^*{}^\Theta_{2,0}(s)},\qquad \text{for } s\in[4m^2,16m^2].
\end{equation}
For the derivation of these relations and their analogues  in higher dimensions see \cite{Karateev:2019ymz,Karateev:2020axc}.

\paragraph{Majorana fermions}
Consider the case of a single Majorana fermion in 2d with mass $m$. Analogously to the bosonic case we can construct the two-particle fermion states \eqref{eq:2PS}. However, now the two particle state must be anti-symmetric under the exchange of the two fermions, thus instead of the normalization condition \eqref{eq:normalization_2PS} we get
\begin{align}
\nn
{}_\text{free fermions}&\<m,\vec k_1;m,\vec k_2|m,\vec p_1;m,\vec p_2\,\>_\text{free fermions}\\
\label{eq:normalization_2PS_fermions}
&=4\sqrt{m^2+\vec p_1^{\,2}}\sqrt{m^2+\vec p_2^{\,2}}\, (2\pi)^{2}\delta^{(1)}(\vec p_1-\vec k_1)\delta^{(1)}(\vec p_2-\vec k_2)-
(\vec p_1 \leftrightarrow \vec p_2).
\end{align}

The scattering amplitude for the Majorana fermion reads as
\begin{multline}
\label{eq:Maj_amp}
\mathcal{S}_\text{fermions}(s) \times (2\pi)^{2}\delta^{(2)}(p_1+p_2-p_3-p_4)\equiv \\
{}_\text{out fermions}\<m,\vec k_1;m,\vec k_2|m,\vec p_1;m,\vec p_2\,\>_\text{in fermions}
\times \theta(\vec p_1-\vec p_2) \theta(\vec k_2-\vec k_1).
\end{multline}
In the case of free fermions the scattering amplitude is simply given by the normalization condition \eqref{eq:normalization_2PS_fermions}. Due to the presence of theta functions only the second term in \eqref{eq:normalization_2PS_fermions} contributes and using the change of variables, one gets
\begin{equation}
\label{eq:free_fermion}
\mathcal{S}_\text{free fermions}(s) = -\mathcal{N}_2,\qquad
\widehat{\mathcal{S}}_\text{free fermions}(s) = -1,
\end{equation}
where we use the hatted amplitude for fermions is defined as in \eqref{eq:Shat}. Analogously to \eqref{eq:Shat} we can extract the interacting part of the fermion scattering
\begin{equation}
\label{eq:scattering_fermions_general}
\widehat{\mathcal{S}}_\text{fermions}(s) = -1+i\mathcal{N}_2^{-1}\mathcal{T}_\text{fermions}(s).
\end{equation}

We finally notice that the scattering amplitude for free Majorana fermions is equivalent to the scattering of bosons with the following interacting part
\begin{equation}
\label{eq:bosons_fermions}
\mathcal{T}_\text{bosons}(s) = 2i \mathcal{N}_2.
\end{equation}
This can be seen by simply plugging \eqref{eq:bosons_fermions} into \eqref{eq:Shat}. One trivially recovers \eqref{eq:free_fermion}.

\section{Analytic Results} 
\label{sec:analytics}

In this section we provide analytic results for the sinh-Gordon, $\phi^4$ and 2d $O(N)$  models defined in section \ref{sec:models}.
The main objects we would like to compute are the $2\rightarrow 2$ scattering amplitudes, the two-particle form factor of the trace of the stress-tensor and its spectral density defined in section \ref{sec:definitions}.
In $d=2$ all these observables are functions of a single variable $s$. It will be often more convenient to use the rapidity variable $\theta$ related to the $s$ variable by 
\begin{equation}
\label{eq:rapidity}
\theta \equiv 2 \text{ArcCosh}\left(\frac{\sqrt{s}}{2m}\right)
\quad\Leftrightarrow\quad
s= 4 m^2\cosh^2 \left(\frac{\theta}{2}\right).
\end{equation}
Under crossing, we have $s\rightarrow 4m^2-s$, which corresponds to $\theta\rightarrow i\pi-\theta$.

\subsection{Sinh-Gordon Model} 
\label{sec:sinh_gordon}
 
We have defined the sinh-Gordon model in \eqref{eq:sinh-Gordon}. In what follows we will review its scattering amplitude and the stress-tensor form factor.\footnote{For a recent extensive study of the sinh-Gordon model see \cite{Konik:2020gdi}.} For aesthetic purposes let us define the following parameter
\begin{equation}
\label{eq:b}
b \equiv \frac{\beta}{\sqrt{8\pi}}.
\end{equation}
 
The spectrum of the sinh-Gordon model consists of a single $Z_2$ odd particle with mass $m$. 
The $2\rightarrow 2$ scattering amplitude was found in \cite{Arinshtein:1979pb}. It reads\footnote{Under analytic continuation this amplitude maps to the scattering of the lightest breathers in the sine-Gordon model. See for example equation (4.18) in \cite{Karateev:2019ymz}. Notice however the slight clash of notation, namely  $\gamma_\text{here}$ is equivalent to $\gamma_\text{there}/8$.}
\begin{equation}
\label{eq:amp_sinhGordon}
\widehat{\mathcal{S}}(\theta) = \frac{\sinh\theta-i\sin\gamma}{\sinh\theta+i\sin\gamma},\qquad
\gamma \equiv \frac{\pi b^2}{1+b^2},
\end{equation}
which is crossing symmetric, since $\sinh\theta$ is invariant when $\theta\rightarrow i\pi-\theta$.
It also possesses the following non-trivial symmetry $b \leftrightarrow b^{-1}$.
We can thus restrict our attention on the following parameter range
\begin{equation}
\label{eq:ranges}
b \in [0,1] \quad\Leftrightarrow\quad \gamma\in[0, \pi/2].
\end{equation}
Using the definition of the non-perturbative quartic coupling \eqref{eq:Lambda_definition}, we conclude that
\begin{equation}
\label{eq:sinh-Gordon_Lambda_gamma}
\Lambda = 8m^2\left(1-\frac{1}{1+\sin\gamma}\right). 
\end{equation}
Due to \eqref{eq:ranges}, the non-perturbative quartic coupling $\Lambda$ in the sinh-Gordon model has the following range
\begin{equation}
\label{eq:range_sinh-Gordon}
\Lambda\in[0, 4m^2].
\end{equation}
One can use the relation \eqref{eq:sinh-Gordon_Lambda_gamma} to eliminate the parameter $\gamma$ and rewrite the scattering amplitude \eqref{eq:amp_sinhGordon} as
\begin{equation}
\label{eq:amp_sinhGordon_rewritten}
\widehat{\mathcal{S}}(\theta) = 1 +
\frac{2i\Lambda}{(\Lambda-8m^2)\sinh\theta-i\Lambda}.
\end{equation}

The form factor of a scalar local operator in the sinh-Gordon model was computed in \cite{Fring:1992pt}. Adjusting the normalization of their result according to \eqref{eq:mass_definition}, we can write the following expression for 2-particle form factor for the trace of the stress-tensor 
\begin{equation}
	\label{eq:form_factor_sinh-Gordon}
\mathcal{F}^\Theta_{2,0}(\theta) = -2m^2 \times\\\exp\left(
8\,\int_0^\infty \frac{dx}{x}
\frac{\sinh\left(\frac{x\gamma}{2\pi}\right) \sinh\left(\frac{x(\pi-\gamma)}{2\pi}\right) \sinh\left( \frac{x}{2}\right)}
{\sinh^2(x)}
\sin^2\left(\frac{x(i \pi -\theta)}{2\pi}\right)
\right).
\end{equation}
In section \ref{sec:s-matrix}, in order to compare the spectral densities of the sinh-Gordon model and the $\phi^4$ model above $s=16m^2$, we will also need the 4-particle form factor for $\Theta$, which we review in appendix \ref{app:sinh-gordonFF}.

Let us now notice that the actual expressions for the scattering amplitude  \eqref{eq:amp_sinhGordon} and for the form factor \eqref{eq:form_factor_sinh-Gordon} are analytic functions of the parameter $\gamma$. They can be thus analytically continued away from the original range of $\gamma$ given by \eqref{eq:ranges}. The resulting amplitude and the form factor are the ones of the so-called staircase model,  which we review next.

\subsection{Staircase Model}
\label{sec:staircase}
 
 Because of the strong-weak duality $b \leftrightarrow b^{-1}$ in the sinh-Gordon model, it is effectively impossible to increase the coupling beyond $b =1$ and as a result $\Lambda$ is restricted to be $\le 4m^2$.  However, by analytically continuing the coupling to complex values, it is formally possible to obtain larger values of $\Lambda$.  The Staircase Model \cite{zamolodchikov2006resonance} is the analytic continuation
 \begin{equation}
 \label{eq:ac}
\gamma = \frac{\pi}{2} + i \theta_0
\end{equation}
with $\theta_0$ real.  Although the Lagrangian is no longer real and it is not clear why such a deformation should correspond to an underlying unitary theory,\footnote{It would be interesting to interpret the theory as a ``Complex CFT'' along the lines of \cite{Gorbenko:2018ncu,Gorbenko:2018dtm}.  In particular, the large amount of RG time that the Staircase Model spends near each minimal model suggests a form of ``walking'' near the minimal model fixed points.} in \cite{zamolodchikov2006resonance}  Zamolodchikov showed that the $c$-function of the theory, defined using the thermodynamic Bethe Ansatz, flows from a free scalar in the UV to the Ising model in the IR and moreover approaches very close to each of the $c<1$ minimal models along this RG flow.  The amount of RG time spent near each minimal model is proportional to $\theta_0$, so that at large $\theta_0$ the $c$-function resembles a staircase.  

Substituting equation (\ref{eq:ac}) into (\ref{eq:sinh-Gordon_Lambda_gamma}), the $\sin \gamma$ in the denominator becomes $\cosh\theta_0$, and now the maximum value that $\Lambda$ can reach is $8m^2$. Solving for $\theta_0$ in terms of $\Lambda/m^2$, we get 
\begin{equation}
\label{eq:theta0}
\theta_0 = \log \left(\frac{2+\sqrt{\frac{\Lambda}{m^2} -4}}{2-\sqrt{\frac{\Lambda}{m^2} -4}}\right),
\end{equation}
which grows logarithmically like $\theta_0 \approx \log\frac{16m^2}{8m^2 - \Lambda}$ as $\Lambda$ approaches the upper limit $8m^2$.  Parametrized in terms of $\Lambda$, the $S$ matrix of the staircase model is given precisely by \eqref{eq:amp_sinhGordon_rewritten} with $\Lambda\in [4m^2,8m^2]$.
Similarly one obtains the form factor of the trace of the stress tensor  in the staircase model by using the analytic continuation \eqref{eq:ac} and \eqref{eq:theta0} in the expression \eqref{eq:form_factor_sinh-Gordon}.
It is interesting to notice that using \eqref{eq:ac}, we can read off the value of $\beta^2$ from \eqref{eq:b} and \eqref{eq:amp_sinhGordon}. Since $\beta=\sqrt{8\pi\frac{\gamma}{\pi-\gamma}}$, one thus has $\left| \beta^2 \right| = 8\pi$ for $m^{-2}\Lambda\geq 4$.

Expanding  the analytically continued amplitude \eqref{eq:amp_sinhGordon_rewritten} around $\Lambda = 8m^2$, one finds 
\begin{equation}
\label{eq:FFCorrection}
\mathcal{T}_\text{bosons} =
2i \mathcal{N}_2 + \frac{s t}{M^2}+{\cal O}\left(\frac{1}{M^4}\right),
\end{equation}
where  $\frac{1}{M^2} \equiv \frac{8m^2 - \Lambda}{4m^4}$. According to the discussion around \eqref{eq:bosons_fermions}, the interacting part of the boson scattering amplitude \eqref{eq:FFCorrection} is equivalent to the scattering of Majorana fermions with the following interacting part
\begin{equation}
\label{eq:FFCorrection_ferm}
\mathcal{T}_\text{fermions} =
\frac{s t}{M^2}+{\cal O}\left(\frac{1}{M^4}\right).
\end{equation}
In other words at $\Lambda = 8m^2$,  $\mathcal{T}_\text{bosons} =
2i \mathcal{N}_2$, which is equivalent to $\mathcal{T}_\text{fermions} = 0$, so the S-matrix approaches that of a free massive fermion (the Ising model).  Moreover, at $\Lambda$ slightly below $8m^2$, the S-matrix additionally has contributions from irrelevant deformations that should capture the approach to Ising from the UV (in this case, from the  next minimal model up, i.e. the tricritical Ising model). In section \ref{sec:IsingTTbar} we will explicitly check this leading correction.

\subsection{$\phi^4$ model}
\label{sec:phi4}
The $\phi^4$ model defined by \eqref{eq:phi4_definition} allows for the presence of one-particle asymptotic states \eqref{eq:1PS} which are $Z_2$ odd. Due to the presence of the $Z_2$ symmetry the ``elastic'' regime in the $\phi^4$ model is extended to $s\in[4m^2,16m^2]$. The relation between the lightcone quantization bare mass $m_0$ and the physical mass $m$ is given by
\begin{equation}
\label{eq:mass0_pert}
m = m_0\left(1 - \frac{\bar\lambda^2}{768} + \frac{\bar\lambda^3}{3072\pi } + O(\bar\lambda^4) \right).
\end{equation}
For higher order corrections see equation (2.14) in \cite{Fitzpatrick:2018xlz}.

Using perturbation theory we compute the two-particle form-factor and the spectral density of the trace of the stress-tensor. The form factor reads
\begin{multline}
\label{eq:FF_pert}
m^{-2}\mathcal{F}^\Theta_{2,0}(s) = -2+\left(\frac{\bar\lambda}{4\pi}\right)\, \Delta(s)+\\
\frac{1}{2}\left(\frac{\bar\lambda}{4\pi}\right)^2
\left(\frac{\pi^2s}{8(s-4m^2)}-\Delta(s)\left(\Delta(s)/2+1\right)\right)+\mathcal{O}(\bar\lambda^3),
\end{multline}
where the function $\Delta(s)$ is defined as
\begin{equation}\label{eq:DeltaDef}
\Delta(s) \equiv-1+\lim_{\epsilon\rightarrow 0^+}
\frac{4m^2\text{ArcTan}\left(\frac{\sqrt{s}}{\sqrt{4m^2-s-i\epsilon}}\right)}{\sqrt{s}\sqrt{4m^2-s-i\epsilon}} .
\end{equation}
The expression \eqref{eq:FF_pert} is valid for any complex value of $s$.
The function $\Delta(s)$ has a single branch cut along the horizontal axis in the $s$ complex plane for $s\in[4m^2,\infty)$. The infinitesimally small $\epsilon$ is present in order to specify the correct side of the branch cut. At times, it is more convenient to use the rapidity variable $\theta$ defined in \eqref{eq:rapidity}, which opens up this branch cut.  In this variable, the $\epsilon$ prescription translates to taking the $\theta>0$ branch for $s>4m^2$, and the function $\Delta(s)$ is simply
\begin{equation}
\Delta(s(\theta)) = \frac{i\pi - \theta}{\sinh \theta} -1.
\end{equation}

The following limits hold true
\begin{equation}
\label{eq:limits}
\Delta(0) = 0,\qquad
\Delta(2m^2) = \frac{\pi}{2} - 1,\qquad
\Delta(4m^2) = 
\infty.
\end{equation}
The first entry in \eqref{eq:limits} implies \eqref{eq:FF_pert} satisfies the normalization condition \eqref{eq:mass_definition}.

The  spectral density of the trace of the stress tensor reads as
\begin{equation}
\label{eq:SD_pert}
\frac{2\pi\mathcal{N}_2}{4m^4}\times\rho_\Theta(s) = 1 +
\frac{\bar\lambda}{4\pi}\times\left(
1+4m^2\mathcal{N}_2^{-1}
\log\left(\frac{\sqrt{s}+\sqrt{s-4m^2}}{\sqrt{s}-\sqrt{s-4m^2}}\right)
\right)
+ \mathcal{O}(\bar\lambda^2).
\end{equation}
It is defined in the region $s\in[4m^2,\infty]$. Fully computing the next correction to the spectral density is quite difficult. We notice however that in the ``elastic'' regime $s\in [4m^2,16m^2]$ with no particle production the next correction to the spectral density simply follows from  \eqref{eq:SD_FF} and \eqref{eq:FF_pert}. 
We derive \eqref{eq:FF_pert} and \eqref{eq:SD_pert} in appendix \ref{app:analytic}. We are not aware of any literature where these results were previously presented, though in principle it should be possible to obtain them from small $\beta$ expansions of results for the corresponding observables in the sinh-Gordon model.
 
For completeness, let us also provide the textbook result for the interacting part of the scattering amplitude. It reads
\begin{equation}
\label{eq:amp_pert}
m^{-2}\mathcal{T}(s) = -\bar\lambda\times \left(1 - \frac{1}{2} \left(\frac{\bar\lambda}{4\pi}\right)\times\big (1+\Delta(s)+\Delta(4m^2-s)\big) + O(\bar\lambda^3) \right).
\end{equation}
It is straightforward to check that \eqref{eq:FF_pert} and \eqref{eq:amp_pert} obey  Watson's equation \eqref{eq:Watson} in the ``elastic'' regime.
Using \eqref{eq:amp_pert} and the second entry in \eqref{eq:limits} we can relate the quartic coupling $\lambda$ and the non-perturbative quartic coupling $\Lambda$ defined in \eqref{eq:Lambda_definition} as follows
\begin{equation}
\label{eq:Lambda_computed}
m^{-2}\Lambda = \bar\lambda -  \frac{\bar\lambda^2(\pi-1)}{8\pi}+\mathcal{O}(\bar\lambda^3).
\end{equation}
Another thing that is important to emphasize is that the perturbative results diverge at the two-particle threshold $s=4m^2$. This divergence is an artifact of perturbation theory and does not appear in the non-perturbative amplitude.
 
By inspecting the perturbative results  \eqref{eq:FF_pert}, \eqref{eq:SD_pert} and \eqref{eq:amp_pert}, we notice that the perturbative expansion parameter is more accurately $\frac{\bar\lambda}{4\pi}$ rather than  $\bar\lambda$. Thus, we expect that the strong coupling regime where perturbation theory breaks down is
\begin{equation}
\label{eq:strong_coupling}
\bar \lambda \sim 4\pi \sim 12.
\end{equation}

\subsection{2d $O(N)$ model in the large $N$ limit}
\label{sec:largeN}
Let us now consider the generalization of the $\phi^4$ model given by \eqref{eq:phi4_definition} where the field $\phi(x)$ has $N$ components and the action is invariant under $O(N)$ symmetry. In such a theory there are three different two-particle states transforming in the trivial, symmetric and antisymmetric representations of the $O(N)$ group. For details see appendix \ref{app:ON}. 

Let us consider here the large $N$ limit $N\rightarrow \infty $. In this limit it is enough to only consider the two-particle states in the trivial representation. In what follows we compute the spectral density and the form factor of the trace of the stress tensor  together with the scattering amplitude for the two-particle states in the trivial representation. Our results are valid to all orders of perturbation theory. The details of all the computations are given in appendix \ref{app:analytic}.

In the large $N$ limit the relation between the physical mass $m$ and lightcone quantization bare mass $m_0$ is extremely simple, namely  
\begin{equation}
m=m_0.
\end{equation}
The two-particle form factor of the stress-tensor in the large $N$ limit reads as 
\begin{equation}
\label{eq:FF_largeN}
m^{-2}\mathcal{F}^\Theta_{2,0}(s) = -2+ \frac{2\bar\lambda \Delta(s)}{8\pi+\bar\lambda\,(1+\Delta(s))}.
\end{equation}
It is important to notice that this form factor does not have a singularity at $s=4m^2$. Using the third entry in \eqref{eq:limits} we conclude that $\mathcal{F}^\Theta_{2,0}(4m^2) = 0$. In the large $N$ limit there is no particle production. As a result the full spectral density is simply given by the two-particle form factor \eqref{eq:FF_largeN} via \eqref{eq:SD_FF}. Concretely speaking
\begin{equation}
2\pi \mathcal{N}_2\rho_\Theta(s) =
|\mathcal{F}^\Theta_{2,0}(s)|^2.
\end{equation}
The full scattering amplitude in the large $N$ limit reads
\begin{equation}
\label{eq:S_largeN}
\widehat{\mathcal{S}}(s) = -\frac{\bar\lambda\, (\pi-i \theta)+8 \pi i\,\text{Sinh}(\theta)}{\bar\lambda\, (\pi+i \theta) -8 \pi i\, \text{Sinh}(\theta)}.
\end{equation}
Note that this S-matrix is not crossing symmetric, contrary to the other models that we consider in this paper. It is straightforward to check that \eqref{eq:FF_largeN} and \eqref{eq:S_largeN} satisfy  Watson's equation \eqref{eq:Watson} for the whole range of energies $s\in[4m^2,\infty)$. Moreover, there is no divergence at the two-particle threshold $s=4m^2$, where  $\widehat{\mathcal{S}}(4m^2)=-1$.
 
Using the definition of the non-perturbative quartic coupling $\Lambda$, the relation between the scattering amplitude and the interacting part of the scattering amplitude and the explicit solution \eqref{eq:S_largeN}, we can evaluate precisely the non-perturbative coupling $\Lambda$ in terms of $\lambda$. It takes the following simple form
\begin{equation}
\label{eq:Lambda_largeN}
m^{-2}\Lambda = \frac{16\bar\lambda}{16+\bar\lambda}.
\end{equation}
One can see that for real positive $\bar \lambda$, we have $\Lambda\in [0,16]$.

\subsection{$T\overline{T}$ deformation of the 2d Ising}
 \label{sec:IsingTTbar}
 
 In the vicinity of the critical point, both $\phi^4$ theory and the Staircase Model flow to the Ising model with a $\mathbb{Z}_2$ symmetry that forbids the magnetic deformation $\sigma$.  In that case, the lowest-dimension deformation around Ising is  the thermal operator $\epsilon$, which is just the fermion mass term of the free Majorana fermion description of the Ising model.  The next-lowest-dimension scalar operator is $T\overline{T}$, which in terms of the left- and right-moving components $\psi$ and $\tilde{\psi}$ of the fermion is
 \be
 \label{eq:TTbarPsi}
 \delta \mathcal{L} = \frac{1}{M^2} \psi \partial_- \psi \tilde{\psi} \partial_+ \tilde{\psi}.
 \ee
Here, $M$ is the scale of the UV cut-off of the low-energy expansion.  In the limit that $M$ is much larger than the mass gap $m$, the contributions to the S-matrix from all other higher dimension operators are suppressed by higher powers of $m/M$, so near $\Lambda/m^2=8$ the S-matrix is well-approximated by the tree-level contribution from (\ref{eq:TTbarPsi}).  The leading contribution to the scattering amplitude is most easily computed in lightcone coordinates, where each $\psi$ contraction with an external fermion produces a factor of $\sqrt{p_-}$ for that fermion, and each $\tilde{\psi}$ contraction produces a factor of $\sqrt{p_-} \frac{\sqrt{2} p_+}{m}$ (the extra factor follows from the fermion equation of motion $\sqrt{2} i \partial_+ \psi = m \tilde{\psi}$).  So, the full tree-level contribution is simply $\frac{2}{m^2}\sqrt{p_{1-} p_{2-} p_{3-} p_{4-}} p_{2-} p_{3+} p_{4+}^2$, anti-symmetrized on all permutations of $p_1, p_2, -p_3,$ and $-p_4$.  Finally, there are only two solutions to the kinematic constraint $p_1+p_2 =p_3 + p_4$; either $p_1 = p_3, p_2=p_4$ or $p_1 = p_4, p_2 =p_3$.  Taking the former, and using $p_+ = \frac{m^2}{2p_-}$, we obtain
\be
\label{eq:deformation_2d_Ising}
\mathcal{T}_\text{fermions} = \frac{m^4 (p_{1-}-p_{2-})^2 (p_{1-}+p_{2-})^2}{M^2 p_{1-}^2 p_{2-}^2}  = \frac{s t}{M^2},
\ee
in agreement with (\ref{eq:FFCorrection_ferm}). 
The Ising model S-matrix has $(\Lambda/m^2, m^2 \Lambda^{(2)})$ = $(8,2)$, and from the above expression we can read off that the leading correction which gives
\begin{equation}
\label{eq:Msq_value}
(\Lambda/m^2, m^2 \Lambda^{(2)}) = \left(8-\frac{4m^2}{M^2}, 2- \frac{2m^2}{M^2}\right).
\end{equation}

\section{Pure S-matrix bootstrap}
\label{sec:s-matrix_pure}

In this section we will construct general non-perturbative bounds on the space of 2d scattering amplitudes of $Z_2$ odd particles (assuming there is no bound state pole). We will define the exact optimization problem in section \ref{sec:optimizationI} and present our numerical results in section \ref{sec:results_I}. The main result of this section is presented in figure \ref{fig:bound}. The amplitudes in the sinh-Gordon and its analytic continuation (the staircase model) saturate the lower edge of this bound.

\subsection{Set-up}
\label{sec:optimizationI}

Let us start by discussing the unitarity constraint. In 2d the scattering amplitude $\widehat{\mathcal{S}}(s)$ must obey the following positive semidefinite condition 
\begin{equation}
\label{eq:unitarity_2x2}
\begin{pmatrix}
1 & \widehat{\mathcal{S}}^*(s)\\
\widehat{\mathcal{S}}(s) & 1
\end{pmatrix}\succeq 0,\qquad \text{for }
s\in [4m^2,\infty).
\end{equation} 
Due to Sylvester's criterion, this condition is equivalent to the more familiar one \eqref{eq:unitarity_Shat}. To see that, one can simply evaluate the determinant of \eqref{eq:unitarity_2x2}.

It was proposed in \cite{Paulos:2016but,Paulos:2017fhb} how to use the constraint \eqref{eq:unitarity_2x2} in practice. One can write the following ansatz for the scattering amplitude which automatically obeys maximal analyticity and crossing
\begin{equation}
\label{eq:ansatz_S}
\widehat{\mathcal{S}}(s) -1=-\frac{\Lambda}{4m^2} +
\sum_{n=1}^{N_\text{max}} a_n\times\left(
\myRho(s;2m^2)^n+\myRho(4m^2-s;2m^2)^n
\right),
\end{equation}
where $\Lambda$ is the non-perturbative quartic coupling defined in \eqref{eq:Lambda_definition}, $a_n$ are some real coefficients and the $\myRho$ variable is defined as
\begin{equation}
\label{eq:rho_variable}
\myRho(s;s_0) \equiv
\lim_{\epsilon\rightarrow 0^+} \frac{\sqrt{4m^2-s_0}-\sqrt{4m^2-s-i\epsilon}}{\sqrt{4m^2-s_0}+\sqrt{4m^2-s-i\epsilon}}.
\end{equation}
Here $s_0$ is a free parameter which can be chosen at  will. For scattering amplitudes it is convenient to choose $s_0=2m^2$; this guaranties that at the crossing symmetric point $s=2m^2$, the $\myRho(s;2m^2)$ variable vanishes. In theory, one should take $N_\text{max}=\infty$. This is impossible in practice however, and one is thus has to choose a large enough but finite value of  $N_\text{max}$ which leads to stable numerical results (stable under the change of $N_\text{max}$).  Alternatively to \eqref{eq:ansatz_S}, one could also parametrize only the interacting part of the scattering amplitude, namely
\begin{equation}
\label{eq:ansatz_T}
\mathcal{T}(s)=-\Lambda +
\sum_{n=1}^{N_\text{max}} \tilde{a}_n\times\left(
\myRho(s;2m^2)^n+\myRho(4m^2-s;2m^2)^n
\right).
\end{equation}
In this ansatz we denote the unknown paramters by $\tilde a$ in order to distinguish them from the parameters $a$ entering in \eqref{eq:ansatz_S}.
Depending on the situation sometimes this choice is more convenient than \eqref{eq:ansatz_S}.

Using SDPB \cite{Simmons-Duffin:2015qma,Landry:2019qug} we can scan the parameter space $(\Lambda, a_0, a_1, a_2, \ldots)$ of the ansatz \eqref{eq:ansatz_S} (or alternatively $(\Lambda, \tilde{a}_0, \tilde{a}_1, \tilde{a}_2, \ldots)$ of the ansatz \eqref{eq:ansatz_T}) by looking for amplitudes with the largest or smallest value of $\Lambda$ which obey \eqref{eq:unitarity_2x2}. Once the allowed range of $\Lambda$ is determined, we can look for example for amplitudes for each allowed value of $\Lambda$ with the largest or smallest value of the parameter $\Lambda^{(2)}$ defined in  \eqref{eq:Lambda_derivative_definition}. Using this definition we can express $\Lambda^{(2)}$ in terms of the paramenters of the ansatz as
\begin{equation}
\Lambda^{(2)}=\frac{\Lambda+m^2 \,(2a_1+a_2)}{4m^4}
\qquad\text{or}\qquad
\Lambda^{(2)}=\frac{2\tilde{a}_1+\tilde{a}_2}{16m^4}.
\end{equation}

\subsection{Numerical Results}
\label{sec:results_I}
\begin{figure}[t]
\centering

\includegraphics[width=0.8\textwidth]{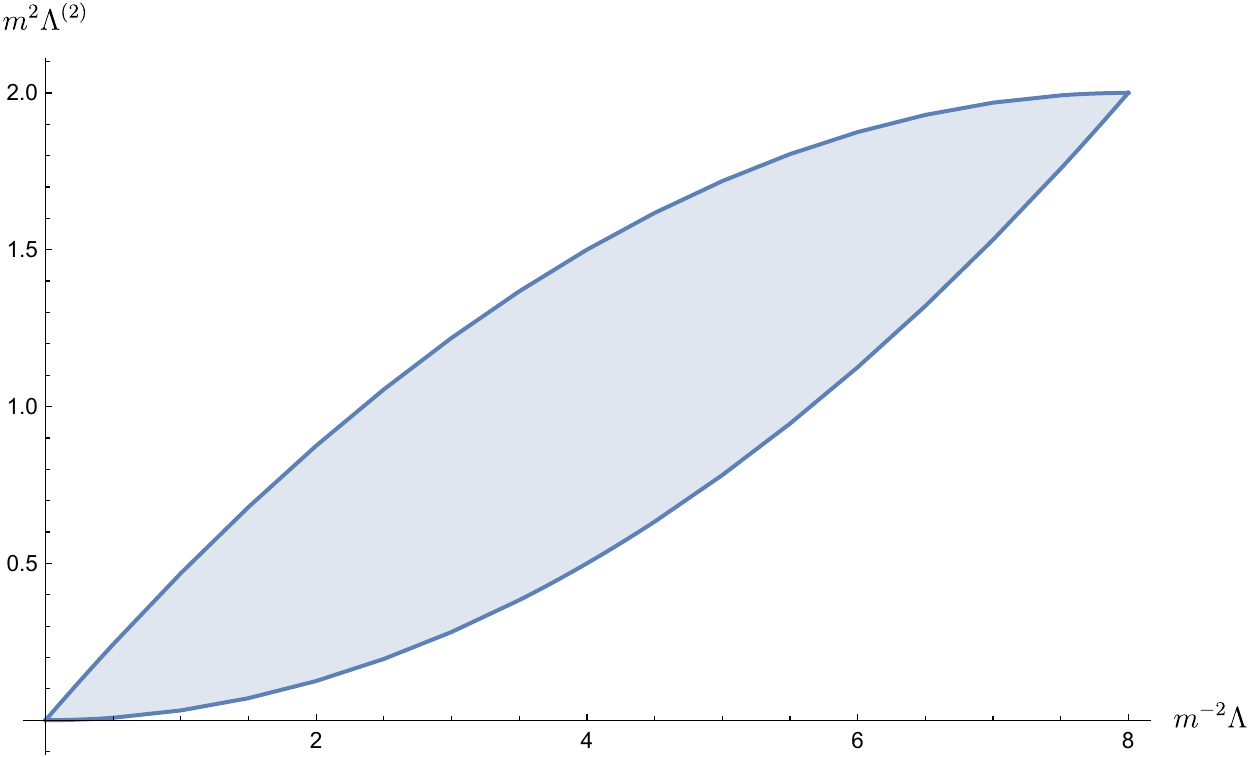}
\caption{Bound on the parameters $\Lambda$ and $\Lambda^{(2)}$. The allowed region is depicted in blue. }
\label{fig:bound}

\end{figure}

\begin{figure}[p]
\centering

\includegraphics[width=0.9\textwidth]{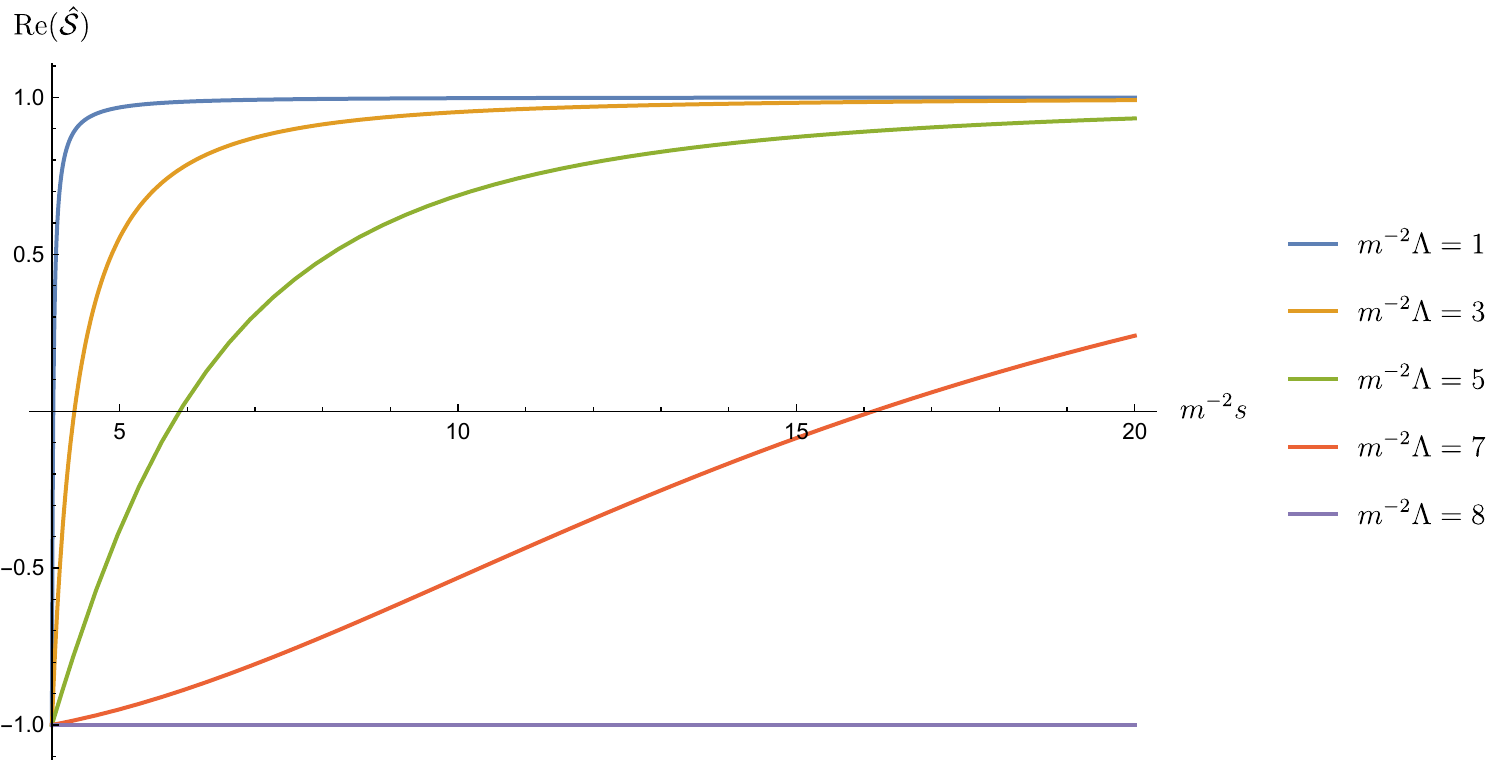}
\caption{Real part of the scattering amplitude obtained by minimizing $\Lambda^{(2)}$ for various values of $\Lambda$. }
\label{fig:plotRe_optI}

\vspace{10mm}

\includegraphics[width=0.9\textwidth]{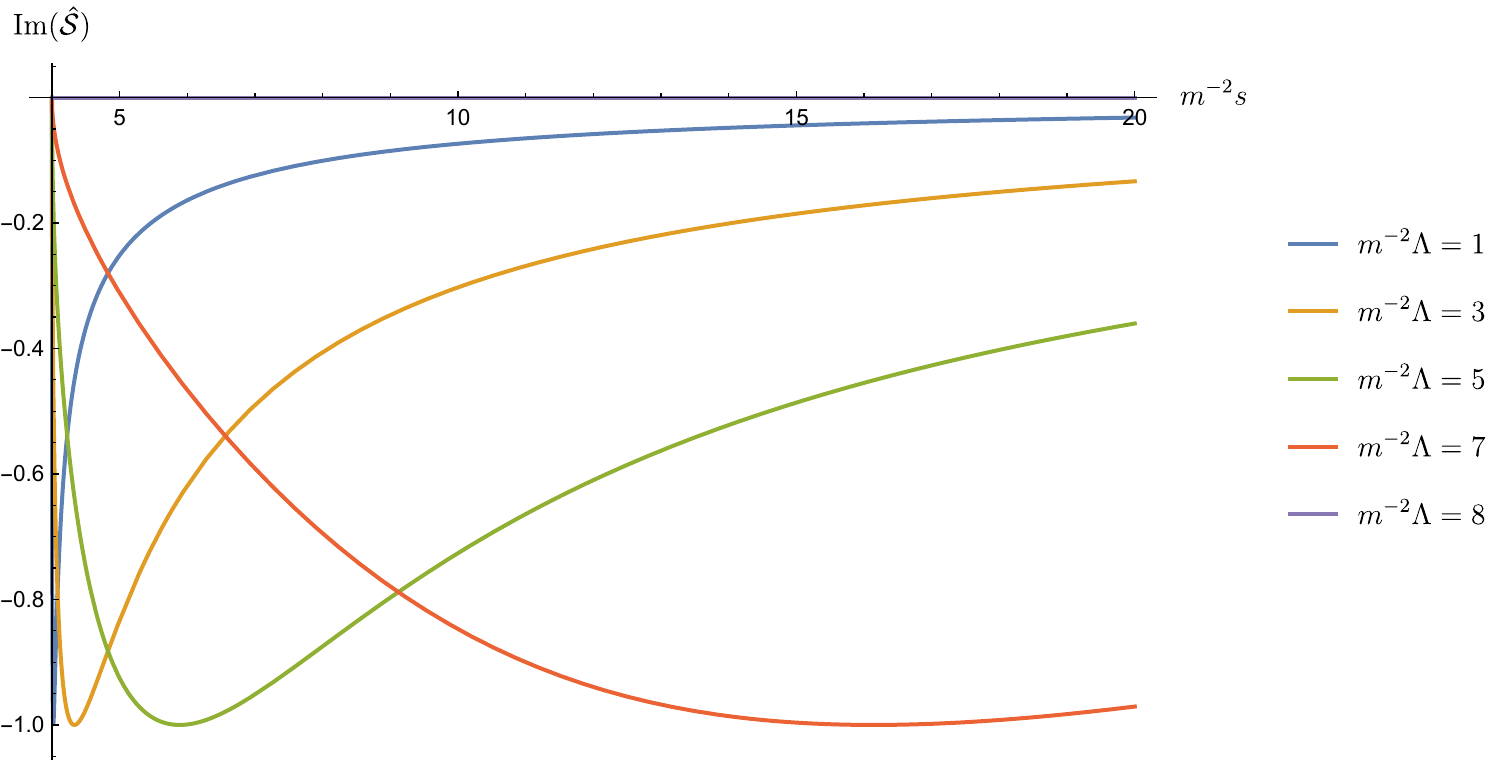}
\caption{Imaginary part of the scattering amplitude obtained by minimizing $\Lambda^{(2)}$ for various values of $\Lambda$.}
\label{fig:plotIm_optI}

\end{figure}

Solving the optimization problem for $\Lambda$ defined in section \ref{sec:optimizationI} we obtain the following bound
\begin{equation}
\label{eq:Lambda_range}
\Lambda \in [0,\, 8m^2].
\end{equation}
For each $\Lambda$ in this range, we can now minimize and maximize the parameter $\Lambda^{(2)}$. As a result we obtain a 2d plot of allowed values which is given in figure \ref{fig:bound}.  On the boundary of the allowed region in figure \ref{fig:bound}, we can extract the numerical expressions of the scattering amplitudes. For instance the scattering amplitudes extracted from the lower edge are presented in figures \ref{fig:plotRe_optI} and \ref{fig:plotIm_optI}. Remarkably they coincide with the analytic expression \eqref{eq:amp_sinhGordon_rewritten} which describes the sinh-Gordon model and its analytic continuation (the staircase model).
In particular, notice that the amplitudes extracted in the vicinity of the tips of the allowed region in figure \ref{fig:bound} approach the following expressions
\begin{equation}
\label{eq:tips}
\widehat{\mathcal{S}}_\text{left tip}(s) = +1
\qquad\text{and}\qquad
\widehat{\mathcal{S}}_\text{right tip}(s) = -1.
\end{equation}
These are the amplitudes of the free boson (lower left corner with $\Lambda=0$) and of the free Majorana fermion (upper right corner with $\Lambda=8m^2$).
Let us also make a fun observation that the amplitudes extracted from the upper edge of the bound in figure \ref{fig:bound} are related to the ones extracted from the lower edge by
\begin{equation}
\widehat{\mathcal{S}}_\text{upper edge}(s;\,\Lambda) = - 
\widehat{\mathcal{S}}_\text{lower edge}(s;\,8m^2-\Lambda).
\end{equation}
Using \eqref{eq:scattering_fermions_general}, we can interpret $\widehat{\mathcal{S}}_\text{upper edge}(s;\,\Lambda)$ as complex conjugated amplitudes of Majorana fermions with the interacting part exactly as in the sinh-Gordon expression with $\Lambda\rightarrow 8m^2-\Lambda$. We do not know any UV complete model from where such amplitudes could originate.

Finally, let us relax the requirement that the amplitude $\mathcal{S}(s)$ is crossing invariant. This means the when writing down the ansatz for the S-matrix, we do not have the second term in the parentheses in equation (\ref{eq:ansatz_S}).  In this case, we obtain the following bound
\begin{equation}
\label{eq:Lambda_range_relaxedCrossing}
\Lambda \in [0, 16m^2].
\end{equation}
We notice that the  large $N$ limit  of the 2d $O(N)$ model with $\phi^4$ potential populates this interval, see \eqref{eq:Lambda_largeN}. We can then minimize $\Lambda^{(2)}$ by fixing the value of $\Lambda$ in the interval \eqref{eq:Lambda_range_relaxedCrossing}. The resulting numerical amplitudes correspond precisely to the large $N$ analytic solution \eqref{eq:S_largeN}. In practice, for this optimization problem it was important to parametrize the interacting part of the scattering amplitude $\mathcal{T}$ instead of $\widehat{\mathcal{S}}$.

\section{S-matrix and Form Factor Bootstrap}
\label{sec:s-matrix}

In this section we will define the numerical optimisation which allows one to compute the two-to-two scattering amplitude and the two-particle form factor of the trace of the stress tensor at  $s>0$  in the 2d $\phi^4$ model given the LCT input obtained in a companion paper \cite{truncffsd}. We begin in section \ref{sec:optimizationII} by quickly reviewing the generalization of the S-matrix bootstrap proposed in \cite{Karateev:2019ymz} which allows one to include local operators. We then explain how one can define an optimization problem which allows one to compute the scattering amplitude and the form factor at  $s>0$. In section \ref{sec:results_II}, we present our numerical findings. The main results are given in figures \ref{fig:T_re} - \ref{fig:FF_im}.
Also in section \ref{sec:results_II}, we will observe that the $\phi^4$ model is very similar to the sinh-Gordon model in the elastic regime. We will investigate this similarity further in section \ref{sec:comparison}.

\subsection{Set-up}
\label{sec:optimizationII}
In \cite{Karateev:2019ymz}, it was shown that unitarity allows to write a more complicated constraint than \eqref{eq:unitarity_2x2} which entangles the scattering amplitude with the two-particle form factor and the spectral density of a scalar local operator $\cO$. In this paper we consider the case where  the local operator is the trace of the stress tensor $\Theta$. The following positive semidefinite condition can be written\footnote{Various entries in this matrix have different mass dimensions. This positivity condition is equivalent however to the one which is obtained from \eqref{eq:unitarity_3x3} by the following rescaling 
\begin{equation*}
\mathcal{F}_\Theta\rightarrow m^{-1}\mathcal{F}_\Theta,\qquad
\rho_\Theta\rightarrow m^{-2}\rho_\Theta.
\end{equation*}
The unitarity condition in the latter form was originally presented in \cite{Karateev:2019ymz}. It contains only dimensionless quantities.
}
\begin{equation}
\label{eq:unitarity_3x3}
\begin{pmatrix}
1 & \widehat{\mathcal{S}}^*(s)
& \mathcal{N}_2^{-1/2}\, \mathcal{F}^*{}^\Theta_{2,0}(s)\\
\widehat{\mathcal{S}}(s) & 1
&  \mathcal{N}_2^{-1/2}\, \mathcal{F}^\Theta_{2,0}(s)\\
\mathcal{N}_2^{-1/2}\, \mathcal{F}^\Theta_{2,0}(s)
&  \mathcal{N}_2^{-1/2}\, \mathcal{F}^*{}^\Theta_{2,0}(s) &
2\pi\rho(s)
\end{pmatrix}\succeq 0,\qquad \text{for }
s\in [4m^2, \infty].
\end{equation}

Analogously to section \ref{sec:optimizationI}, one can define various numerical optimization problems which utilize \eqref{eq:unitarity_3x3} instead of \eqref{eq:unitarity_2x2}. For that we should write an ansatz for all the ingredients entering in \eqref{eq:unitarity_3x3}. For the scattering amplitude we use the ansatz \eqref{eq:ansatz_S} or \eqref{eq:ansatz_T}. In practice, we will use \eqref{eq:ansatz_T} in this section. For the form factor we can write instead
\begin{equation}
\label{eq:ansatz_FF}
\mathcal{F}^\Theta_{2,0}(s) =
-2m^2 + \sum_{n=1}^{N_\text{max}} b_n\times \myRho(s;0)^n,
\end{equation}
where $b_n$ are some real parameters.
By construction it is an analytic function in $s$ with a single branch cut on the real axis between $4m^2$ and $+\infty$. The $\myRho$ variable was defined in \eqref{eq:rho_variable}. Here, we have chosen the parameter $s_0$ to be 0, such that $\myRho(s;0)$ vanishes at $s=0$. This is convenient since this ansatz automatically satisfies the normalization condition \eqref{eq:relation_FF} at $s=0$. If we also write an ansatz for the spectral density (which is simply a real function), one can then bound for example the UV central charge \eqref{eq:central_charge} for various values of $\Lambda$. Although this may be an interesting problem, we do not pursue it in this paper.

Instead of parametrizing the spectral density in this section, we will use its explicit form in the 2d $\phi^4$ model found in the companion paper \cite{truncffsd}, see figure 4 there. We use the superscript LCT in order to denote these spectral densities, namely
\begin{equation}
\label{eq:SD_LCT}
s\in[4m^2, s_\text{max}]:\qquad \rho_\Theta^\text{LCT}(s).
\end{equation}
Here $s_\text{max}$ is the maximal value of $s$ for which we trust the results of \cite{truncffsd}. In \cite{truncffsd}, see figure 12, we have also computed the two-particle form factor of the trace of the stress tensor at $s\le 0$. We also use the LCT superscript to denote these form factors, namely
\begin{equation}
\label{eq:FF_LCT}
s\in[s_\text{min},0]:\qquad \mathcal{F}^{\Theta \text{LCT}}_{2,0}(s).
\end{equation}
Here $s_\text{min}<0$ is the minimal value of $s$ for which we trust the results of \cite{truncffsd}.
We refer to \eqref{eq:SD_LCT} and \eqref{eq:FF_LCT} as the input data.

Let us now precisely define our optimization problem. Given the value of the physical mass $m$ (which is obtained by the LCT method), determine the unknown coefficients $\Lambda$, $a_n$ and $b_n$ in the ansatze \eqref{eq:ansatz_T} and \eqref{eq:ansatz_FF}, such that $\Lambda$ has the maximal/minimal value and the following constraints are satisfied
\begin{align}
\label{eq:constraint_1}
&\begin{pmatrix}
1 & \widehat{\mathcal{S}}^*(s)
& \mathcal{N}_2^{-1/2}\, \mathcal{F}^*{}^\Theta_{2,0}(s)\\
\widehat{\mathcal{S}}(s) & 1
&  \mathcal{N}_2^{-1/2}\, \mathcal{F}^\Theta_{2,0}(s)\\
\mathcal{N}_2^{-1/2}\, \mathcal{F}^\Theta_{2,0}(s)
&  \mathcal{N}_2^{-1/2}\, \mathcal{F}^*{}^\Theta_{2,0}(s) &
2\pi\rho_\Theta^\text{LCT}(s)
\end{pmatrix}\succeq 0,\quad
s\in [(4+\sigma)m^2, s_\text{max}],\\
\label{eq:constraint_2}
&\begin{pmatrix}
1 & \widehat{\mathcal{S}}^*(s)\\
\widehat{\mathcal{S}}(s) & 1
\end{pmatrix}\succeq 0,\quad 
s\in [4m^2,(4+\sigma)m^2)\cup (s_\text{max},\infty)
\end{align} 
The first constraint \eqref{eq:constraint_1} allows one to inject information about the LCT spectral density \eqref{eq:SD_LCT} in the set-up. The second constraint \eqref{eq:constraint_2} can be seen as the reduced version of the first one in the region where no information about the spectral density is available.
In the above equations we have introduced an addition small parameter $\sigma \ll 1$. The numerical bootstrap set-up is sensitive to numerical errors in the LCT data, and the presence of $\sigma$ mitigates the effect of these errors in the spectral density near threshold $s=4m^2$ and the uncertainty in the value of the physical mass itself.
In addition to equation \eqref{eq:constraint_1} and \eqref{eq:constraint_2}, we require that the Ansatz for the form factor match the one obtained by the LCT method. We can impose this by demanding
\begin{equation}
\label{eq:condition_FF}
|\mathcal{F}^\Theta_{2,0}(s)-\mathcal{F}^{ \Theta \text{LCT}}_{2,0}(s)|\leq \epsilon,\quad \text{for }
s\in[s_\text{min},\, 0],
\end{equation}
where $\epsilon \geq 0$ is a small positive parameter. We have introduced the $\epsilon$ parameter in the set-up in order to accommodate the numerical errors in the LCT input data. The constraint \eqref{eq:condition_FF} can be equivalently rewritten in the semi-positive form as
\begin{equation}
\label{eq:condition_FF_SDP}
\begin{pmatrix}
\epsilon & \mathcal{F}^{*\Theta}_{2,0}(s)-\mathcal{F}^{*\Theta \text{LCT}}_{2,0}(s)\\
\mathcal{F}^\Theta_{2,0}(s)-\mathcal{F}^{ \Theta \text{LCT}}_{2,0}(s) & \epsilon
\end{pmatrix}\succeq 0,\quad \text{for }
s\in[s_\text{min},\, 0].
\end{equation} 
In practice, we parameterize $\epsilon$ in terms of the exponent $\delta$ defined in the following way
\begin{equation}
\label{eq:delta}
m^{-2}\epsilon = 10^{-\delta}.
\end{equation}
The larger the value of $\delta$, the stronger the constraint \eqref{eq:condition_FF_SDP} becomes.

When we present our numerical results in section \ref{sec:results_II}, we will see that given a large enough value of $\delta$ in \eqref{eq:delta}, we find a unique solution to the optimisation problem described in this section, namely the upper and lower bounds lead to almost the same result. Moreover, we will see that the unitarity conditions \eqref{eq:constraint_1} and \eqref{eq:constraint_2} tend to get saturated in the ``elastic'' regime $s\in [4m^2,16m^2]$.
As a result, the obtained form factors obey equation \eqref{eq:SD_FF_rel}  and the scattering amplitudes obey equation \eqref{eq:Watson} as they should. In the non-elastic regime $s\geq 16m^2$, the LCT spectral density contains four- and higher- particle contributions, however we do not include four- and higher-particle form factors in the set-up. Therefore, conservatively speaking, this means that for $s\geq 16m^2$ the behaviour of the obtained scattering amplitude and the form factor has nothing to do with the $\phi^4$ model.

Formulating the above paragraph in different words, one can roughly  say that the above optimization procedure determines the coefficients of the form factor ansatz in equation \eqref{eq:ansatz_FF} given two constraints:  that the Ansatz matches the LCT form factor result \eqref{eq:FF_LCT} for $s\le0$, and the square of its norm saturates the LCT spectral density \eqref{eq:SD_LCT} for $4m^2\le s\le 16m^2$ via \eqref{eq:SD_FF_rel}. The scattering amplitude is then obtained by solving  Watson's equation \eqref{eq:Watson}.

\subsection{Numerical Results}
\label{sec:results_II}

We present now the solutions of the optimization problem defined in section \ref{sec:optimizationII}. As a demonstration of our approach, in section \ref{sec:demonstration}, instead of using the LCT input data (which obviously contains numerical errors), we use the input data obtained from the analytic solution for the 2d $O(N)$ model in the large $N$ limit given in section \ref{sec:largeN}. We stress however that we use only the part of the analytic data which is computable with the LCT methods.
The reason for this exercise  is to show how the optimization problem works in the presence of high accuracy data. We present our optimization for the $\phi^4$ model using the LCT data in section \ref{sec:results_phi4}. For small values of the quartic coupling constant $\bar \lambda$, our results are in agreement with perturbation theory. For large values of $\bar \lambda$ our results are novel.

In order to proceed, let us provide some details on the choice of the optimization parameters used in SDPB.
We use the following range for the input data
\begin{equation}
s_\text{min}= -80m^2,\qquad s_\text{max}= 100m^2.
\end{equation} 
We use the following size of the ansatzes in equation \eqref{eq:ansatz_T} and equation \eqref{eq:ansatz_FF}
\begin{equation}
\label{eq:value_Nmax}
N_\text{max} = 50,
\end{equation}
which is large enough in practice.
We impose the conditions \eqref{eq:constraint_1}, \eqref{eq:constraint_2} and \eqref{eq:condition_FF_SDP} at a finite number of points $s$. Let us denote by $N_\mathcal{F}$ the number of $s$ values picked in the interval $[s_\text{min},0]$ where the condition \eqref{eq:condition_FF_SDP} is imposed, and by $N_\rho$ the number of $s$ values picked in the interval $[(4+\sigma)m^2,s_\text{max}]$ where the condition \eqref{eq:constraint_1} is imposed.
For the choice of $N_\text{max}$ in \eqref{eq:value_Nmax}, we chose the following values
\begin{equation}
N_\mathcal{F} = 1000,\qquad
N_\rho              = 2500.
\end{equation}
We impose the condition  \eqref{eq:constraint_2} at about 100 points in the range $s> s_\text{max}$. We use the Chebyshev grid to distribute the above points. In practice for the LCT data with $\bar\lambda\leq 13$  we use $\sigma=0.001$ and for $\bar\lambda>13$ we use $\sigma=0.01$. This indicates that the LCT data for higher values of $\bar\lambda$ contains larger errors. For smaller values of $\sigma$ the optimization problem often simply does not converge.

Our strategy is then as follows. We run the optimization routine for different values of $\epsilon$ or equivalently $\delta$, see \eqref{eq:delta}. For small values of $\delta$, the problem is not constraining enough. For too large values of $\delta$, the problem becomes unfeasible. In order to find the optimal value for $\delta$ we perform the binary scan in the range
\begin{equation}
\delta\in[0,10].
\end{equation}
We then pick the largest value of $\delta$ where the optimization problem is still feasible. The binary scan is performed until the difference between the feasible and unfeasible values of $\delta$ drops below some threshold value. We pick this threshold value to be 0.1.

\subsubsection{Infinite Precision Example} 
\label{sec:demonstration}

We study here the 2d $O(N)$ model in the large $N$ limit where the exact analytic solution exists, see section \ref{sec:largeN}. We pick the following value of the quartic coupling
\begin{equation}
\bar \lambda = 10
\end{equation}
as an example. At large $N$ we keep only the singlet component of the scattering amplitude, which therefore loses its crossing symmetry $s \leftrightarrow 4m^2-s$, see appendix \ref{app:ON}. As a result, in the ansatz \eqref{eq:ansatz_T}, we relax crossing symmetry by dropping the last term in the sum.

\begin{figure}[p]
\centering

\includegraphics[width=0.5\textwidth]{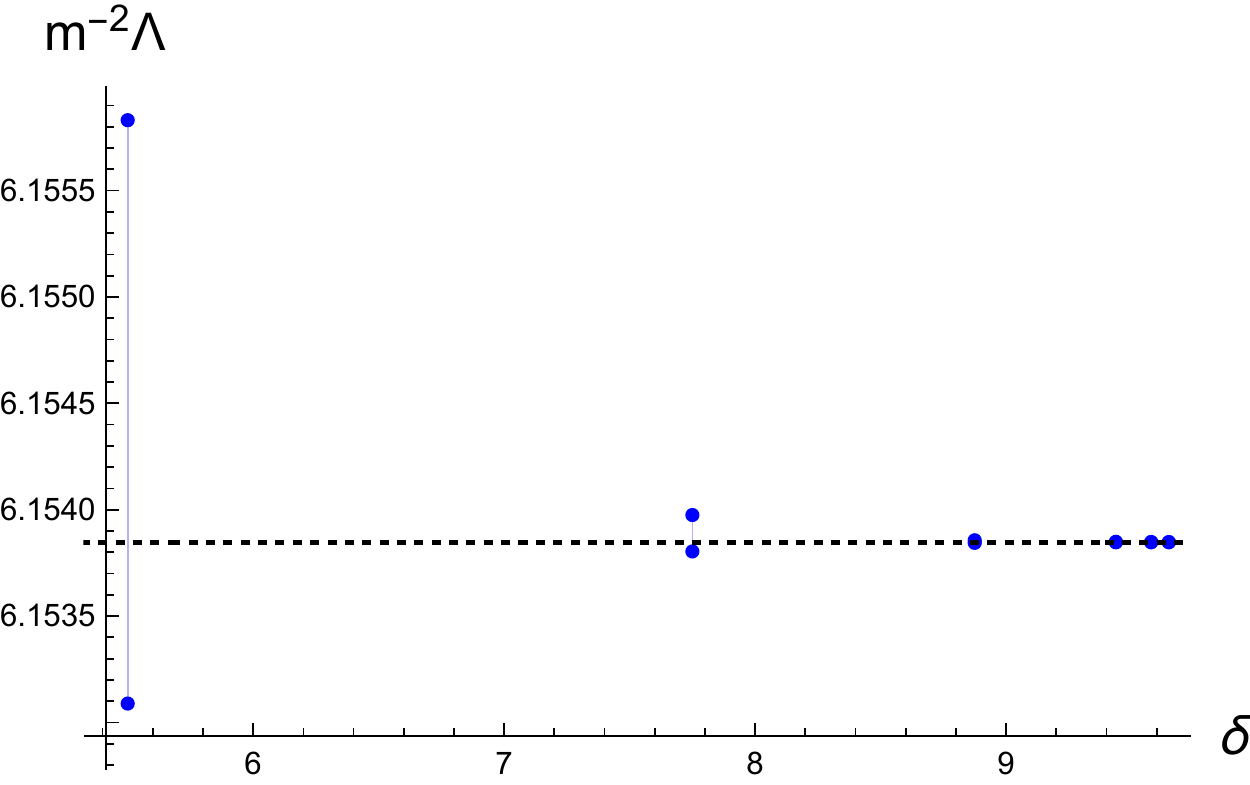}
\caption{Lower and upper bounds on the non-perturbative quartic coupling constant $\Lambda$. Blue dots are the numerical data. The blue vertical lines indicate the allowed region for $\Lambda$ for each value of $\delta$. The horizontal dashed line indicates the analytic value of $\Lambda \approx 6.1538$, which the numerical lower and upper bounds are expected to converge to. Here we use the analytic large $N$ data of the 2d $O(N)$ model to mimic the LCT input data.}
\label{fig:boubnd_largeN_analytic}

\includegraphics[width=0.4\textwidth]{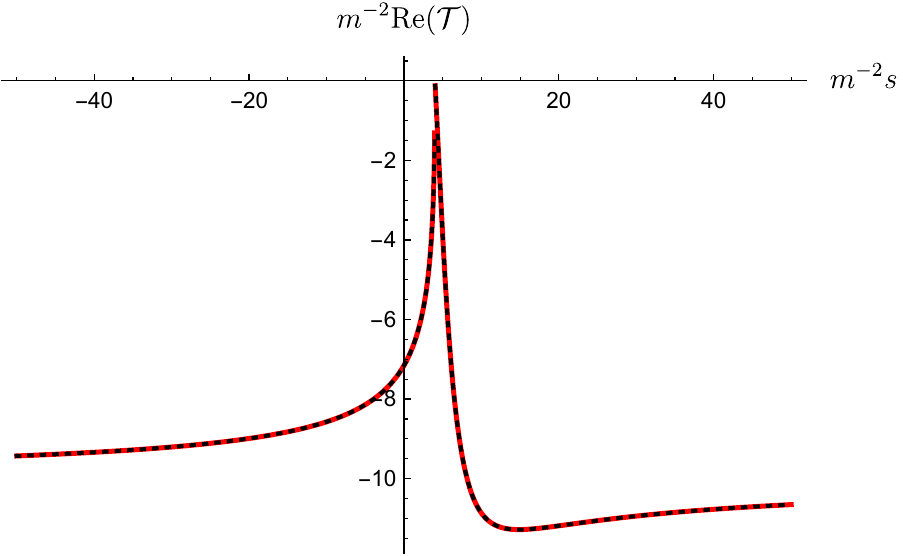}
\includegraphics[width=0.45\textwidth]{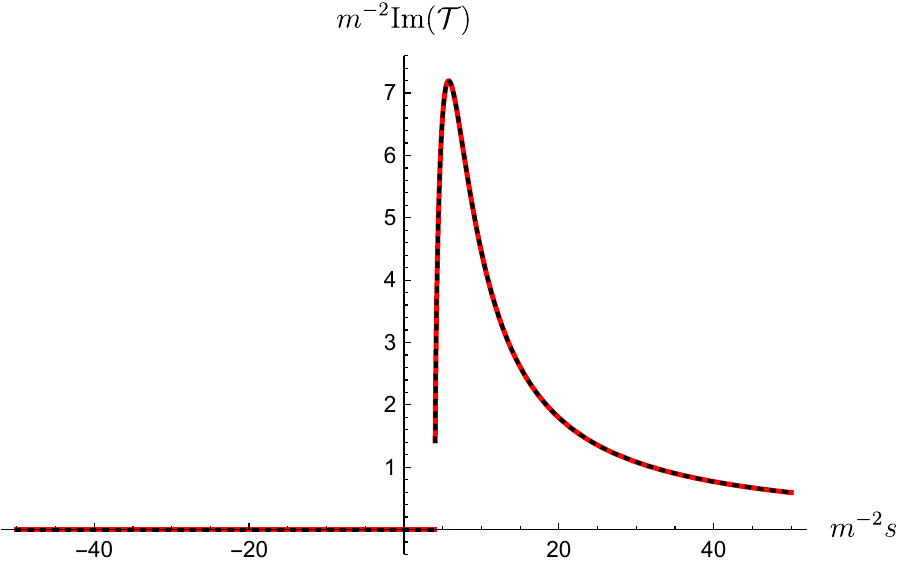}
\caption{The real and imaginary part of the interacting part of the scattering amplitude $\mathcal{T}(s)$. The solid red line is the numerical result. The dashed black line is the expected result from perturbation theory (\ref{eq:S_largeN}). Here we use the analytic large $N$ data of the 2d $O(N)$ model to mimic the LCT input data.}
\label{fig:T_argeN_analytic}

\includegraphics[width=0.45\textwidth]{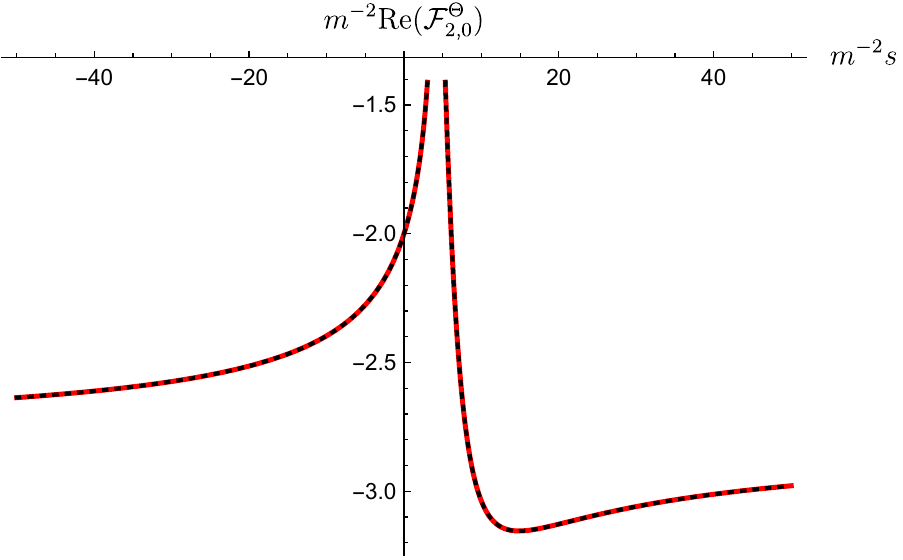}
\includegraphics[width=0.45\textwidth]{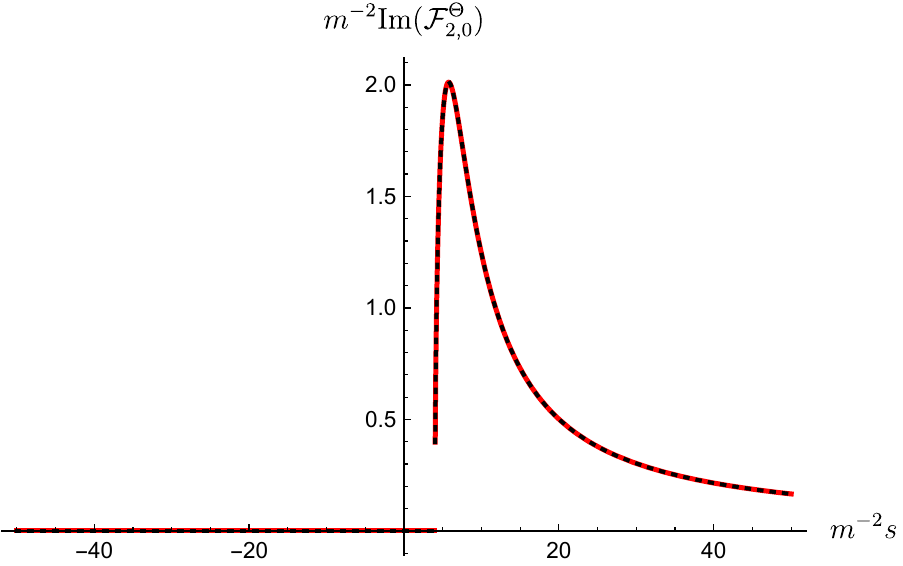}
\caption{The real and imaginary part of the form factor of the trace of the stress tensor  $\mathcal{F}^\Theta_{2,0}(s)$. The solid red line is the numerical result. The dashed black line is the expected result from perturbation theory (\ref{eq:FF_largeN}). Here we use the analytic large $N$ data of the 2d $O(N)$ model to mimic the LCT input data.}
\label{fig:F_argeN_analytic}

\end{figure}

The bound on the non-perturbative quartic coupling $\Lambda$ is given on figure \ref{fig:boubnd_largeN_analytic}. We see that the upper and lower bounds quickly converge to the analytic value of $\Lambda$ and starting from $\delta\gtrsim 9$ basically coincide. 
We pick the ``lower bound'' solution with the largest value of $\delta$ and extract the interacting part of the scattering amplitude and the form factor. The result is given in figures \ref{fig:T_argeN_analytic} and \ref{fig:F_argeN_analytic} respectively. The optimization problem result is given by the red solid line and the analytic results are given by the black dashed line. Both are in a perfect agreement.

\subsubsection{$\phi^4$ model}  
\label{sec:results_phi4}
\begin{figure}[th]
\centering

\includegraphics[width=0.95\textwidth]{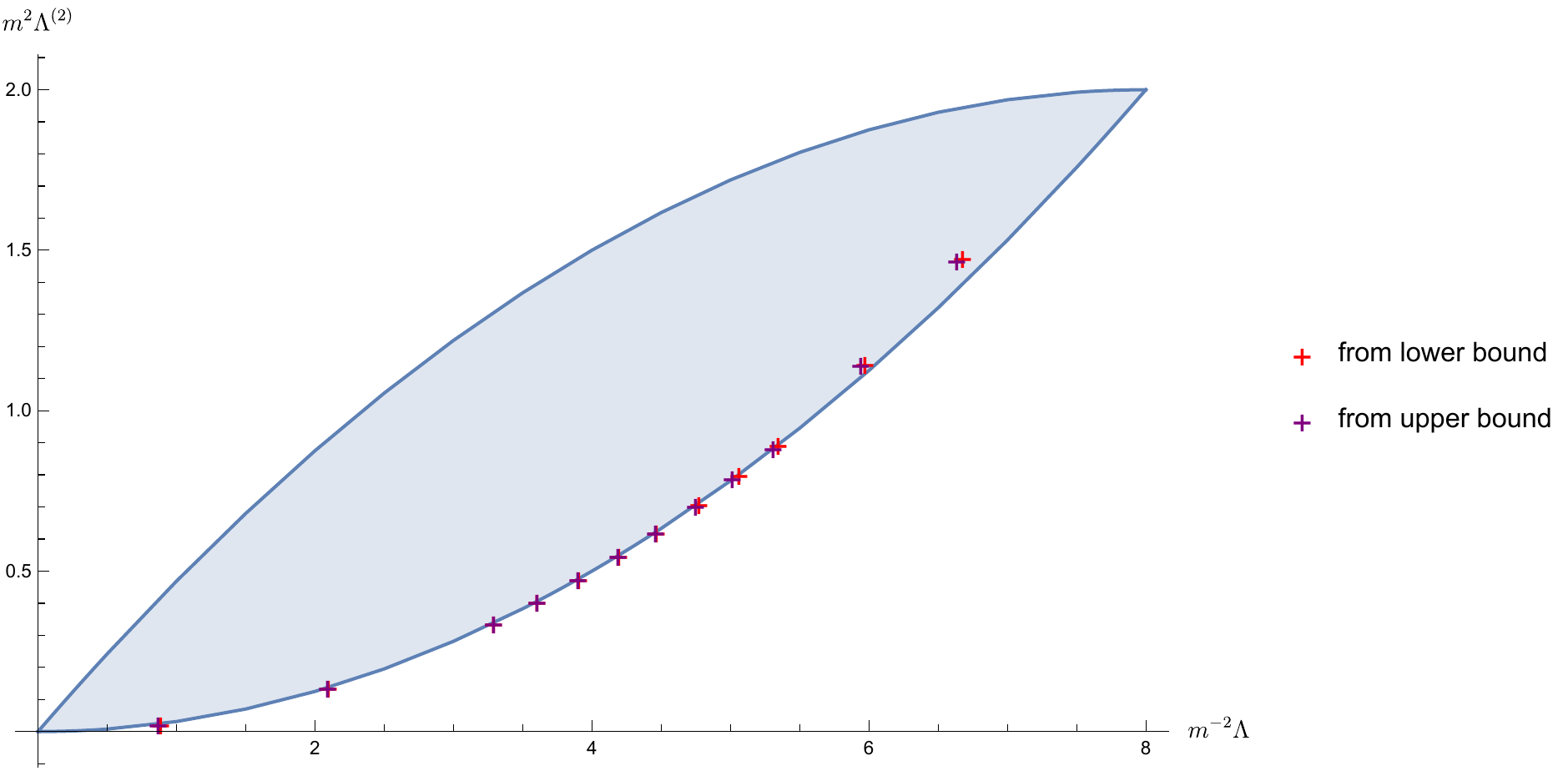}
\caption{Bound on the parameters $\Lambda$ and $\Lambda^{(2)}$. The allowed region is depicted in blue. The obtained numerical values for the $\phi^4$ model using the LCT data from table \ref{tab:phi4_parameters} are indicated by red and purple crosses. These crosses correspond to lower and upper bounds respectively.}
\label{fig:boundWithData}

\end{figure}

Let us now address the optimization problem with the $\phi^4$ LCT data as an input. In what follows we will denote the data obtained by maximization of $\Lambda$ by the subscript ``upper'' and the data obtained by minimization of $\Lambda$ by the subscript ``lower''.
The obtained numerical values of $\Lambda$ and $\Lambda^{(2)}$ are presented in table  \ref{tab:phi4_parameters}. Looking at this table one can see that both optimization problems lead to almost the same numerical values. This indicates that our procedure converges to the unique solution. The relative difference between $\Lambda_\text{upper}$ and $\Lambda_\text{lower}$ can be taken as a rough error estimate.
It is illuminating to place the data of table \ref{tab:phi4_parameters} on figure \ref{fig:bound}. We display the result in figure \ref{fig:boundWithData}. Remarkably the $\phi^4$ model lies very close to the boundary of the allowed region and almost coincides with the sinh-Gordon/staircase model. We address the similarity between the two models in detail in the next section.

\begin{table}

\centering
\begin{tabular}{|c|c|c|c|c|c|c|c|}
\hline 
$\bar \lambda$& 1 & 3 & 6 & 7 & 8 & 9  \\ 
\hline 
$m^{-2}\Lambda_\text{upper}$ 
& 0.903 & 2.102 & 3.292 & 3.608 & 3.909 & 4.196 \\ 
\hline 
$m^{-2}\Lambda_\text{lower}$ 
& 0.878 & 2.093 & 3.290 & 3.604 & 3.904 & 4.189 \\ 
\hline 
$m^{+2}\Lambda^{(2)}_\text{upper}$ 
& 0.029 & 0.140 & 0.340 & 0.409 & 0.479 & 0.551  \\ 
\hline 
$m^{+2}\Lambda^{(2)}_\text{lower}$ 
& 0.027 & 0.140 & 0.340 & 0.408 & 0.478 & 0.550 \\ 
\hline 
$1-\Lambda_\text{lower}/\Lambda_\text{upper}$ 
& 0.028 & 0.004 & 0.0004 & 0.001 & 0.001 & 0.002\\ 
\hline 
\end{tabular} 

\vspace{5mm}

\begin{tabular}{|c|c|c|c|c|c|c|}
\hline 
$\bar \lambda$ &10 & 11 & 12 & 13 & 16 & 18 \\ 
\hline 
$m^{-2}\Lambda_\text{upper}$ 
&4.465 &4.773 & 5.062 & 5.347 & 5.974 & 6.681 \\ 
\hline 
$m^{-2}\Lambda_\text{lower}$ 
&4.462 & 4.753 & 5.018 & 5.310 & 5.941 & 6.635 \\ 
\hline 
$m^{+2}\Lambda^{(2)}_\text{upper}$ 
&0.624 & 0.713 & 0.804 & 0.897 & 1.149 & 1.479 \\ 
\hline 
$m^{+2}\Lambda^{(2)}_\text{lower}$ 
&0.623 & 0.708 & 0.792 & 0.887 & 1.146 & 1.472 \\ 
\hline 
$1-\Lambda_\text{lower}/\Lambda_\text{upper}$ 
&0.001 & 0.004 & \,0.009\; & 0.007 & 0.006 & 0.007 \\ 
\hline 
\end{tabular} 
\caption{The numerical values of the non-perturbative couplings $\Lambda$ and $\Lambda^{(2)}$ describing the $\phi^4$ model computed for various values of $\bar\lambda$. We present the numerical values for both the upper and the lower bound. We also indicate a relative difference between the upper and the lower values of $\Lambda$. Analogous values are shown for the sinh-Gordon model and its analytic continuation (staircase model) in table \ref{tab:sinh_gordon_parameters}.}
\label{tab:phi4_parameters}

\end{table}

\begin{figure}[p]
\centering

\includegraphics[width=1\textwidth]{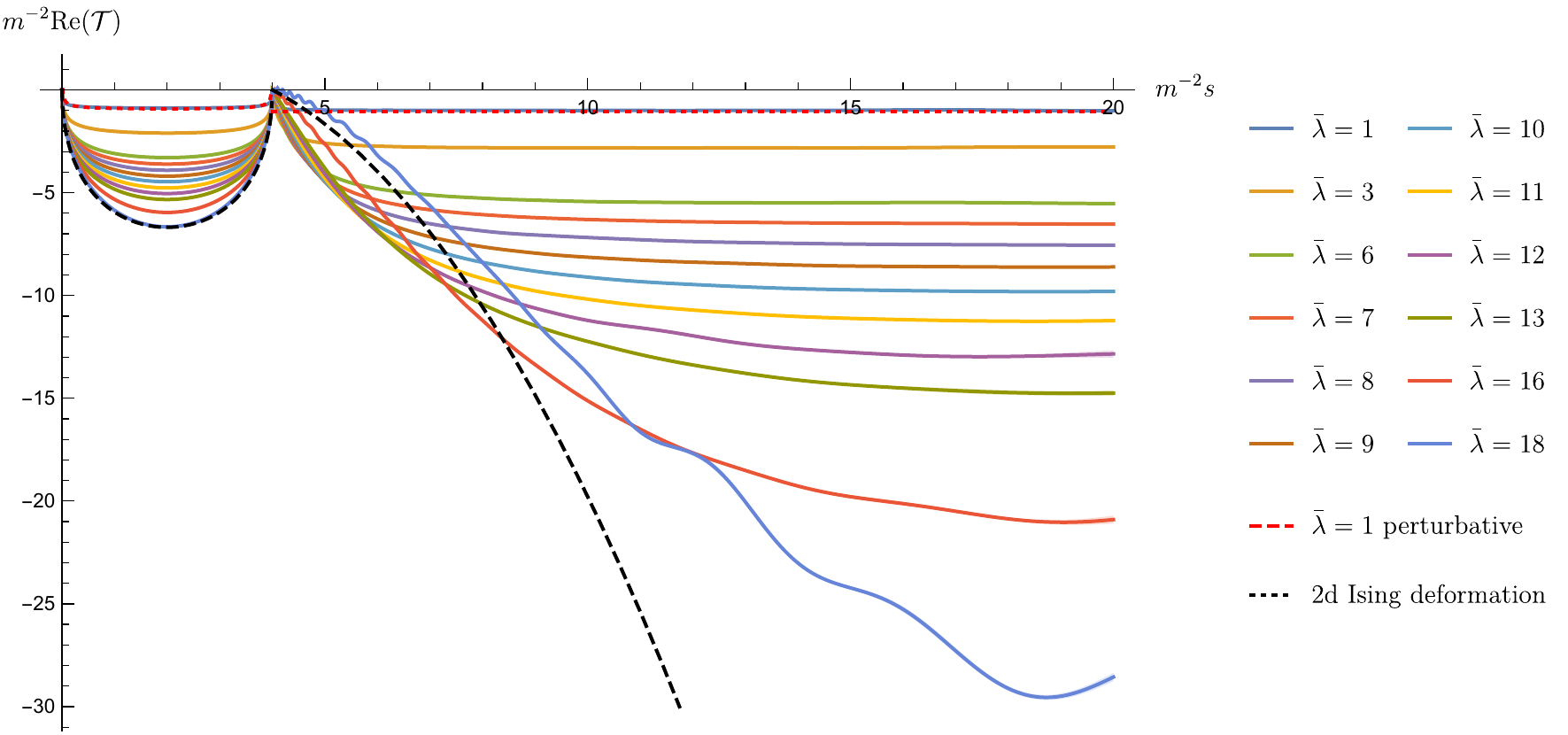}
\caption{Real part of the interacting scattering amplitude in the $\phi^4$ theory computed using the LCT data as an input to the S-matrix/form factor bootstrap problem for various values of $\bar\lambda$. As a consistency check, we also plotted the real part of the perturbative two-loop scattering amplitude  (equation (\ref{eq:amp_pert})) with $\bar\lambda=1$ (red dotted line).}
\label{fig:T_re}

\vspace{10mm}

\includegraphics[width=1\textwidth]{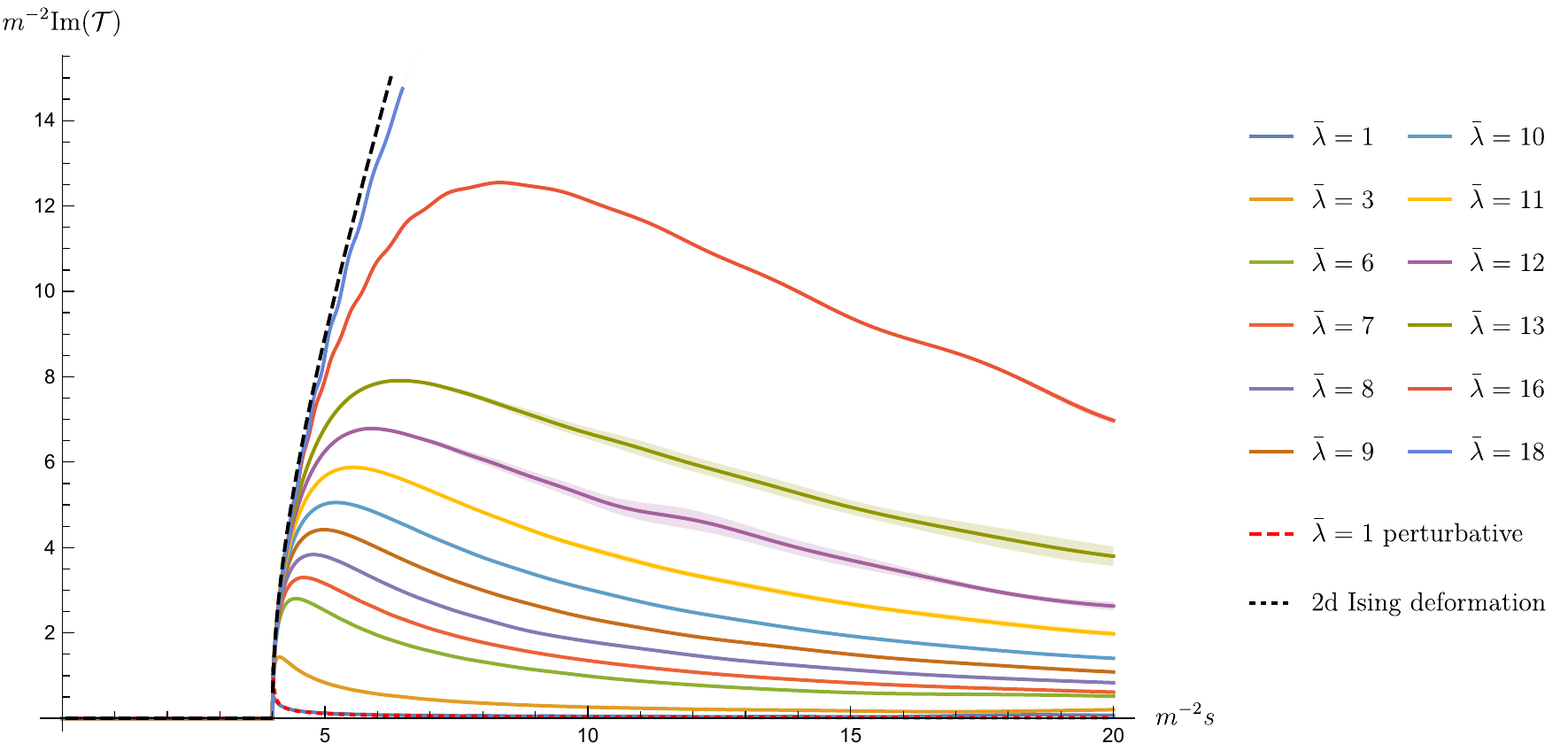}
\caption{Imaginary part of the interacting scattering amplitude in the $\phi^4$ theory computed using the LCT data as an input to the S-matrix/form factor bootstrap problem for various values of $\bar\lambda$. As a consistency check, we also plotted the imaginary part of the perturbative two-loop scattering amplitude  (equation (\ref{eq:amp_pert})) with $\bar\lambda=1$ (red dotted line).}
\label{fig:T_im}

\end{figure}

\begin{figure}[p]
\centering

\includegraphics[width=1\textwidth]{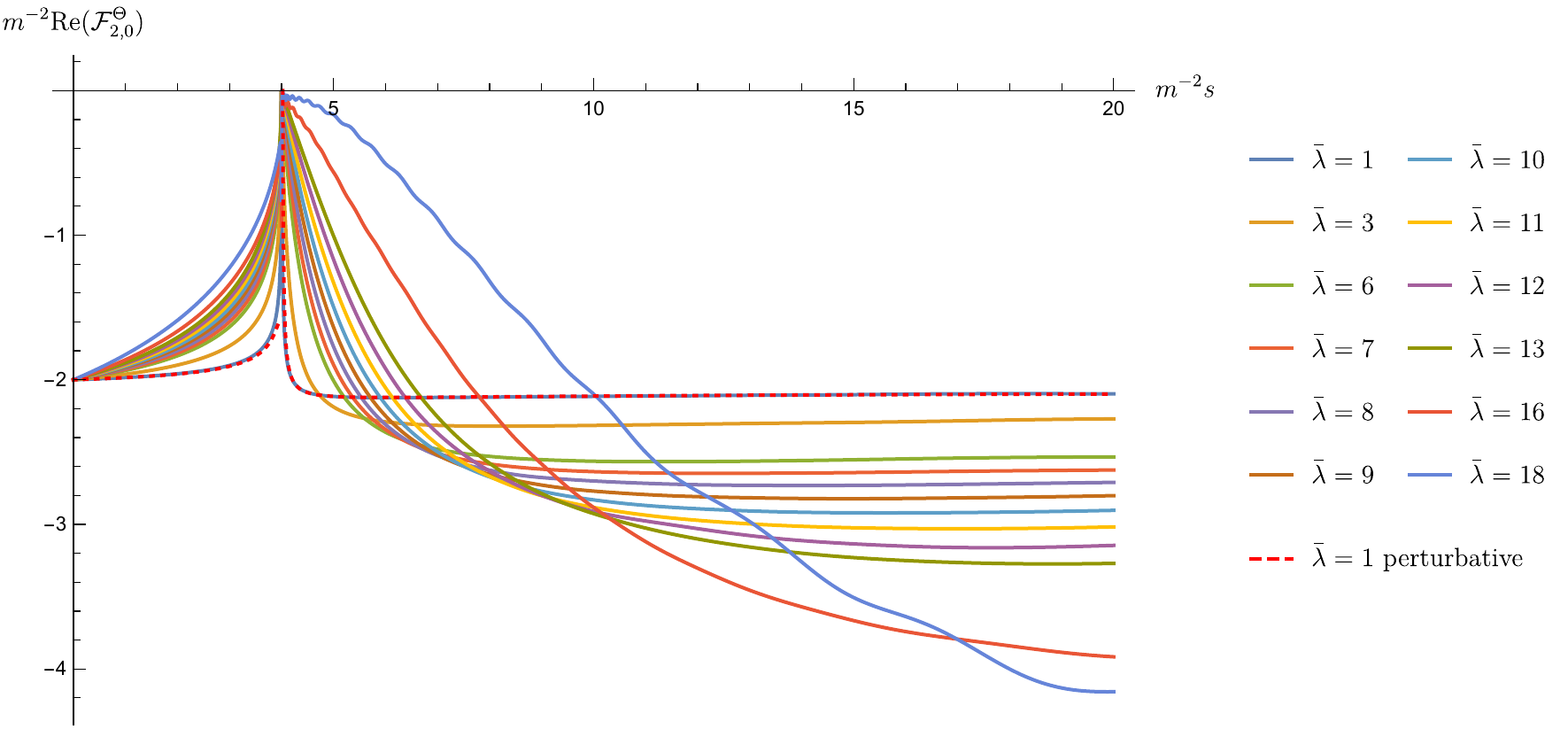}
\caption{Real part of the form factor of the trace of the stress tensor  in the $\phi^4$ theory computed using the LCT data as an input to the S-matrix/form factor bootstrap problem for various values of $\bar\lambda$. As a consistency check, we also plotted the real part of the perturbative two-loop form factor  (equation (\ref{eq:FF_pert})) with $\bar\lambda=1$ (red dotted line).}
\label{fig:FF_re}

\vspace{10mm}

\includegraphics[width=1\textwidth]{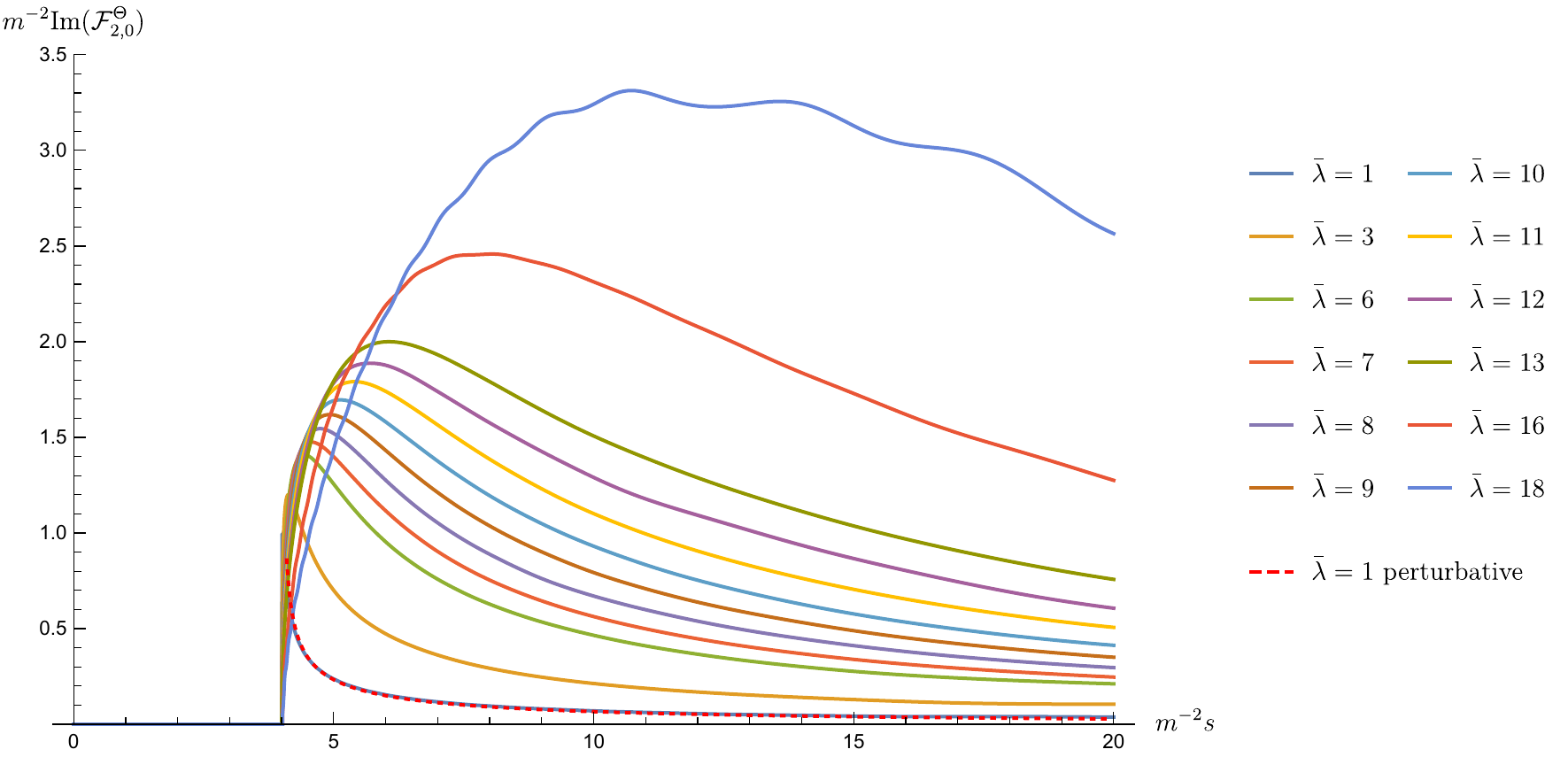}
\caption{Imaginary part of the form factor of the trace of the stress tensor  in the $\phi^4$ theory computed using the LCT data as an input to the S-matrix/form factor bootstrap problem for various values of $\bar\lambda$. As a consistency check, we also plotted the imaginary part of the perturbative two-loop form factor   (equation (\ref{eq:FF_pert})) with $\bar\lambda=1$ (red dotted line).}
\label{fig:FF_im}

\end{figure}

As a solution of our optimization problem we obtain not only the data of table \ref{tab:phi4_parameters} but also all the coefficients in the ansatz \eqref{eq:ansatz_T} and \eqref{eq:ansatz_FF}. Taking these coefficients as averages between the upper and lower bound results we obtain numerical expressions for the interacting part of the scattering amplitude and the  form factor of the trace of the stress tensor. The results are presented in figures \ref{fig:T_re} - \ref{fig:FF_im} for different values of $\bar\lambda$.
For $\bar\lambda=1$ we can compare our result with the perturbative amplitude \eqref{eq:amp_pert}. It is depicted by the red dashed lines in figure \ref{fig:T_re} and \ref{fig:T_im}. We find an excellent agreement. For completeness we provide here the perturbative value of $m^{-2}\Lambda$ for $\bar\lambda=1$ using
equation \eqref{eq:Lambda_computed}. It reads
\begin{equation}
m^{-2}\Lambda = 0.914 \pm 0.006.
\end{equation}
This value is rather close to the one of table \ref{tab:phi4_parameters} for the upper bound which is $0.903$. For $\bar\lambda=18$ we could try to compare our result with the scattering amplitude of the deformed 2d Ising model. It is given by equation \eqref{eq:bosons_fermions} and \eqref{eq:deformation_2d_Ising} and reads
\begin{equation}
\label{eq:deformed_amplitude}
\mathcal{T}(s) = 4i\sqrt{s}\sqrt{s-4m^2}+\frac{s(4m^2-s)}{M^2}.
\end{equation}
The value of $M^2$ can be estimated from equation \eqref{eq:Msq_value} by plugging there the value $m^{-2}\Lambda=6.681$ found in table \ref{tab:phi4_parameters}. The amplitude \eqref{eq:deformed_amplitude} is depicted in figure \ref{fig:T_re} and \ref{fig:T_im} by the black dashed line. We see that the $\bar\lambda=18$ result has a similar shape to the amplitude \eqref{eq:deformed_amplitude}. Notice however that this comparison is rather crude since the $\bar\lambda=18$ amplitude is still far away from the critical point (its mass gap in unit of $m_0$ is $m/m_0\simeq 0.6186$).

\begin{figure}[hpt]
\centering

\includegraphics[width=1\textwidth]{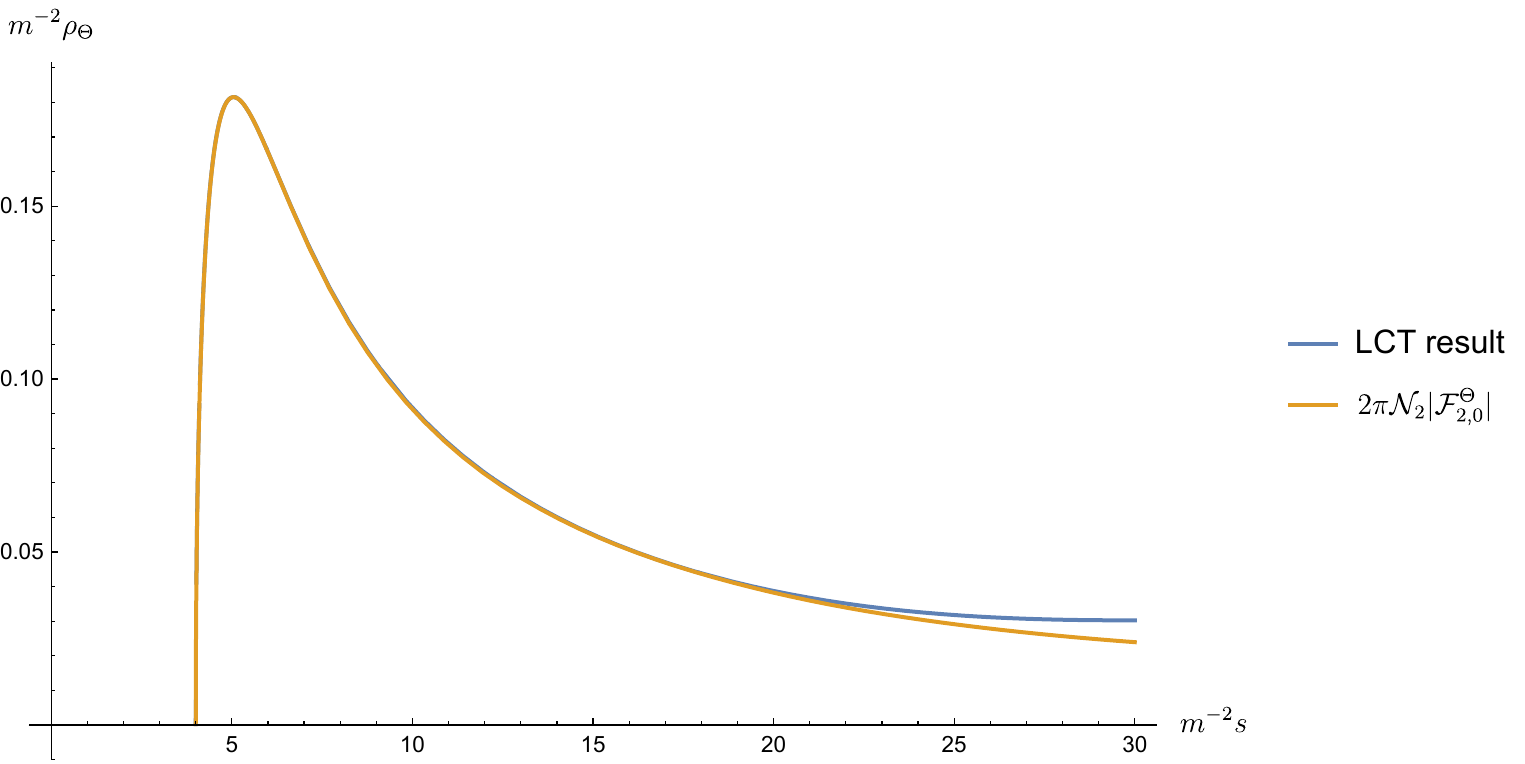}
\caption{Comparison between the LCT spectral density and the spectral density reconstructed from the obtained form factor for $\bar \lambda = 10$}
\label{fig:comparison}

\end{figure}

\begin{figure}[p]
\centering

\includegraphics[width=1\textwidth]{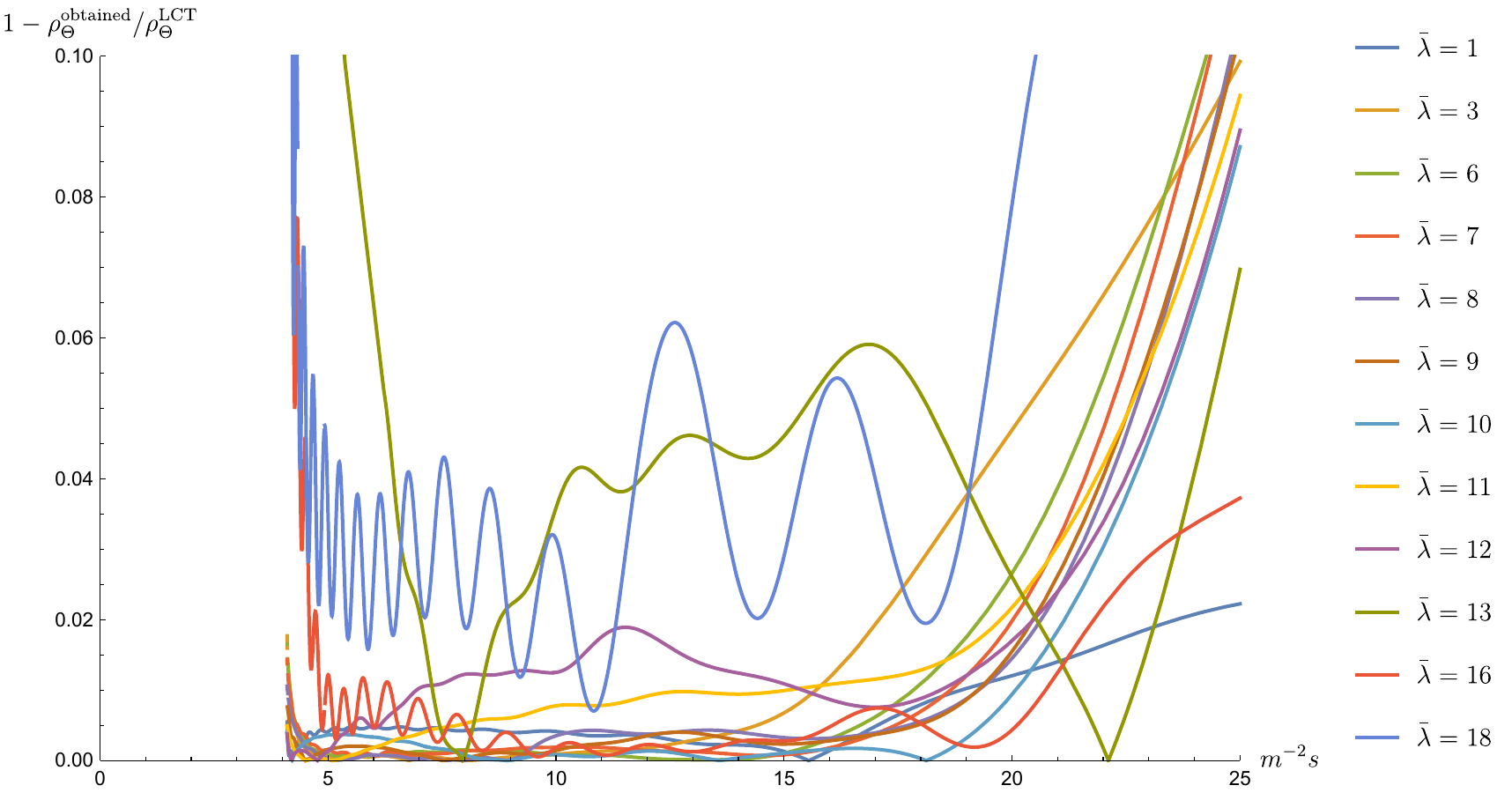}
\caption{Relative error between the LCT spectral density and the spectral density obtained from the obtained two-particle form factor for various values of $\bar\lambda$.}
\label{fig:checkRho}

\vspace{10mm}

\includegraphics[width=1\textwidth]{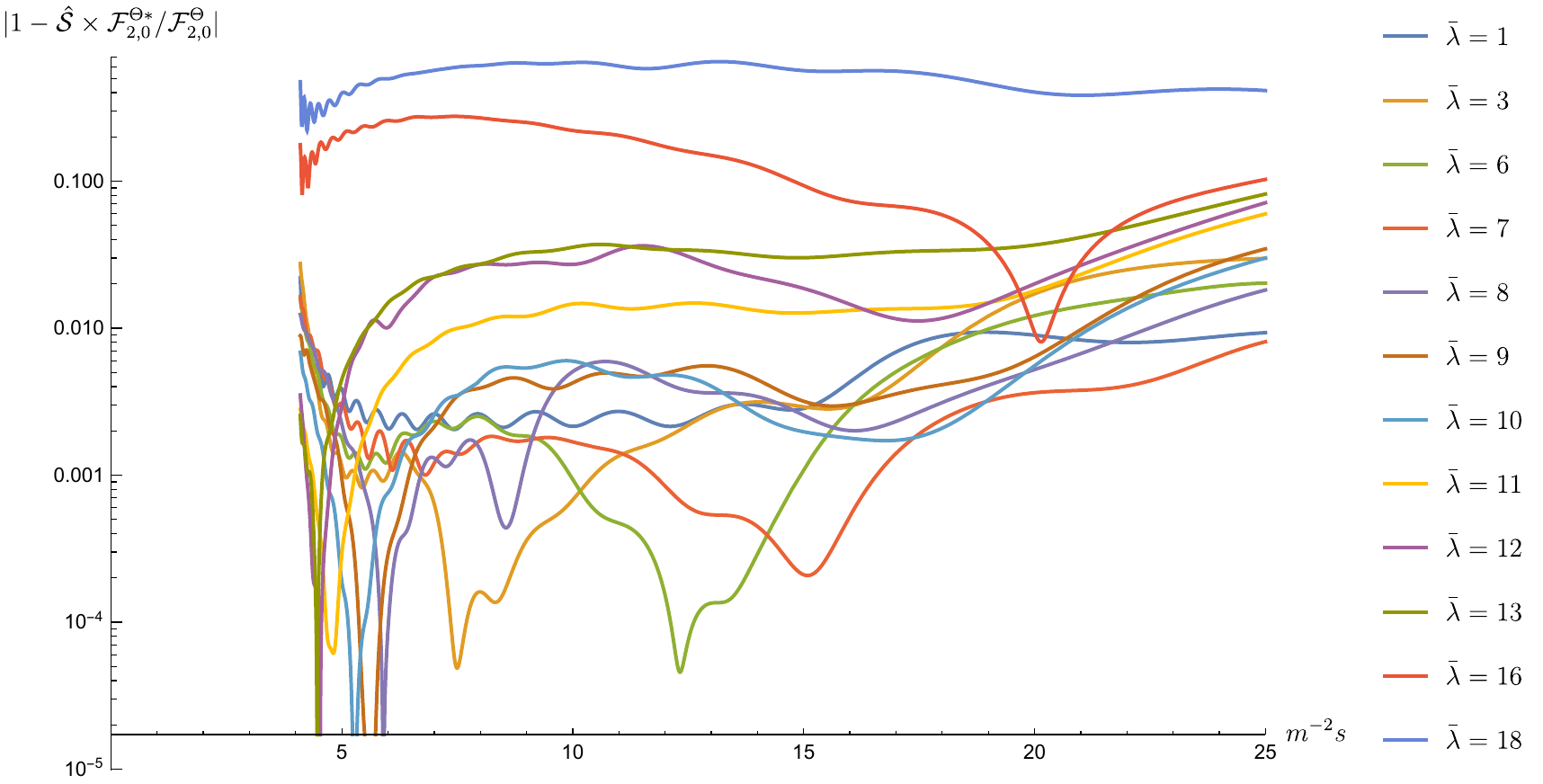}
\caption{Check of  Watson's equation \eqref{eq:Watson} using the obtained expressions of the form factor and the spectral density for various values of $\bar\lambda$. The vertical axis is given in the log scale.}
\label{fig:checkWatson}

\end{figure}

In the ``elastic'' regime $s\in[4m^2, 16m^2]$, one can reconstruct the spectral density from the obtained two particle form factor using equation \eqref{eq:SD_FF_rel}. For $\bar\lambda = 10$ we explicitly compare the reconstructed two-particle part of the spectral density with the LCT result (which was used as part of the input data to the optimization problem) in figure \ref{fig:comparison}. We see that in the elastic regime they basically coincide. We present relative error between the reconstructed two-particle part of the spectral density and the LCT result for different values of $\bar\lambda$ in figure \ref{fig:checkRho}. The relative errors become large at the threshold $s=4m^2$ (since $\rho_\Theta^\text{LCT}$ is approaching 0 as $s$ goes to $4m^2$, and a small uncertainty in the form factor can cause a somewhat large relative error), but stay relatively low in the ``elastic'' region. 
This provides a solid check for our bootstrap results for the form factor. The obtained scattering amplitude must obey  Watson's equation \eqref{eq:Watson} in the ``elastic'' regime. As presented in figure \ref{fig:checkWatson}, our bootstrap results indeed satisfy it well. This provides validation of our bootstrap results for the obtained scattering amplitudes. Outside of the ``elastic'' regime the presence of four- and higher-particle form factors becomes necessary for the bounds from unitarity to be tight. Since we do not have them in our bootstrap set-up, a potential concern is that the bootstrap algorithm tries to saturate unitarity in this regime by letting the two-particle form factor grow larger than it should be.  So it is not clear if our results for the form factor and the scattering amplitude in the $s\geq 16m^2$ regime are relevant to the $\phi^4$ model. From figure \ref{fig:checkRho} and \ref{fig:checkWatson}, one can also see that generally for larger $\bar\lambda$, the relative errors are larger. Therefore, we expect the uncertainties in the results from the S-matrix/form factor bootstrap in figure \ref{fig:T_re} - \ref{fig:FF_im} to be relatively larger for larger $\bar\lambda$.

\subsection{Comparison of the sinh-Gordon model and $\phi^4$ model}
\label{sec:comparison}

Figure \ref{fig:boundWithData} nicely summarises the results of sections \ref{sec:results_I} and \ref{sec:results_II}. It provides the allowed region in the space of consistent quantum field theories (blue region) and indicates the position of the $\phi^4$ model in this region (red and purple crosses). Remarkably, the $\phi^4$ model lies super close to the lower boundary of the allowed region where the sinh-Gordon model and its analytic continuation (the staircase model) lie. In this section we discuss the plausibility of this result.

To begin with, let us write explicitly the parameters $\Lambda$ and $\Lambda^{(2)}$ in the sinh-Gordon model (and its analytic continuation) for various values of $\beta^2$. The results are summarised in table \ref{tab:sinh_gordon_parameters}. We chose the values of $\beta^2$ in these tables in such a way that the sinh-Gordon (and its analytic continuation) has the same values of $\Lambda$ as in table \ref{tab:phi4_parameters}. As already expected from figure \ref{fig:boundWithData}, the values of $\Lambda^{(2)}$ of the $\phi^4$ model and the (analytically continued) sinh-Gordon model are almost the same. Tables \ref{tab:phi4_parameters} and \ref{tab:sinh_gordon_parameters} quantify this similarity.

The comparison of tables \ref{tab:phi4_parameters} and \ref{tab:sinh_gordon_parameters} can be summarized as follows: given some value of $\bar\lambda$ in the $\phi^4$ model, there is always some value $\beta^2$ in the (analytically continued) sinh-Gordon model which results in the $\Lambda$ and $\Lambda^{(2)}$ values similar to the ones in the $\phi^4$ model. Only in the preturbative regime when $\bar\lambda\ll4\pi$ we have $\beta^2\approx \bar\lambda$. (For example, the $\phi^4$ model with $\bar\lambda=1$ is similar to the sinh-Gordon model with $\beta^2=1.043$.)

\begin{figure}
\centering

\includegraphics[height=4.1cm]{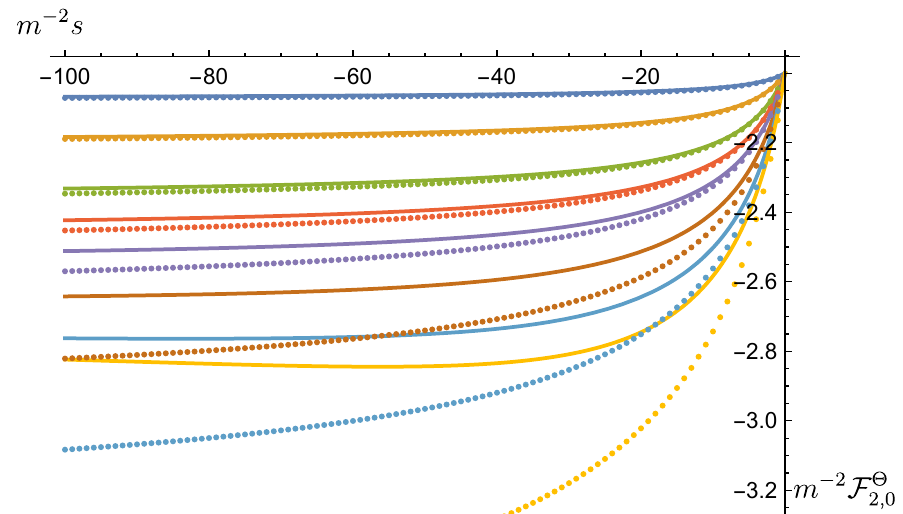}\quad
\includegraphics[height=4.1cm]{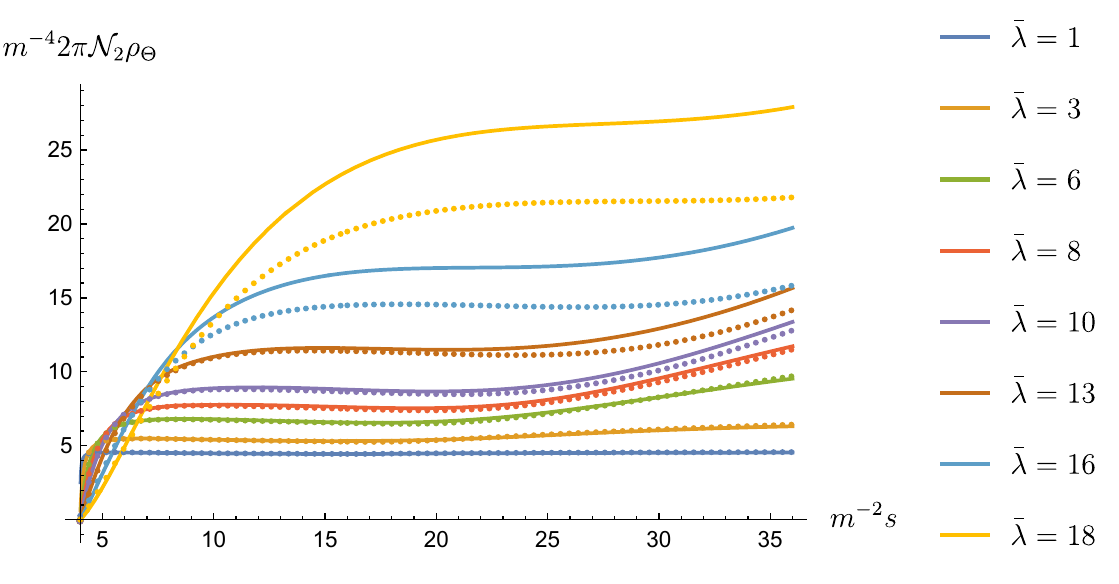}
\caption{Comparison of the  form factor of the stress tensor at $s\le 0$ (left plot) and its spectral density (right plot) in the $\phi^4$ model (solid lines) and the (analytically continued) sinh-Gordon model (dotted lines) for $\Lambda=\{0.903, 2.102, 3.292, 3.909, 4.465, 5.347, 5.974, 6.681\}$ which correspond to $\bar\lambda =\{1,3,6,8,10,13,16,18\}$ in the $\phi^4$ model according to table \ref{tab:phi4_parameters}. The solid lines for the $\phi^4$ model are from light-cone truncation computation, while the dotted lines for the sinh-Gordon model are from the analytic  form factor formulas (\ref{eq:form_factor_sinh-Gordon}) and (\ref{eq:4particleFF}). Note that for the sinh-Gordon spectral densities, we included the four-particle form factor contribution, so that the result shown above is exact up to $s=36m^2$.}
\label{fig:SGandPhi4Comparison_1}

\vspace{3ex}

\includegraphics[height=4.1cm]{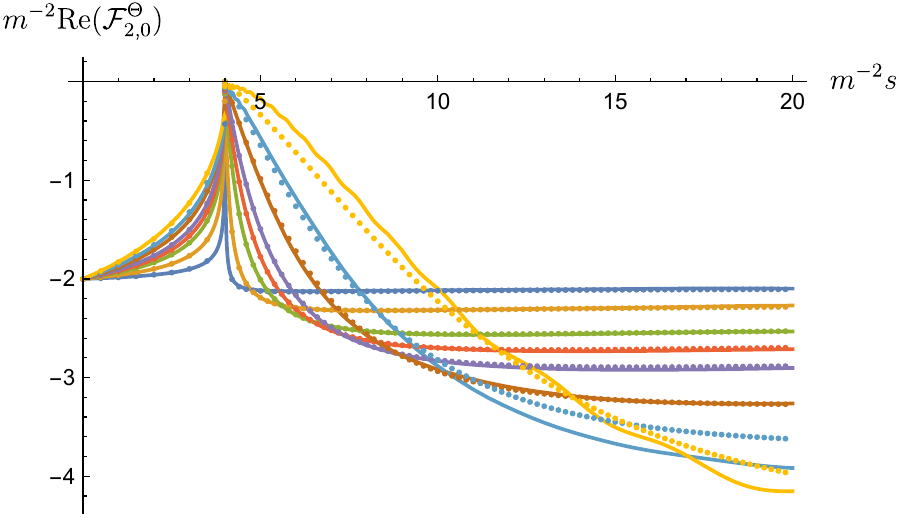}\quad
\includegraphics[height=4.1cm]{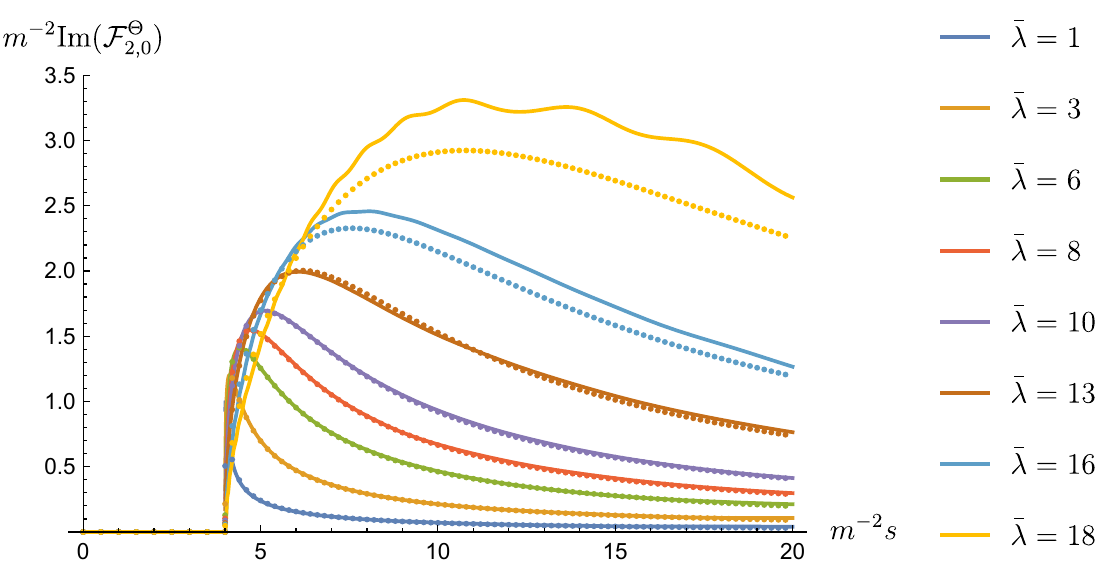}
\caption{Comparison of the real part (left plots) and imaginary part (right plot) of the form factor of the stress tensor at $s>0$ in the $\phi^4$ model (solid lines) and the (analytically continued) sinh-Gordon model (dotted lines) for $\Lambda=\{0.903, 2.102, 3.292, 3.909, 4.465, 5.347, 5.974, 6.681\}$ which correspond to $\bar\lambda =\{1,3,6,8,10, 13,16,18\}$ in the $\phi^4$ model according to table \ref{tab:phi4_parameters}. The solid lines for the $\phi^4$ model are from the S-matrix/form factor bootstrap, while the dotted lines for sinh-Gordon are from the analytic two-particle form factor formula (\ref{eq:form_factor_sinh-Gordon}).}
\label{fig:SGandPhi4comparison_2}

\end{figure}

\begin{figure}
\centering

\includegraphics[height=4.1cm]{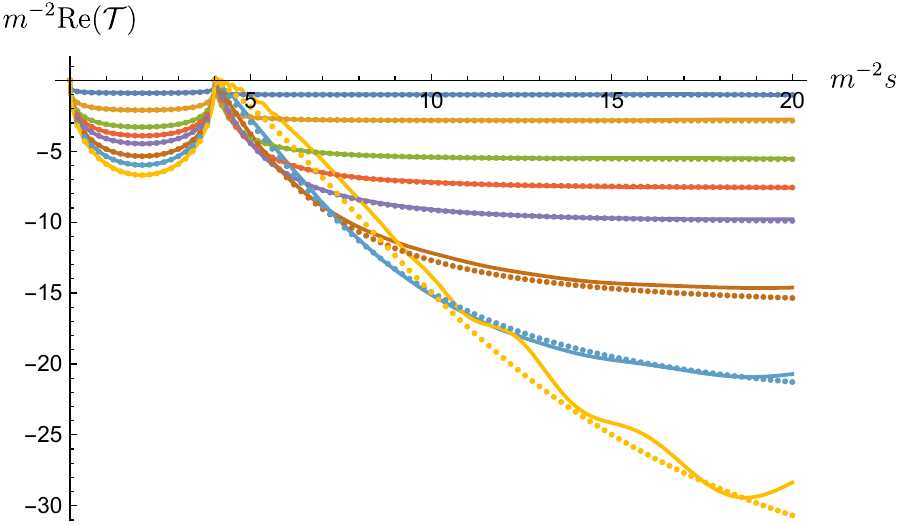}\quad
\includegraphics[height=4.1cm]{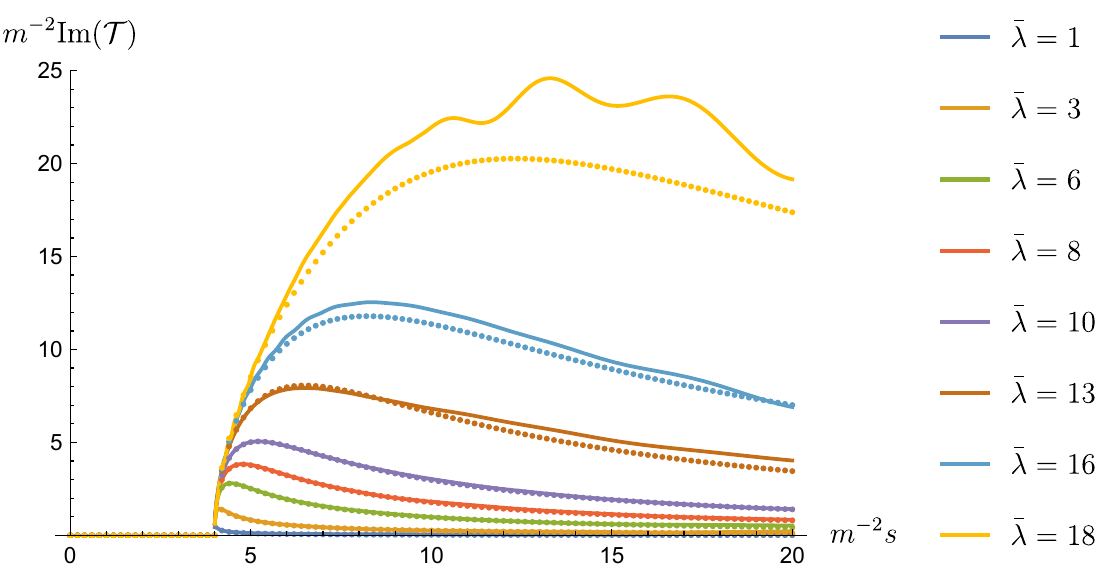}
\caption{Comparison of the two-to-two scattering amplitudes in the $\phi^4$ model (solid lines) and the (analytically continued) sinh-Gordon model (dotted lines) for $\Lambda=\{0.903, 2.102, 3.292, 3.909, 4.465, 5.347, 5.974, 6.681\}$, which correspond to $\bar\lambda =\{1,3,6,8,10,13,16,18\}$ in the $\phi^4$ model according to table \ref{tab:phi4_parameters}. The solid lines for the $\phi^4$ model are from the S-matrix/form factor bootstrap, while the dotted lines for the sinh-Gordon model are from the analytic  S-matrix formula (\ref{eq:amp_sinhGordon})}
\label{fig:phi4VsSGTPlots}
\end{figure}

Using the values of $\Lambda$ in table \ref{tab:phi4_parameters}, one can compute the scattering amplitude, the form factor of the trace of the stress tensor  and spectral density in the sinh-Gordon model (and its analytic continuation) using the results of section \ref{sec:sinh_gordon}. We compare them with our LCT expressions for the form factor at  $s\le 0$ and the spectral density in figure \ref{fig:SGandPhi4Comparison_1}. We observe that the two models have a very similar behaviour in a wide range of values $s$ even at strong coupling. Since the $\phi^4$ LCT data is so close to the sinh-Gordon model, it is not surprising that  the form factor at $s>0$ and the scattering amplitudes we obtain from the numerical optimization will also be similar to those of the sinh-Gordon model. 
To be concrete, we compare the form factors at  $s>0$ in the two models in figure \ref{fig:SGandPhi4comparison_2} and the scattering amplitudes in figure \ref{fig:phi4VsSGTPlots}. Notice especially that the form factors and scattering amplitudes for these two theories are almost the same at $0<s<4m^2$, even for large $\bar\lambda$. This also explains what we saw in figure \ref{fig:boundWithData}. 

It is important to stress that the amplitude and the form factor in the $\phi^4$ and sinh-Gordon models must differ in the non-elastic regime $s>16m^2$, however our bootstrap method does not allow us to compute the $\phi^4$ observables in this regime reliably to see the difference. 
It is interesting that there exist amplitudes belonging to different models which are very similar in the elastic regime and differ significantly in the non-elastic regime. See \cite{Tourkine:2021fqh} for a related discussion, where the authors studied the question of how sensitive the elastic part of the amplitude is to the inelastic regime, if one regards the latter as an input to the S-matrix bootstrap and the former as an output.  In particular, it would likely shed light on the similarity of the $\phi^4$ and sinh-Gordon elastic amplitudes by studying how much our S-matrix bounds vary under changes of the inelastic amplitudes, using the framework of \cite{Tourkine:2021fqh}.

\begin{table}
\centering

\begin{tabular}{|c|c|c|c|c|c|}
\hline 
$\beta^2$ & 1.043 & 3.298 & 8.219 & 11.1466 & 17.183 \\ 
\hline 
$m^{-2}\Lambda$ 
& 0.908 & 2.102 & 3.292 & 3.610 & 3.912 \\ 
\hline 
$m^{2}\Lambda^{(2)}$ 
& 0.026 & 0.138 & 0.339 & 0.407 & 0.478  \\ 
\hline 
\end{tabular} 

\vspace{5mm}
\begin{footnotesize}
\begin{tabular}{|c|c|c|c|c|c|c|c|}
\hline 
$\beta^2$ & $8\pi \,e^{-0.560 i}$ & $8\pi \,e^{-0.850 i}$ & $8\pi \,e^{-1.081 i}$ & $8\pi \,e^{-1.255 i}$ & $8\pi \,e^{-1.402 i}$ & $-8\pi \,e^{+1.466 i}$ &  $-8\pi \,e^{+1.196 i}$\\ 
\hline 
$m^{-2}\Lambda$ 
& 4.197 & 4.466 & 4.772 & 5.062 & 5.346 & 5.974 & 6.681 \\ 
\hline 
$m^{2}\Lambda^{(2)}$ 
& 0.551 & 0.623 & 0.712& 0.801 & 0.893 & 1.115 & 1.395 \\ 
\hline 
\end{tabular} 
\end{footnotesize}

\caption{The numerical values of non-perturbative couplings $\Lambda$ and $\Lambda^{(2)}$ describing the sinh-Gordon model and its analytic continuation (staircase model) computed for various values $\beta^2$. Analogous values are shown for the $\phi^4$ model in table \ref{tab:phi4_parameters}.   }
\label{tab:sinh_gordon_parameters}

\end{table}

\newpage
\section{Discussion and Future Directions}
\label{sec:discussion}

The main purpose of this paper was to start with some nonperturbative data for a specific model, in this case for $\phi^4$ theory in 2d, and to inject that data into the S-matrix/form factor bootstrap in order to compute additional observable quantities.  Ideally, one might hope that with a finite amount of such data, the constraints of crossing, analyticity, and unitarity completely determine the rest of the theory.  Less ambitiously, the S-matrix bootstrap/form factor might simply provide a robust method to extract additional results from some initial data.  In our specific application, our `initial data' was the spectral density of the stress tensor, and its form factor with two-particle states in a certain kinematic regime $s\le 0$, computed using lightcone Hamiltonian truncation methods from our companion paper \cite{truncffsd}. Roughly, the bootstrap can take this data and obtain the form factor in a different kinematic regime, at $s>0$, after which the form of the elastic scattering amplitude follows from Watson's theorem.  An important part of the challenge was that the input data itself is determined numerically, so that simply analytically continuing between different kinematic regimes is not straightforward.\footnote{In a system at finite volume, Luscher's method \cite{Luscher:1985dn,Luscher:1986pf} provides another handle on the elastic scattering amplitudes, which could be used to verify or improve the S-matrix bootstrap results.  The work \cite{bajnok2016truncated} applied Luscher's method to equal-time Hamiltonian truncation in finite volume in the broken phase of $\phi^4$ theory, and it should be possible to repeat their analysis in the unbroken phase that we have studied in this work.  One could instead try to obtain the finite volume spectrum from lattice Monte Carlo rather than from truncation methods. }

One of the surprises of our analysis is that the elastic 2-to-2 S-matrix in $\phi^4$ theory is extremely close to that of the sinh-Gordon model and its analytic continuation (the staircase model), after the couplings of both models are adjusted to have the same value of $\Lambda$ (the interacting part of the scattering amplitude value at the crossing-symmetric point $s=2m^2$).  The fact that the scattering amplitudes in both models are are somewhat close is perhaps not  very surprising.  As we have emphasized, the 2-to-2 S-matrices for the two theories are identical in perturbation theory around $\Lambda=0$ until $\CO(\Lambda^4)$, which is the first order in perturbation theory where $ \phi^4$ has particle production.  Moreover, both theories reach a critical point at the upper limit $\Lambda=8m^2$ where they describe the Ising model S-matrix, and perturbation theory around this upper limit is described at $\CO( 8m^2-\Lambda)$ by the leading irrelevant deformation $T \bar{T}$, so the first difference between the theories arises at $\CO( (8m^2-\Lambda)^2)$.  So one could reasonably expect the S-matrices  to be quite similar in between these two limits.  Nevertheless, the degree to which they agree even at intermediate strongly coupled values is still remarkable.  One might worry that this agreement is an artifact of the S-matrix bootstrap itself, which tends to push theories to saturate unitarity conditions and therefore tends to find integrable models.  In fact, we have shown that a pure S-matrix bootstrap analysis, without any injection of dynamical data from LCT, exactly finds the sinh-Gordon/staircase model S-matrix.  However, we emphasize that our $ \phi^4$ S-matrix bootstrap analysis used a different optimization condition from our pure S-matrix bootstrap analysis.  In the former, we fixed the data from LCT and maximized $\Lambda$, whereas in the latter we fixed $\Lambda$ and maximized the second derivative $\Lambda^{(2)}$ of the S-matrix at the crossing-symmetric point.  Moreover, $\phi^4$ theory really should saturate unitarity in the elastic regime $4m^2 < s < 16m^2$ due to kinematics, so one cannot think of this saturation as an artifact of the S-matrix bootstrap. Rather, in practical terms it appears that the origin of this close agreement is that even at strong coupling, the stress tensor form factor at $s\le 0$ and the spectral density at $4m^2 < s < 16m^2$, which we compute in LCT, is very similar to that of sinh-Gordon/staircase model.\footnote{We also compute the stress tensor spectral density at $s> 16m^2$, and here we do see a significant deviation between $\phi^4$ and sinh-Gordon.   However, the S-matrix bootstrap result for the elastic scattering amplitude does not seem to be very sensitive to the detailed behavior of the spectral density in this regime. } 

Although 2-to-2 elastic scattering appears to be very similar in $ \phi^4$ theory and the sinh-Gordon model, we do not expect it to be similar at large $s$ and it certainly cannot be similar for 2-to-$2+n$ since particle production exactly vanishes in sinh-Gordon.  The S-matrix bootstrap with both two- and four-particle external states would therefore be particularly illuminating in this case since it would uncover more of the qualitative difference between the two models.  In $d>2$, including higher multiplicities in the S-matrix bootstrap is likely quite challenging due to the large kinematic parameter space, but in $d=2$ we are optimistic that it would be practical.  If one wanted to use the S-matrix bootstrap in combination with UV CFT operators, as we have done in this work, then the inclusion of four-particle external states would necessitate the appearance of four-particle form factors $\CF_{4,0}^\Theta$ in the unitarity condition which is very hard to compute in the LCT framework. Perhaps, one could simply parameterize it and try to obtain it as one of the outputs of the S-matrix bootstrap.

Finally, we end by mentioning possible generalizations of the method.   There are many other models in 2d that would be interesting to analyze using this approach.  LCT can be applied to theories with  more general field content in 2d, including gauge fields and fermions, and 2d QCD at finite $N_c$ would be a particularly interesting application.\footnote{See e.g. \cite{Dempsey:2021xpf,Katz:2014uoa,Katz:2013qua} for recent LCT and DLCQ applications to 2d QCD.}   Our approach here is similar in spirit to that of \cite{Gabai:2019ryw}, which studied Ising Field theory with both a $\sigma$ and $\epsilon$ deformation using TFFSA and Luscher's method \cite{Luscher:1985dn,Luscher:1986pf}, but it would be interesting to see if any more mileage could be gained by also including form factors and spectral densities in a generalized unitarity condition as we did in this paper.
  More ambitiously, our method in principle can be applied to higher dimensions, the main challenge being that it is difficult to obtain the input data. LCT has been applied to the $\phi^4$ model in 3d, and the stress tensor spectral density was obtained in \cite{Anand:2020qnp}.\footnote{Both lightcone and equal-time Hamiltonian truncation have seen important recent progress for $\phi^4$ theory in $d>2$ \cite{Hogervorst:2014rta,Katz:2016hxp,Elias-Miro:2020qwz,Anand:2020qnp}. One of the main challenges has been dealing with state-dependent counterterms for divergences. The recent works \cite{Elias-Miro:2020qwz,Anand:2020qnp} developed systematic methods to handle this issue and specifically applied their work in the context of 3d $\phi^4$ theory.} One would have to generalize  our treatment of form factors to 3d, but the basic idea would be the same.  In $d>2$ there are two stress-tensor two-particle form factors $\mathcal{F}_{2,0}^\Theta(s)$ and $\mathcal{F}_{2,0}^{(2)}(s)$, as  well as two spectral densities $\rho_\Theta(s)$ and $\rho_2(s)$, and the scattering amplitude  $\mathcal{S}(s,t)$ can be decomposed into partial amplitudes $\mathcal{S}_j(s)$ with $j=0,2,4,\ldots$.  The generalization of the unitarity constraint \eqref{eq:unitarity_3x3} was worked out  in \cite{Karateev:2020axc}.\footnote{See equations (6.36) and (6.41) there.}  So although generalizing our work to 3d would involve significant work, at least all the pieces have already been assembled and are waiting to be used.

\begin{center}
\subsection*{Acknowledgments}
\end{center}

We thank Ami Katz, Alexander Monin, Giuseppe Mussardo, Jo\~ao Penedones, Balt van Rees, Matthew Walters, for helpful conversations, and in particular Ami Katz and Matthew Walters for comments on a draft. ALF and HC were supported in part by the US Department of Energy Office of Science under Award Number DE-SC0015845 and the Simons Collaboration Grant on the Non-Perturbative Bootstrap, and ALF in part by a Sloan Foundation fellowship. 

\

\appendix

\section{Kinematics of 2d Scattering}
\label{app:kinematics_2d}
Consider the scattering of two identical scalar particles in two space-time dimensions. We denote the initial two-momenta of two particles (before the scattering) by $p_1^\mu$ and $p_2^\mu$ and the final two-momenta of two particles (after the scattering) by $k_1^\mu$ and $k_2^\mu$. The two particles obey the mass-shell condition
\begin{equation}
\label{eq:mass_shell}
p_i^2 = k_i^2 = -m^2,
\end{equation}
where $i=1,2$. The conservation of two-momenta leads to the requirement
\begin{equation}
\label{eq:conservation}
p_1^\mu+p_2^\mu-k_1^\mu-k_2^\mu =0.
\end{equation}
Due to the mass-shell condition \eqref{eq:mass_shell}, there are only two different solutions for the two momenta after the scattering, namely
\begin{equation}
\label{eq:two_solutions}
\vec k_1 = \vec p_1,\quad \vec k_2= \vec p_2
\qquad\text{or}\qquad
\vec k_1 = \vec p_2,\quad \vec k_2= \vec p_1.
\end{equation}
Let us recall that the Mandelstam variables are defined as
\begin{align}
\label{eq:Mandelstam}
s\equiv - (p_1+p_2)^2,\qquad
t\equiv - (p_1-k_1)^2,\qquad
u\equiv - (p_1-k_2)^2.
\end{align}
Plugging the two solutions \eqref{eq:two_solutions} into the definition of the Mandelstam variables we see that they correspond to two different situation
\begin{equation}
t=0
\qquad\text{or}\qquad
u=0.
\end{equation}

The two solutions \eqref{eq:two_solutions} are related by the discrete $Z_2$ symmetry $\vec k_1 \leftrightarrow \vec k_2$. The scattering in $d=2$ happens on the line. It is standard to work with the convention when two particle states are defined in such a way that particle 1 (with momentum $\vec p_1$) is to the left of particle 2 (with momentum $\vec p_2$) on the line. Then the in two-particle states are required to obey $\vec p_1>\vec p_2$. This condition forces the trajectories of two particles to cross as time goes by. Instead the out two-particle states are required to obey $\vec p_1<\vec p_2$ condition which ensures that the particles will never meet in the future. When considering the scattering process $p_1 p_2\rightarrow k_1k_2$, the above convention is imposed by adding the following product of step-functions
\begin{equation}
\theta(\vec p_1-\vec p_2)\theta(\vec k_2-\vec k_1)
\end{equation}
into the definition of 2d scattering amplitudes. Plugging here the solution \eqref{eq:two_solutions} we see that in this convention the $t=0$ solution vanishes and we are left only with the $u=0$ solution.

Let us now derive a very useful relation. Consider the Dirac $\delta$-function which encodes the conservation condition \eqref{eq:conservation}, namely
\begin{equation}
\label{eq:Diract_delta_function}
\delta^2(p_1+p_2-k_1-k_2)=
\delta(p_1^0+p_2^0-k_1^0-k_2^0)
\delta(\vec p_1+\vec p_2-\vec k_1-\vec k_2).
\end{equation}
Here the energies $p_i^0$ and $k_i^0$ are fixed in term of the momenta $\vec p_i$ and $\vec k_i$ due to the mass-shell condition \eqref{eq:mass_shell}. Given the initial values of $\vec p_i$, this Dirac $\delta$-function restricts the values of $\vec k_i$ to their allowed range, in 2d this restriction is severe and leads only to two possibilities \eqref{eq:two_solutions}. Let us now imagine that we would like to integrate \eqref{eq:Diract_delta_function} with some kernel over all possible values of $\vec k_i$, namely
\begin{equation}
\int_{-\infty}^{+\infty} d\vec k_1\int_{-\infty}^{+\infty} d\vec k_2\, f(\vec k_1, \vec k_2) \delta^2(p_1+p_2-k_1-k_2).
\end{equation}
In order to perform this integration we need to perform several steps which we explain below.

Due to the second Dirac $\delta$-function in the right-hand side of \eqref{eq:Diract_delta_function}, we have $\vec k_2 = \vec p_1+\vec p_2-\vec k_1$. Thus, we can fully eliminate the integral over $\vec k_2$. Plugging this restriction back into \eqref{eq:Diract_delta_function} and using the mass-shell condition \eqref{eq:mass_shell}, we get
\begin{equation}
\delta(p_1^0+p_2^0-k_1^0-k_2^0)=
\delta\Big(g(\vec k_1)\Big),
\end{equation}
where we have defined
\begin{equation}
g(\vec k_1) \equiv
p_1^0+p_2^0-\sqrt{m^2+\vec k_1^{\,2}}-\sqrt{m^2+(\vec p_1+\vec p_2-\vec k_1)^{\,2}}.
\end{equation}
Let use now use the standard property of the Dirac $\delta$-functions and the fact that $g(\vec k_1)=0$ has only two solutions given by \eqref{eq:two_solutions}. We have then
\begin{equation}
\delta(p_1^0+p_2^0-k_1^0-k_2^0)=
\frac{\delta(\vec k_1-\vec p_1)}{\Big| g'(\vec p_1) \Big|}+
\frac{\delta(\vec k_1-\vec p_2)}{\Big| g'(\vec p_2) \Big|}.
\end{equation}
Evaluating the derivatives we finally obtain
\begin{equation}
\label{eq:result}
\delta(p_1^0+p_2^0-k_1^0-k_2^0)=
\frac{p_1^0p_2^0}{\left|\vec p_1 p_2^0-\vec p_2 p_1^0\right|}\times\left( \delta(\vec k_1-\vec p_1)+\delta(\vec k_1-\vec p_2) \right).
\end{equation}
Plugging \eqref{eq:result} into \eqref{eq:Diract_delta_function}, we get the final relation
\begin{multline}
\label{eq:final_relation_Dirac}
\delta^2(p_1+p_2-k_1-k_2)=\\
\frac{p_1^0p_2^0}{\left|\vec p_1 p_2^0-\vec p_2 p_1^0\right|}\times\left( \delta(\vec k_1-\vec p_1)\delta(\vec k_2-\vec p_2)+\delta(\vec k_1-\vec p_2)\delta(\vec k_2-\vec p_1) \right).
\end{multline}
Equivalently we could write it as
\begin{multline}
\label{eq:final_relation_Dirac_2}
4\left|\vec p_1 p_2^0-\vec p_2 p_1^0\right|\times\delta^2(p_1+p_2-k_1-k_2)=\\
4p_1^0p_2^0\times\left( \delta(\vec k_1-\vec p_1)\delta(\vec k_2-\vec p_2)+\delta(\vec k_1-\vec p_2)\delta(\vec k_2-\vec p_1) \right).
\end{multline}

Let us now evaluate the expression \eqref{eq:final_relation_Dirac_2} in the center of mass frame defined as $\vec p_2 = -\vec p_1$. Plugging this condition into \eqref{eq:Mandelstam} we conclude that in the center of mass frame
\begin{equation}
|\vec p_1| = \frac{1}{2}\,\sqrt{s-4m^2},\qquad
p_1^0=\frac{1}{2}\,\sqrt{s}.
\end{equation}
Plugging these into the left-hand side of \eqref{eq:final_relation_Dirac_2} we get
\begin{equation}
\label{eq:COM_N2_factor}
4\left|\vec p_1 p_2^0-\vec p_2 p_1^0\right| = 2\sqrt{s}\sqrt{s-4m^2}.
\end{equation}
We then notice that the quantity $4\left|\vec p_1 p_2^0-\vec p_2 p_1^0\right|$ is Lorentz invariant, thus \eqref{eq:COM_N2_factor} holds in a generic frame!
The result \eqref{eq:final_relation_Dirac_2} together with \eqref{eq:COM_N2_factor} gives precisely  \eqref{eq:normalization_2PS_2d}.

\section{$O(N)$ model}
\label{app:ON}

Let us consider the case when the system has a global $O(N)$ symmetry.  We will require our asymptotic states to transform in the vector representation of  $O(N)$. They will thus carry an extra label $a=1\ldots N$. The one particle states are normalized as before with an addition of the Kronecker delta due to the presence of the $O(N)$ vector indicies
\begin{align}
\label{eq:normalization_global}
{}_b\<m,\vec p_2|m,\vec p_1\>_a
=2p^0 \delta_{ab}\times 2\pi\delta(\vec p_2-\vec p_1).
\end{align}
The full scattering amplitude can be decomposed into three independent scattering amplitudes $\sigma_i(s)$, $i=1,2,3$. In the notation of \cite{Zamolodchikov:1978xm} we have
\begin{align}
\nn
{}_{cd}\< m,\vec p_3;m,\vec p_4|S|m,\vec p_1;m,\vec p_2\>_{ab}=&
(2\pi)^{2}\delta^{(2)}(p_1+p_2-p_3-p_4)\times\\
&\big(\sigma_1(s)\delta_{ab}\delta_{cd}+
\sigma_2(s)\delta_{ac}\delta_{bd}+
\sigma_3(s)\delta_{ad}\delta_{bc}
\big).
\label{eq:scattering_amplitudes_ZZ}
\end{align}
Crossing $1\leftrightarrow 3$ implies the following relations
\begin{equation}
\label{eq:crossing}
\sigma_1(s) = \sigma_3(4m^2-s),\quad
\sigma_2(s) = \sigma_2(4m^2-s).
\end{equation}

Let us discuss unitarity now.
The two-particle states transform in the reducible $O(N)$ representation and can be further decomposed into three irreducible representations as
\begin{equation}
\label{eq:global_decomposition}
|m,\vec p_1;m,\vec p_2\>_{ab}=
\frac{\delta_{ab}}{\sqrt{N}}|m,\vec p_1;m,\vec p_2\>^{\bullet}+
|m,\vec p_1;m,\vec p_2\>^{\textbf{S}}_{(ab)}+
|m,\vec p_1;m,\vec p_2\>^{\textbf{A}}_{[ab]},
\end{equation}
where we have defined
\begin{align}
\label{eq:bullet}
|m,\vec p_1;m,\vec p_2\>^{\bullet} &\equiv \frac{1}{\sqrt{N}}\,\sum_{a=1}^N|m,\vec p_1;m,\vec p_2\>_{aa},\\
|m,\vec p_1;m,\vec p_2\>^{\textbf{S}}_{(ab)} &\equiv \frac{1}{2}\,\Big(
|m,\vec p_1;m,\vec p_2\>_{ab}+|m,\vec p_1;m,\vec p_2\>_{ba}\Big)-\frac{\delta_{ab}}{\sqrt{N}}|m,\vec p_1;m,\vec p_2\>^{\bullet},\\
|m,\vec p_1;m,\vec p_2\>^{\textbf{A}}_{[ab]} &\equiv \frac{1}{2}\,\Big(
|m,\vec p_1;m,\vec p_2\>_{ab}-|m,\vec p_1;m,\vec p_2\>_{ba}\Big).
\end{align}
The labels $\bullet$, $\textbf{S}$ and $\textbf{A}$ stand for trivial, symmetric traceless and antisymmetric representations. Using the normalization condition \eqref{eq:normalization_global} we find that
\begin{align}
\label{eq:normalization_trivial}
{}^{\bullet}\langle m,\vec p_3;m,\vec p_4|m,\vec p_1;m,\vec p_2\>^{\bullet}
&=\mathcal{N}_2\times(2\pi)^{2}\delta^{(2)}(p_1+p_2-p_3-p_4),\\
{}_{(cd)}^{\textbf{S}}\langle m,\vec p_3;m,\vec p_4|m,\vec p_1;m,\vec p_2\>^{\textbf{S}}_{(ab)}
&=\mathcal{N}_2\, T^{ab,cd}_{\textbf{S}}\times(2\pi)^{2}\delta^{(2)}(p_1+p_2-p_3-p_4),\\
{}_{[cd]}^{\textbf{A}}\langle m,\vec p_3;m,\vec p_4|m,\vec p_1;m,\vec p_2\>^{\textbf{A}}_{[ab]}
&=\mathcal{N}_2\, T^{ab,cd}_{\textbf{A}}\times(2\pi)^{2}\delta^{(2)}(p_1+p_2-p_3-p_4).
\end{align}
Notice that the normalization condition for the trivial representation is exactly the one used in the main text, see \eqref{eq:normalization_2PS_2d}.

 Taking into account \eqref{eq:global_decomposition} alternatively to \eqref{eq:scattering_amplitudes_ZZ} we can rewrite the full scattering amplitude in terms of independent scattering amplitudes $S_{\bullet}(s)$, $S_{\textbf{S}}(s)$ and $S_{\textbf{A}}(s)$, as
\begin{align}
\nn
{}_{cd}\< m,\vec p_3;m,\vec p_4|S|m,\vec p_1;m,\vec p_2\>_{ab} =&
(2\pi)^{2}\delta^{(2)}(p_1+p_2-p_3-p_4)\times\\
&\big(S_{\bullet}(s)T^{ab,cd}_{\bullet}+S_{\textbf{S}}(s)T^{ab,cd}_{\textbf{S}}+S_{\textbf{A}}(s)T^{ab,cd}_{\textbf{A}}\big),
\label{eq:scattering_amplitudes}
\end{align}
where the tensor structures associated to the three irreducible representations are defined as
\begin{equation}
T^{ab,cd}_{\bullet}\equiv \frac{1}{N}\delta_{ab}\delta_{cd},\quad
T^{ab,cd}_{\textbf{S}}\equiv \frac{\delta_{ac}\delta_{bd}+\delta_{ad}\delta_{bc}}{2}-\frac{1}{N}\delta_{ab}\delta_{cd},\quad
T^{ab,cd}_{\textbf{A}}\equiv \frac{\delta_{ac}\delta_{bd}-\delta_{ad}\delta_{bc}}{2}.
\end{equation}
The relation between two sets of amplitudes $\sigma_1$, $\sigma_2$, $\sigma_3$ and $S_{\bullet}$, $S_{\textbf{S}}$, $S_{\textbf{A}}$ simply reads as
\begin{align}
\nn
S_{\bullet}(s) &=\sigma_2(s)+\sigma_3(s)+N \sigma_1(s),\\
\label{eq:definitions}
S_{\textbf{S}}(s) &=\sigma_2(s)+\sigma_3(s),\\
\nn
S_{\textbf{A}}(s) &=\sigma_2(s)-\sigma_3(s),
\end{align}
In section 3.2 of \cite{Karateev:2019ymz} it was shown that using the states in the irreducible representation of the $O(N)$ one can formulate the unitarity constraints in the simple form. For the trivial representation we have
\begin{equation}
\label{eq:unitarity_t}
\begin{pmatrix}
1 & \mathcal{N}_2^{\,-1}{\mathcal{S}}_{\bullet}^*(s)
& \mathcal{N}_2^{\,-1/2}\, \mathcal{F}_{2,0}^{*\Theta}(s)\\
\mathcal{N}_2^{\,-1}{\mathcal{S}}_{\bullet}(s) & 1
&  \mathcal{N}_2^{\,-1/2}\, \mathcal{F}_{2,0}^{\Theta}(s)\\
\mathcal{N}_2^{\,-1/2}\, \mathcal{F}_{2,0}^{\Theta}(s)
&  \mathcal{N}_2^{\,-1/2}\, \mathcal{F}_{2,0}^{*\Theta}(s) &
2\pi\rho_\Theta(s)
\end{pmatrix}\succeq 0,
\end{equation}
where the form factor is defined as
\begin{align}
\label{eq:form_factor_trivial}
\mathcal{F}_{2,0}^{\Theta}(s) &\equiv 
\<0|\Theta(0)|m,\vec p_1;m,\vec p_2\>^{\bullet}\\
&=\sqrt{N}\;\<0|\Theta(0)|m,\vec p_1;m,\vec p_2\>_{11}.
\nn
\end{align}
For the symmetric and antisymmetric representations we have instead
\begin{equation}
\label{eq:unitarity_sa}
\begin{pmatrix}
1 & \mathcal{N}_2^{\,-1}{\mathcal{S}}_{\textbf{S}}^*(s)\\
\mathcal{N}_2^{\,-1}{\mathcal{S}}_{\textbf{S}}(s) & 1
\end{pmatrix}\succeq 0,\qquad
\begin{pmatrix}
1 & \mathcal{N}_2^{\,-1}{\mathcal{S}}_{\textbf{A}}^*(s)\\
\mathcal{N}_2^{\,-1}{\mathcal{S}}_{\textbf{A}}(s) & 1
\end{pmatrix}\succeq 0.
\end{equation}
The crossing equations \eqref{eq:crossing} in the new basis read as
\begin{equation}
\label{eq:crossing_ON}
\begin{pmatrix}
{\mathcal{S}}_{\bullet}(s) \\
{\mathcal{S}}_{\textbf{S}}(s)\\
{\mathcal{S}}_{\textbf{A}}(s)
\end{pmatrix}=
\begin{pmatrix}
\frac{1}{N}  && \frac{1}{2}-\frac{1}{N}+\frac{N}{2} && \frac{1}{2}-\frac{N}{2} \\
\frac{1}{N}  && \frac{1}{2}-\frac{1}{N} &&  \frac{1}{2}\\
-\frac{1}{N}  && \frac{1}{2}+\frac{1}{N} && \frac{1}{2}
\end{pmatrix}
\begin{pmatrix}
{\mathcal{S}}_{\bullet}(4m^2-s) \\
{\mathcal{S}}_{\textbf{S}}(4m^2-s)\\
{\mathcal{S}}_{\textbf{A}}(4m^2-s)
\end{pmatrix}.
\end{equation}

Let us now consider the 2d $O(N)$ model with $\phi^4$ potential.
In the large $N$ limit $N\rightarrow \infty$ limit using perturbation theory it is straightforward to show that
\begin{equation}
\sigma_i(s) = \frac{\bar\sigma_i(s)}{N} + O(N^{-2}),\quad
i=1,2,3,
\end{equation}
where $\bar\sigma_i(s)$ is the finite part in the large $N$ limit. Using \eqref{eq:definitions} we conclude that
\begin{equation}
S_{\bullet}(s) = \bar \sigma_1(s),\qquad
NS_{\textbf{S}}(s) =\bar\sigma_2(s)+\bar\sigma_3(s),\qquad
NS_{\textbf{A}}(s) =\bar\sigma_2(s)-\bar\sigma_3(s).
\end{equation}
Using these we can read off from \eqref{eq:crossing_ON} the crossing equation for the trivial scattering amplitude. It reads
\begin{equation}
S_{\bullet}(s) = \bar \sigma_1(s) = \bar \sigma_3(4m^2-3).
\end{equation}
Clearly this crossing equation does not close if we consider only the trivial scattering amplitude.

\section{Perturbative Computations}
\label{app:analytic}

In this appendix, we will detail various analytic computations of the form factors and scattering amplitudes in solvable limits (large $N$, non-relativistic, and perturbative $\lambda$) that we use throughout the paper.   

\subsection{Feynman Diagrams}
\label{app:Feynman}

\begin{figure}
\centering
\begin{subfigure}
 \centering
\includegraphics[width=0.2\textwidth]{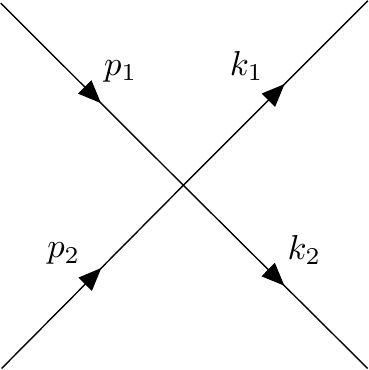}	
\end{subfigure}\qquad\qquad\qquad
\begin{subfigure}
 \centering
\includegraphics[width=0.3\textwidth]{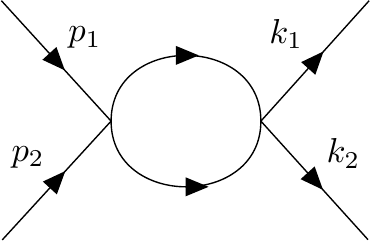}	
\end{subfigure}
\caption{Feynman diagrams for the 2-to-2 $S$-matrix up to one loop (plus crossed diagrams).}
\label{fig:feynmanSmatrix}
\end{figure}

\subsubsection{$\phi^4$ theory}
\begin{figure}
\centering
\begin{subfigure}
 \centering
\includegraphics[width=0.2\textwidth]{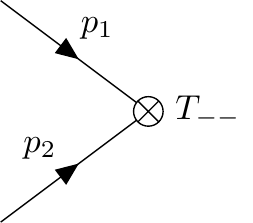}	
\end{subfigure}\quad\quad
\begin{subfigure}
\centering
\includegraphics[width=0.3\textwidth]{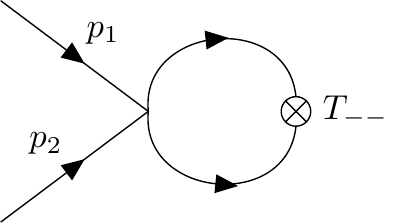}
\end{subfigure}

\begin{subfigure}
\centering
\includegraphics[width=0.4\textwidth]{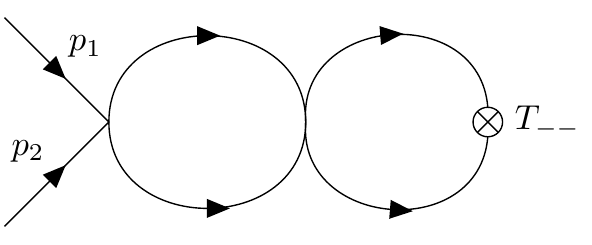}
\end{subfigure}\quad\quad
\begin{subfigure}
\centering
\includegraphics[width=0.2\textwidth]{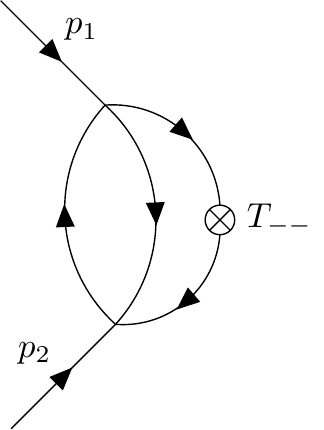}
\end{subfigure}
\caption{Feynman diagrams for the two-particle form factor of $\Theta$ up to two loops.}
\label{fig:feynmanFF}
\end{figure}

\begin{figure}
\centering
\begin{subfigure}
 \centering
\includegraphics[width=0.3\textwidth]{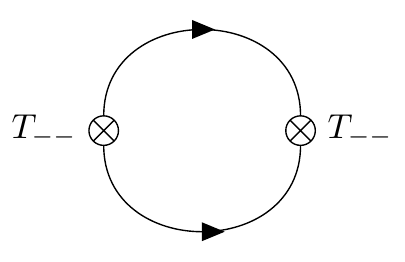}	
\end{subfigure}\qquad\qquad
\begin{subfigure}
 \centering
\includegraphics[width=0.45\textwidth]{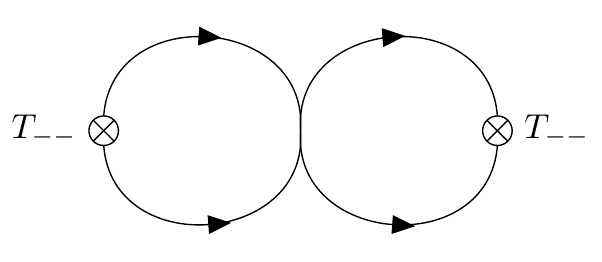}	
\end{subfigure}
\caption{Feynman diagrams for the $T_{--}$ two-point function up to two loops.}
\label{fig:feynmanTheta2pt}
\end{figure}

We begin with the form factors and amplitudes in a loop expansion, in powers of the coupling $\lambda$.  The leading order $\CO(\lambda^0)$ free theory expressions are
\begin{equation}
\hat{{\cal S}} = 1 , \quad
\CF_{2,0}^\Theta = -2m^2 ,  \quad
\pi \rho_\Theta =  2 m^4 \omega^2 \theta(s-4m^2),
\end{equation}
where $\omega \equiv \frac{1}{2\sqrt{s(s-4m^2)}} = {\cal N}_2^{-1} $.

To compute the form factors and spectral densities of $\Theta$, it is in general easier to compute those of $T_{--}$ first and then use the Ward identity than it is to compute those of $\Theta$ directly.  The reason is that $T_{--}$ is simply $(\partial_- \phi)^2$, independent of the interaction and mass terms, and so involves fewer Feynman diagrams.  At tree-level,
\begin{equation}
\< m^2, p_1; m^2, p_2 | T_{--}(0)\> = 2 p_{1-} p_{2-},
\end{equation}
The Ward identity implies
\begin{equation}
\< m^2, p_1; m^2, p_2 | \Theta(0)\> = -\frac{s}{p_-^2} \< m^2, p_1; m^2, p_2 | T_{--}(0) \>
\ee
where 
\begin{equation}
s  = (p_1 + p_2)^2 = m^2 \frac{p_-^2}{p_{1-} p_{2-}}, \qquad p_- \equiv p_{1-} + p_{2-},
\end{equation}
so at $\CO(\lambda^0)$ we obtain
\begin{equation}
\< m^2, p_1; m^2, p_2 | \Theta(0) \> = - 2m^2,
\end{equation}
as claimed, and as can easily be verified by a direct computation with $\Theta$.

At the next order, $\CO(\lambda)$, the S-matrix is given by a tree diagram, the form factor involves an one-loop computation, and the spectral density involves a two-loop diagram, as shown in the corresponding diagrams in figure \ref{fig:feynmanSmatrix},
\ref{fig:feynmanFF}, and \ref{fig:feynmanTheta2pt}.  The S-matrix is simply
\begin{equation}
\hat{{\cal S}} = 1 - i \lambda \omega^2. 
\end{equation}
 The form factor one-loop diagram (top right diagram in figure \ref{fig:feynmanFF}) can be computed by standard methods, 
\begin{equation}
\< m^2, p_1; m^2 , p_2| T_{--}(0)\> = -\frac{\lambda}{4\pi} \int_0^1 dx \frac{x(1-x) p_-^2}{m_0^2 - x(1-x)s}.
\end{equation}
The integral over the Feynman parameter $x$ can be done in closed form to obtain the expression given in equation (\ref{eq:FF_pert}), which for reference we write here as
\begin{equation}
\< m^2, p_1; m^2 , p_2| \Theta(0)\> = -2m^2 + \frac{\lambda}{4\pi}  \Delta(s)  + \CO(\lambda^2), \quad \Delta(m^2 x) \equiv-1+\lim_{\epsilon\rightarrow 0^+}
\frac{4\text{ArcTan}\left(\frac{\sqrt{x}}{\sqrt{4-x-i\epsilon}}\right)}{\sqrt{x(4-x-i\epsilon)}} .
\end{equation}
We have used $m_0 = m + \CO(\lambda^2)$, so $m$ and $m_0$ are interchangeable at this order.

The $\CO(\lambda)$ (i.e. two-loop) diagram (second diagram in figure \ref{fig:feynmanTheta2pt}) for the $T_{--}$ time-ordered two-point function factors into a product of two one-loop diagrams. The Ward identity can again be used to obtain the correlator with $T_{--}$s replaced by $\Theta$s, so by evaluating a couple of one-loop diagrams we obtain
\begin{small}
\begin{equation}
\pi \rho_\Theta(s) = \textrm{Re} \int d^2 x e^{- i p \cdot x} \< {\rm vac} | \Theta(x) \Theta(0) | {\rm vac}\>_T = \theta(s-4m^2) \left[ 2 m^4 \omega^2 + \frac{\lambda}{(4\pi)^2}  \textrm{Im}( \Delta^2 (s) ) + \CO(\lambda^2) \right] .
\end{equation}
\end{small}
At $s>4m^2$, the result for $\rho_\Theta$ can be written a bit more explicitly with the following expressions for the real and imaginary parts of $\Delta(s)$: 
\begin{equation}
\Delta(s) \stackrel{s>4m^2}{=} 
- \left[ 4 m^2\frac{{\rm ArcCosh} \left( \sqrt{\frac{s}{4m^2}} \right)}{\sqrt{s(s-4m^2)}}  \right]  - 4 i \pi m^2 \omega^2 . 
\end{equation}

One can also perform these computations directly with $\Theta$; in that case, it is crucial to include a subtle contribution $\propto \lambda \phi^2$ in the definition of $\Theta$ itself:
\begin{equation}
\Theta = m^2 \phi^2 + \frac{\lambda}{12} \phi^4 + \frac{\lambda}{8 \pi}  \phi^2,
\end{equation}
see e.g. \cite{Anand:2017yij}  for details.\footnote{One way to ``discover'' the  contribution  $\frac{\lambda}{8\pi} \phi^2$ to $\Theta$ is that   the relation $\CF_{2,0}^{\Theta}(0)=-2m^2$ is not satisfied at one-loop if it is not included.}

At the next order, $\CO(\lambda^2)$, the perturbative diagrams for $\CF_{2,0}^\Theta$ and $\rho_\Theta$ become more difficult to evaluate, involving a two-loop and three-loop computation, respectively.  Here we will only derive the $\CO(\lambda^2)$ contribution to $\CF_{2,0}^\Theta$.  As a check, in the next subsection we will rederive the $\CO(\lambda^2)$ contribution to $\CF^\Theta_{2,0}$ using dispersion relations.   Since particle production is kinematically forbidden for $s<16m^2$, we can actually obtain the three-loop spectral density in this regime from the two-loop form factor.   

First, the $\CO(\lambda^2)$ contribution to the S-matrix involves only an one-loop diagram (second diagram in figure \ref{fig:feynmanSmatrix})
 that can be easily evaluated:
\begin{equation}
{\cal T} = -\lambda + \frac{\lambda^2}{8 \pi m^2} \left( 1+ 4 \pi i m^2 \omega^2 \right) + \CO(\lambda^3).
\end{equation}
To compute the $\CO(\lambda^2)$ correction to $\CF_{2,0}^\Theta$, we again compute $\< m^2, p_1; m^2, p_2 | T_{--}(0)\>$ and use the Ward identity.  There are two two-loop diagrams that must be evaluated, as shown in Fig. \ref{fig:feynmanFF}. The first is a simple product of two one-loop diagrams, and is easily evaluated to be
\begin{equation}
F_{2,0}^{\Theta} \supset - \frac{\lambda^2}{2(4\pi)^2 m^2} \Delta(s) (\Delta(s)+1) .
\end{equation}
 The second two-loop diagram in Fig. \ref{fig:feynmanFF} involves the integral\footnote{See e.g. (10.57) in \cite{Peskin:1995ev}, which is easily generalized to the diagram we are considering.}
 \begin{small}
\begin{equation}
{\cal I} \equiv \int_0^1 dx \int_0^1 dy \int_0^1 dw \int \frac{d^2 k}{(2\pi)^2} \frac{(1-w)k_-(k+p)_-}{(w[x(1-x)(k+p_1)^2] + (1-w) [k^2 + 2 y k\cdot p+ y p^2] + m^2 )^3} ,
\end{equation}
\end{small}
(where $p= p_1+p_2$) plus a symmetric contribution with $p_1 \leftrightarrow p_2$. 
With some effort, these integrals can be evaluated and massaged into the closed form result in equation (\ref{eq:FF_pert}). In (\ref{eq:FF_pert}),  we have also had to adjust for an $\CO(\lambda^2)$ 
wavefunction renormalization \cite{Serone:2018gjo},\footnote{The wavefunction renormalization factor is given by $b_2^{(1)}$ from Table 8 of \cite{Serone:2018gjo}; we have used the fact that their numeric value for $b_2^{(1)}$ is equal to $\frac{9}{2\pi^2} -\frac{3}{8}$.}
\begin{equation}
 Z^{-1} = 1 - \left( \frac{\lambda}{4! m_0^2} \right)^2 \left( \frac{9}{2\pi^2} - \frac{3}{8} \right) + \CO(\lambda^2),
\end{equation}
after which the tree-level contribution to $\< m^2, p_1; m^2, p_2 | \Theta(0) \>$ becomes
\begin{equation}
\< m^2, p_1; m^2, p_2 | \Theta(0) \> \supset -2 m^2 Z^{-1} = - 2 m^2 \left( 1 - \left( \frac{\lambda}{4! m^2} \right)^2 \left( \frac{9}{2\pi^2} - \frac{3}{8} \right) + \CO(\lambda^2) \right).
\end{equation}
This wavefunction renormalization contribution has the effect of canceling out the $s=0$ contribution from the other two-loop diagrams, so that the Ward identity $\CF_{2,0}^\Theta(0) = -2 m^2$ is preserved.

\subsubsection{2d $O(N)$ model in the large $N$ limit}

The S-matrix, form factor $\CF_{2,0}^\Theta$, and spectral density $\rho_\Theta$ in the $O(N)$ theory at large $N$ simply involve diagrams we have just computed, together with a standard resummation of higher loop diagrams that factorize and form a geometric series.

  The $\Theta$ form factor is, in units with $m=1$,
  \be
 \CF_{2,0}^{\Theta}(s) = -2 \left( 1- \frac{\lambda \Delta(s)}{8 \pi + \lambda (1+\Delta(s))}\right),
  \ee
  where $\Delta(s)$ is the function given in (\ref{eq:DeltaDef}).  The S-matrix is simplest in the rapidity variable $\theta$:
  \be
  S = - \frac{(- i \theta + \pi) \lambda  \csch(\theta) + 8 i \pi}{(i \theta+\pi) \lambda \csch(\theta) - 8 i \pi}.
  \ee
  Finally, the time-ordered two-point function $\mathbf{\Delta}_\Theta(s)$ is 
  \be
  \mathbf{\Delta}_\Theta(s) = 4i\frac{\frac{s}{6} -\Delta(s) +\frac{\lambda}{8\pi}  \left(\frac{s}{6} + \Delta(s)  \left(\frac{s}{6}-1\right)\right)}{8 \pi + \lambda(\Delta(s) +1) },
   \ee
   and the spectral density $\rho_{\Theta}(s)$ can be obtained either by taking the real part of $\mathbf{\Delta}_\Theta(s)$ or by using the form factor together with  the fact that the large $N$ limit theory saturates the inequality (\ref{eq:SD_FF}).

\subsection{Dispersion Relations}

As a check of our previous two-loop formulas for the $\phi^4$ model, we will see how to rederive the one- and two-loop contributions using unitarity, Watson's equation, and dispersion relations.  

We begin with the S-matrix,
\be
\hat{S} = 1 + i \omega^2 \CT
\ee
 By definition, up to $\CO(\lambda)$, it is
\be
\CT=-\lambda + \CO(\lambda^2) \qquad \omega^2 = \frac{1}{2\sqrt{s(s-4m^2)}}
\ee
Let us also divide $\CT$ into real and imaginary parts as follows
\be
\CT = \CT_R + i \CT_I
\ee
From on-shell unitarity $S S^*=1$, we infer that at $s>4$,
\be
\CT_I  = \lambda^2\frac{\omega^2}{2} + \CO(\lambda^3)
\ee
Then, we can reconstruct $\CT$ at $\CO(\lambda^2)$ from its imaginary part using dispersion relations:
\be
\CT(s) = \CT_\infty -\frac{1}{\pi} \int_{4m^2}^\infty d\mu^2 \CT_I(\mu^2) \left( \frac{1}{s-\mu^2} - \frac{1}{s-(4m^2-\mu^2)} \right)  = \CT_{\infty}+ \frac{\lambda ^2}{4 \sqrt{(4m^2-s) s}} .
 \ee
It is easy to see that the imaginary part of $\CT(s)$ is indeed $\CT_I(s)$.  The  constant ``subtraction'' piece  $\CT_\infty$ depends on the definition of the theory and cannot be determined by dispersion relations.  If we define the theory to have a bare quartic coupling ${\cal L} \supset - \frac{\lambda }{4!} \phi^4$ without additional counterterms, then a one-loop computation shows $\CT_\infty = -\lambda + \frac{\lambda^2}{8\pi m^2} + \CO(\lambda^3)$. 

Next, we apply Watson's equation to obtain the form factor for $\Theta$. In the rest of this appendix, for notational convenience, we will denote $\mathcal{F}^\Theta_{2,0}$ simply as $\CF$. At $\CO(\lambda^0)$, we have $\CF=-2m^2$.  Expanding $\CF$ in powers of $\lambda$,\footnote{Note that the subscripts in $\CF$ in equation (\ref{eq:FFExpansion}) have different meanings from those in other parts of this paper.}
\be\label{eq:FFExpansion}
\CF = -2m^2 \left(1+ \frac{\lambda}{m^2} ( \CF_{1,R} + i \CF_{1,I}) + \frac{\lambda^2}{m^4} (\CF_{2,R} + i \CF_{2,I}) + \dots \right)
\ee
and  imposing
\be
\frac{\CF(s)}{\CF^*(s)} = \hat{S}(s),
\ee
we immediately find
\be
\CF_{1,I} = -\frac{1}{2} m^2 \omega^2 = -\frac{m^2}{4 \sqrt{s(s-4m^2)}}
\ee
Applying dispersion relations, we have
\be
\CF_1(s) = \CF_{1,\infty} - \frac{1}{\pi} \int_4^\infty d\mu^2 \frac{\CF_{1,I}(\mu^2)}{s-\mu^2}  = \CF_{1,\infty} -\frac{m^2\sec ^{-1}\left(\frac{2m}{\sqrt{4m^2-s}}\right)}{2 \pi  \sqrt{(4m^2-s) s}}
 \ee
 where  $\CF_{1,\infty}$ is another constant subtraction.  We can fix its value by demanding that $\CF(0)=-2m^2$, which implies
 \be
 \CF_{1,\infty} = \frac{1}{8\pi}
 \ee
 Putting this together, we obtain
 \be
 \CF_1(s) = - \frac{1}{8\pi} \Delta(s) .
 \ee
At the next order, using our expression for $\CT$ up to $\CO(\lambda^2)$, we find from Watson's equation that
\be
\label{eq:ImF2}
\CF_{2,I}(s) =-\frac{m^2\text{csch}^{-1}\left(\frac{2m}{\sqrt{s-4m^2}}\right)}{8 \pi  (s-4m^2) s} - \frac{\pi}{32} m^2\delta(s-4m^2) .
    \ee
    There is a subtle $\delta(s-4m^2)$ contribution here that arises from taking the difference between $\frac{1}{s-4m^2}$ and $\left( \frac{1}{s^-4m^2}\right)^*$, which differ by a $\delta$ function at $s=4m^2$ due to the change in $i \epsilon$ prescription under complex conjugation.  The clearest way to see this difficult term is by studying the non-relativistic limit $s \sim 4m^2$ directly, as we will do in the next subsection.

Finally, we can reconstruct the full form factor at this order:
\be
\CF_2(s) = \CF_{2,\infty} - \frac{1}{\pi} \int_{4m^2}^\infty d\mu^2 \frac{\CF_{2,I}(\mu^2)}{s-\mu^2} =- \frac{1}{(8\pi)^2} \left( \frac{\pi^2 s}{8(s-4m^2)} -  \Delta(s)(\Delta(s)/2+1) \right),
 \ee
 which agrees with the result (\ref{eq:FF_pert}).
We again fixed the subtraction term $\CF_{2,\infty}$ by demanding that $\CF(0)=-2m^2$.

\subsection{Nonrelativistic Limit}

Scattering of two particles near the threshold $s = 4m^2$ is simply a one-dimensional non-relativistic quantum mechanics problem that can be solved, as we review briefly.  The interaction $\lambda \phi^4$ is a $\delta$ function potential in position space. We take the scattering wavefunction to be
\be
\psi(x) = \left\{ \begin{array}{cc} e^{i k x} + S e^{- i k x}, & x < 0 \\ S e^{i k x} + e^{- i k x}, & x>0 \end{array} \right. ,
\label{eq:nonrelPsi}
\ee
which is even as a function of $x$ since the two particles are identical.  The Schrodinger equation is
\be
- \frac{\psi''(x)}{2(m/2)} + \frac{\lambda}{8}  \delta(x) \psi(x) = E \psi(x)
\ee
where $E= \frac{s-4m^2}{4m} = \frac{k^2}{m}$. 

  Integrating the Schrodinger equation around $x=0$, we find
\be\label{eq:nonrelaS}
S = \frac{16 k-i m \lambda}{16  k+i m \lambda} \approx   - \frac{\lambda+ 8 i \theta}{\lambda- 8i \theta}
\ee
where we have written the result in terms of rapidity $\theta$ and taken the limit $\theta \rightarrow 0$ with $\lambda/\theta$ fixed. 
This is regular at small $\theta$, but if we first take a small $\lambda$ limit then each power in $\lambda$ is individually singular:
\be
S(\theta) \stackrel{\theta \rightarrow 0}{\approx}1-\frac{i \lambda }{4
   \theta }-\frac{\lambda ^2}{32 \theta ^2} + \frac{i \lambda ^3}{256 \theta ^3} + \dots
   \ee
   where we have kept only the most singular terms at each order.    
So the small $\lambda$ limit and the small $\theta$ limit do not commute, and in fact perturbative loop computations, which are an expansion in powers of $\lambda$, should not be trusted below about $\theta \lesssim \frac{\lambda}{4}$.

From the scattering wavefunction, we can also extract the form factor in this limit. Each individual particle has energy $m+E/2$, so the time-dependent wavefunction is
\be
\psi_2(x_1, x_2, t_1, t_2) = e^{i (m+\frac{E}{2}) (t_1+t_2)} \psi(\frac{x_1-x_2}{2})
\ee
where $\psi(x)$ is from eq. (\ref{eq:nonrelPsi}). We will take $m=1$.  In second quantization, $\psi_2$ is
\be
\psi_2(x_1, x_2, t_1 , t_2) = \< \phi(x_1, t_1) \phi(x_2, t_2) | p_1, p_2\>
\ee
where $|p_1, p_2\>$ is the two-particle state, with momentum $p_1 = -p_2 = k/2$ since we are in the rest frame.  Then, we can easily calculate the overlap with $T_{--}$ by taking
\be
\< T_{--}(0) | p_1, p_2\> = \lim_{x_i \rightarrow 0, t_i \rightarrow 0} (\partial_{t_1} - \partial_{x_1})(\partial_{t_2}-\partial_{x_2}) \psi_2(x_1, x_2 , t_1, t_2) . 
\ee
In the nonrelativistic limit, both $E = k^2$ and $k$ go to zero, and we obtain
\be
\< T_{--}(0) | p_1, p_2\> = -(1+S).
\ee
with $S$ given in equation (\ref{eq:nonrelaS}). This result has the correct phase according to Watson's equation, since one can easily check that $\frac{1+S}{1+S^*} = S$.  Expanding in small $\lambda$ we have
\be
-\frac{1 }{2}\< T_{--}(0) | p_1, p_2\> =1-\frac{i
   \lambda }{8 \theta }-\frac{\lambda ^2}{64 \theta ^2} + \dots,
   \ee
which agrees with the perturbative  result (\ref{eq:FF_pert}) if we expand (\ref{eq:FF_pert}) in small $\theta$.  From this expression, we see that $\CF_{2,0}^\Theta$ at $\CO(\lambda^2)$ has a pole $\sim \frac{\lambda^2}{32 (s-4)}$, which implies that the imaginary part of $\CF_{2,0}^\Theta$ at $\CO(\lambda^2)$ contains a $\delta$ function of the form
   \be
   \textrm{Im}(\CF_{2,0}^\Theta) \supset - \lambda^2 \frac{\pi}{32(s-4)}
   \ee
   as claimed in eq (\ref{eq:ImF2}).

\section{Sinh-Gordon Form Factors and $C$-function}
\label{app:sinh-gordonFF}
In this appendix, we provides some details about the four-particle
form factor of the trace of the stress-tensor $\Theta$ and the computation
of the spectral density in section 5 in the sinh-Gordon/staircase model. The
result of the $2n$-particle form factor for $\Theta$ is given in
\cite{Fring:1992pt} (the form factors with an odd number of particles vanish for
$\Theta$). In terms of the minimal form factor 
\begin{equation}
\CF_{\text{min}}\left(\theta\right)=\cn\exp\left(8\int_{0}^{\infty}\frac{dx}{x}\frac{\sinh\left(\frac{x\gamma}{2\pi}\right)\sinh\left(\frac{x(\pi-\gamma)}{2\pi}\right)\sinh\left(\frac{x}{2}\right)}{\sinh^{2}(x)}\sin^{2}\left(\frac{x(i\pi-\theta)}{2\pi}\right)\right),
\end{equation}
with the normalization constant 
\begin{equation}
\mathcal{N}=\exp\left[-4\int_{0}^{\infty}\frac{dx}{x}\frac{\sinh\left(\frac{x\gamma}{2\pi}\right)\sinh\left(\frac{x}{2}\left(1-\frac{\gamma}{\pi}\right)\right)\sinh\frac{x}{2}}{\sinh^{2}x}\right],
\end{equation}
the 2-particle form factor in equation (\ref{eq:form_factor_sinh-Gordon}) in our convention is given by 
\begin{equation}
\CF_{2,0}^{\Theta}\left(\theta\right)=-2m^{2}\frac{\CF_{\text{min}}\left(\theta\right)}{\cn}.
\end{equation}
And the expression for the four-particle form factor is \begin{equation}
\CF_{4,0}^{\Theta}(\theta_1,\theta_2,\theta_3,\theta_4)=\frac{8\pi m^{2}\sin\gamma}{\cn^{2}}\sigma_{1}^{(4)}\sigma_{2}^{\left(4\right)}\sigma_{3}^{(4)}\prod_{0<i<j\le4}\frac{\CF_{\text{min}}\left(\theta_{ij}\right)}{x_{i}+x_{j}},\label{eq:4particleFF}
\end{equation}
where $\theta_{ij}=\theta_i-\theta_j$,  $x_{i}=e^{\theta_{i}}$, and $\sigma_{k}^{\left(4\right)}$s
are degree $k$ symmetric polynomials of $x_{i}$. Specifically, we
have 
\begin{align}
\sigma_{1}^{\left(4\right)} & =x_{1}+x_{2}+x_{3}+x_{4},\nonumber \\
\sigma_{2}^{\left(4\right)} & =x_{1}x_{2}+x_{1}x_{3}+x_{1}x_{4}+x_{2}x_{3}+x_{2}x_{4}+x_{3}x_{4},\\
\sigma_{3}^{\left(4\right)} & =x_{1}x_{2}x_{3}+x_{1}x_{2}x_{4}+x_{1}x_{3}x_{4}+x_{2}x_{3}x_{4}.\nonumber 
\end{align}
As mentioned in \cite{Fring:1992pt}, to numerically evaluate the form factors,
it is easier to use the following expression for $\CF_{\text{min}}$:
\begin{align}
\nn
\CF_{\text{min}}\left(\theta\right) & =\cn I_N(\theta)\\
&\times\prod_{k=0}^{N-1}\left[\frac{\left(1+\left(\frac{\hat{\theta}/2\pi}{k+\frac{1}{2}}\right)^{2}\right)\left(1+\left(\frac{\hat{\theta}/2\pi}{k+\frac{3}{2}-\frac{\gamma}{2\pi}}\right)^{2}\right)\left(1+\left(\frac{\hat{\theta}/2\pi}{k+1+\frac{\gamma}{2\pi}}\right)^{2}\right)}{\left(1+\left(\frac{\hat{\theta}/2\pi}{k+\frac{3}{2}}\right)^{2}\right)\left(1+\left(\frac{\hat{\theta}/2\pi}{k+\frac{1}{2}+\frac{\gamma}{2\pi}}\right)^{2}\right)\left(1+\left(\frac{\hat{\theta}/2\pi}{k+1-\frac{\gamma}{2\pi}}\right)^{2}\right)}\right]^{k+1}\label{eq:FminExpression2}
\end{align}
where the integral $I_N(\theta)$ is defined as
\begin{multline}
I_N(\theta)\equiv
\exp\Bigg[8\int_{0}^{\infty}\frac{dx}{x}\frac{\sinh\left(\frac{x\gamma}{2\pi}\right)\sinh\left(\frac{x}{2}\left(1-\frac{\gamma}{\pi}\right)\right)\sinh\frac{x}{2}}{\sinh^{2}x}
\times\\
\left(N+1-Ne^{-2x}\right)e^{-2Nx}\sin^{2}\left(\frac{x\hat{\theta}}{2\pi}\right)\Bigg].
\end{multline}
Here $\hat{\theta}\equiv i\pi-\theta$. The integral in the exponent of $I_N(\theta)$ is
approaching 0 as one increases $N$. And for large $N$, the the contribution
from the integral is actually negligible. For example, for $N=1000$,
in the cases we considered in this paper, the integral is order $\co\left(10^{-8}\right)$.
In the actual computation for getting the spectral density, we simply
take large enough $N$ and discard the integral part in (\ref{eq:FminExpression2}).
We then use a rational function to fit the result for $\CF_{\text{min}}$
for each values of $\gamma$ (which only introduces an uncertainty of order $\CO(10^{-8}$)), in order for Mathematica to be able to evaluate the 4-particle form factor contribution to the spectral density quickly later on. 

The spectral density
is given exactly by 
\begin{equation}
\label{eq:spectral-SHG}
\rho_{\Theta}(s)=\rho_{\Theta,2}(s)\theta(s-4m^2)+\rho_{\Theta,4}(s) \theta(s-16m^2)
+\rho_{\Theta,6}(s) \theta(s-36m^2)+\ldots,
\end{equation}
where 
\begin{multline}
\rho_{\Theta,2n}\left(s\right)=\frac{1}{2\pi}\frac{1}{(2n)!}\int\frac{d\theta_{1}\ldots d\theta_{2n}}{(2\pi)^{2n}}\left|F_{2n,0}^{\Theta}\left(\theta_{1},\ldots,\theta_{2n}\right)\right|^{2}
\\
\delta\left(\sum_{i=1}^{2n}m\sinh\theta_{i}\right)\delta\left(\sum_{i=1}^{2n}m\cosh\theta_{i}-\sqrt{s}\right),
\end{multline}
In what follows we will consider only two- and four-particle contributions to the spectral density only. This means that $\rho_{\Theta}(s)$ remains exact up to $s=36m^{2}$ and then will start deviating from the exact answer due to six- and higher particle states.

The comparison of the sinh-Gordon spectral density (with two- and four-particle states) and the $\phi^{4}$ spectral density is given in figure \ref{fig:SGandPhi4Comparison_1}. They happen to be very similar in a wide range of values of $s$.
Using \eqref{eq:spectral-SHG} one can also compute the $C$-function. We show the contributions
to the change of the central charge $\Delta C=12\pi\int_{0}^{\infty}ds\frac{\rho_{\Theta}(s)}{s^{2}}$
from the two-particle and four-particle form factors in figure \ref{fig:SinhGordonDeltaC}. As expected we reproduce the value of the free boson $\Delta C=1$ very well (at least for $\Lambda\le 4$).  For small values of $\Lambda $, the free boson central charge is mostly given by the two-particle part of the spectral density. When $\Lambda$ increases, four- and then higher-particle contributions become important.

\begin{figure}
\begin{center}
\includegraphics[width=0.9\textwidth]{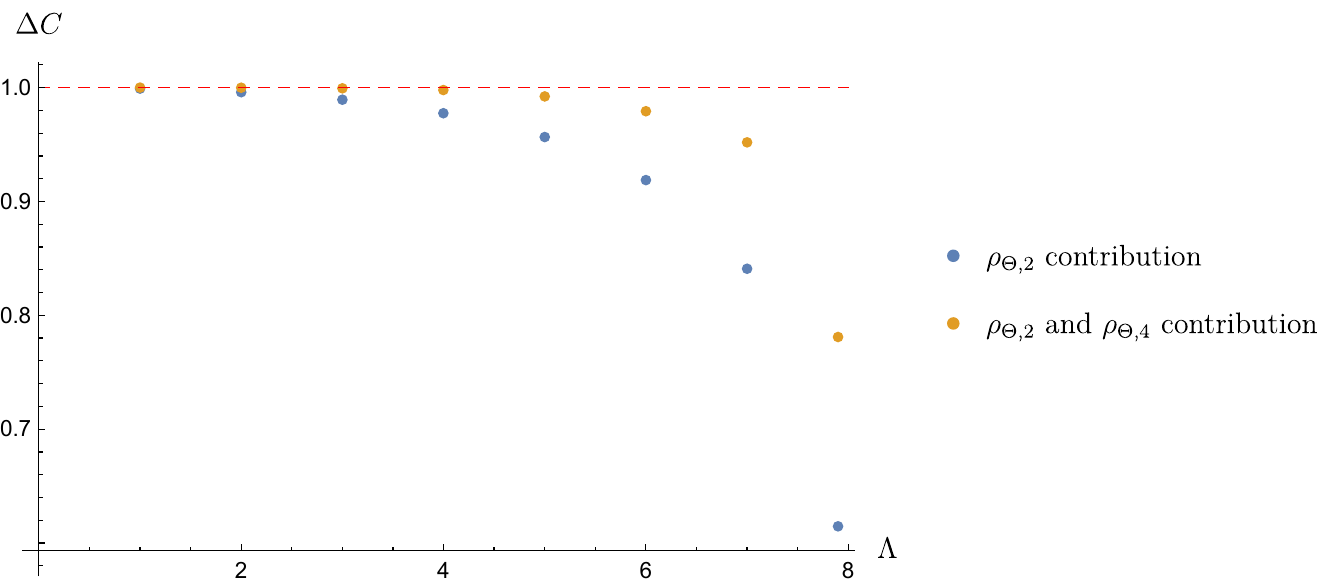}
\caption{Contributions to $\Delta C$ from the two-particle and four-particle form factors in the sinh-Gordon/staircase model for various values of the non-perturbative quartic coupling $\Lambda$. The red dashed line is $\Delta C=1$ for comparison.}
\label{fig:SinhGordonDeltaC}
\end{center}
\end{figure}

\bibliographystyle{JHEP}
\bibliography{refs}


\end{document}